\documentclass[useAMS,usenatbib]{mn2e}
\usepackage{color}
\usepackage{verbatim}
\usepackage{array} 
\usepackage{amsmath}
\usepackage{graphicx}
\usepackage{epstopdf}
\usepackage[labelfont=bf,labelsep=period,justification=raggedright,singlelinecheck=false,font=small]{caption}
\usepackage{wrapfig}
\usepackage{float}
\usepackage{amssymb}
\usepackage[normalem]{ulem}
\usepackage{hyperref}
\usepackage{undertilde}
\usepackage[T1]{fontenc}
\usepackage[utf8]{inputenc}

\usepackage[belowskip=-15pt,aboveskip=0pt]{caption}
 
\def\apj{ApJ}
\def\aap{A\& A}
\def\apjl{ApJL}
\def\apjs{ApJS}
\def\mnras{MNRAS}

\def\araa{ARA\& A}
\def\nat{Nature}
\def\prd{Phys. Rev. D}
\def\ssr{Space Sci. Rev.}

\def \jqsrt{J. Quant. Spec. Radiat. Transf.}
\def \physrep{Phys. Rep}

\newcommand\altaffilmark[1]{$^{#1}$}
\newcommand\altaffiltext[1]{$^{#1}$}

\newcommand{\msun}{{\rm M}_{\odot}}

\restylefloat{table}

\begin{document}
\defcitealias{Lack11}{L11}

\title[Cosmic Rays in FIRE-2]{Cosmic ray feedback in the FIRE simulations: constraining cosmic ray propagation with GeV gamma ray emission}

\author[T. K. Chan et al.]
  {T. K. ~Chan\altaffilmark{1} \thanks{Email: (TKC) tkc004@physics.ucsd.edu},
  D. ~Kere\v{s}\altaffilmark{1} \thanks{Email: (DK) dkeres@physics.ucsd.edu},
  P. F.~Hopkins\altaffilmark{2},
  E. ~Quataert\altaffilmark{3}, K.-Y. Su\altaffilmark{2},\newauthor
  C. C. ~Hayward\altaffilmark{4}, \& C.-A. Faucher-Gigu\`ere\altaffilmark{5} \\
  \altaffiltext{1}{Department of Physics, Center for Astrophysics and Space Sciences,University of California at San Diego,}\\ {9500 Gilman Drive, La Jolla, CA 92093, USA}\\
  \altaffiltext{2}{TAPIR, Mailcode 350-17, California Institute of Technology, Pasadena, CA 91125, USA} \\
  \altaffiltext{3}{Department of Astronomy and Theoretical Astrophysics Center, University of California Berkeley, Berkeley, CA 94720, USA}\\
  \altaffiltext{4}{Center for Computational Astrophysics, Flatiron Institute, 162 Fifth Avenue, New York, NY 10010, USA}\\
  \altaffiltext{5}{Department of Physics and Astronomy and CIERA, Northwestern University, 2145 Sheridan Road, Evanston, IL 60208, USA}
}

\maketitle

\begin{abstract}
We present the implementation and the first results of cosmic ray (CR) feedback in the Feedback In Realistic Environments (FIRE) simulations. We investigate CR feedback in non-cosmological simulations of dwarf, sub-$L\star$ starburst, and $L\star$ galaxies with different propagation models, including advection, isotropic and anisotropic diffusion, and streaming along field lines with different transport coefficients. We simulate CR diffusion and streaming simultaneously in galaxies with high resolution, using a two moment method. We forward-model and compare to observations of $\gamma$-ray emission from nearby and starburst galaxies. We reproduce the $\gamma$-ray observations of dwarf and $L\star$ galaxies with constant isotropic diffusion coefficient $\kappa \sim 3\times 10^{29}\,{\rm cm^{2}\,s^{-1}}$. Advection-only and streaming-only models produce order-of-magnitude too large $\gamma$-ray luminosities in dwarf and $L\star$ galaxies. We show that in models that match the $\gamma$-ray observations, most CRs escape low-gas-density galaxies (e.g.\ dwarfs) before significant collisional losses, while starburst galaxies are CR proton calorimeters. While adiabatic losses can be significant, they occur only after CRs escape galaxies, so they are only of secondary importance for $\gamma$-ray emissivities. Models where CRs are ``trapped'' in the star-forming disk have lower star formation efficiency, but these models are ruled out by $\gamma$-ray observations. For models with constant $\kappa$ that match the $\gamma$-ray observations, CRs form extended halos with scale heights of several kpc to several tens of kpc.
\end{abstract}


\label{firstpage}

\section{Introduction}
\label{introduction}

Cosmic rays (CRs) are charged particles moving with relativistic speeds, mainly generated through shock acceleration of supernova remnants (SNRs)  \citep{Bell04} (and possibly also in active galactic nuclei in massive galaxies). Unlike thermal energy, they can propagate through the galactic interstellar medium (ISM) rapidly via advection, diffusion and streaming \citep{Stro07,Zwei13,Gren15}, and transfer energy to gas via Coulombic and hadronic interactions \citep{Mann94,Enss07,Guo08}. Their energy density is comparable to thermal and magnetic energies in the solar neighborhood \citep{Ginz85,Boul90}, so CRs are believed to be dynamically important in galaxy evolution.

The impacts of CRs on galaxy evolution have been studied with analytic models \citep{Ipav75,Brei91,Brei93,Zira96,Socr08,Ever08,Dorf12,Mao18} and idealized and cosmological simulations (e.g. \citealt{Jube08,Uhli12,Boot13,Wien13,Hana13,Sale14,Sale14cos,Chen16,Simp16,Giri16,Pakm16,Sale16,Wien17,Rusz17,Buts18,Farb18,Jaco18,Giri18}). These studies show CRs can drive multiphase winds, reduce star formation rates in low mass galaxies, thicken gaseous disks, and modify the phase structure of the circum-galactic medium (CGM). It has also been suggested that CRs may play an important role in the galactic dynamo \citep{Park92,Hana09,Kulp11,Kulp15}. 

Despite its importance, the details of CR propagation are uncertain. The most popular CR propagation models are self confinement and extrinsic turbulence (\citealt{Zwei13}, and reference herein). In the former picture, CRs interact with a series of Alfven waves, which results in random scattering in pitch angles. The waves are then amplified via the streaming instability of CRs, increasing the scattering and trapping CRs in a background medium. These ``self confinement'' interactions effectively transfer energy from the CRs to thermal plasma. In the extrinsic turbulence model, CRs propagate through random magnetic field lines and are scattered by the background turbulent magnetic fields. This mechanism is especially important for high energy CRs, since there are too few high energy CRs amplifying the Alfven waves and the self-confinement mechanism fails \citep{Zwei13}. These two mechanisms confine and isotropise the CR distribution explaining the remarkably low CR anisotropy observed from the Earth (see, e.g. \citealt{Hill75}) and the long residence time ($>10^7\;{\rm yr}$) inferred from the ratios between stable primary and secondary nuclei \citep{Stro07}. Their long confinement time and small anisotropy imply that CRs have short mean free paths ($\sim {\rm pc}$) and their propagation can therefore be approximated by a random walk, so CRs can be well described as a diffusive fluid, obeying an advection diffusion equation (see, e.g. \citealt{Zwei17}, for arguments for the CR fluid theory).
 
Most of the studies of CR propagation have focused on an approximate picture of the Milky Way described by the flat halo diffusion model \citep{Ginz76}. This model consists of a cylindrical gaseous halo with a radius around 20 kpc and a height larger than 1 kpc, and a thinner yet more dense cylindrical internal disk with CR sources. CRs are assumed to diffuse isotropically (averaged over the scale of hundreds of pc) with a spatially constant but energy dependent diffusion coefficient, and ``escape'' through the halo boundaries to intergalactic space. Extensions of this model are commonly used in numerical CR propagation codes, e.g.\ {\small GALPROP}\footnote{ \url{https://galprop.stanford.edu/}}, which attempt to synthesize observational constraints on the MW gas density distribution, CR abundances and spectra, $\gamma$-ray and radio emission, and theoretical models for e.g. galactic winds and diffusive re-acceleration \citep{Stro98,Stro01}. These models are commonly used to constrain the ``effective'' isotropic-equivalent diffusion coefficient of CRs averaged over the whole MW disk \citep[e.g.][]{Stro07,Trot11}. However, there are still large uncertainties in the role of gas dynamics and small-scale gas density fluctuations (``clumpiness''), magnetic field geometries on small scales, the spatial and temporal distribution of CR sources, the size and mass distribution of the gaseous galactic halo, and the CR propagation model. To make progress, self-consistent modeling of galaxy evolution that includes CR propagation together with hydrodynamics or magneto-hydrodynamics is required.

In addition to the CR energy density and abundance of nuclei, high energy $\gamma$-ray emission can serve as an independent constraint on CR propagation \citep{Acke12diffusegamma,Stro00,Stro04}. High energy ($>{\rm GeV}$) CRs collide with nuclei in the interstellar medium (ISM) and produce pions, which decay into ${\rm GeV}$ $\gamma$-rays. Since pionic $\gamma$-ray luminosity is proportional to CR density and most of the energy density of the CRs is at energies around {\rm GeV} (from the direct CR observations, e.g. in \citealt{AMS02}), CR distribution and propagation can be constrained with high energy $\gamma$-ray observations.

Recently, $\gamma$-ray emission was observed from Local Group \citep{Abdo10LMC,Abdo10M31,Abdo10SMC} and starburst galaxies \citep{Acer09,Acci09,Abdo10SBG,Abda18}, which can be used as a probe of CRs beyond the solar system and the Milky Way (MW). \citet{Abdo10M31} found a correlation between $\gamma$-ray emission and star formation rate (SFR) with a steeper than linear drop at low SFRs ($E_\gamma \propto {\rm SFR}^{1.4\pm 0.3}$; summarized in \citealt{Acke12}).

To explain this correlation, \cite{Lack11} (hereafter \citetalias{Lack11}) constructed one-zone leaky box model of galaxies where a fixed fraction of SN energy is injected as CRs. They assume CRs escape with an energy dependent escape time and that the CR energy density and spectral distribution are in a steady state (the injections and losses are balanced).  Constrained with the observed CR abundances and the far infrared (FIR)-radio correlation \citep{Lack10}, the model was used to estimate pionic $\gamma$-ray luminosities of galaxies. They found that in order to explain the correlation between $\gamma$-ray emission and SFR, in starburst galaxies, most CR protons are required to lose their energy via collisions with the ISM (i.e. that starbursts are ``CR proton calorimeters'', as in the earlier calculations of \citealt{TQW}; see also \citealt{Abra12,Yoas13,Wang18}), while in dwarf galaxies, most of CR protons should escape. The main drivers of this effect are that SFR drops with decreasing gas surface density \citep{Kenn98}, and that lower gas densities enable CRs to escape before heavy pionic losses. Subsequent observational studies have reached the same conclusion regarding efficient escape in galaxies like the MW, Andromeda (M31), the Large and Small Magellanic Clouds (LMC, SMC; see, e.g. \citealt{Lope18}).

In this study, we investigate the impact of CRs on dwarf, sub-$L\star$ starburst, and $L_\star$ galaxies, using idealized simulations of galaxy evolution. We run galaxy simulations with both CR diffusion and streaming with high spatial resolution and diffusivity thanks to the newly developed two-moment method \citep[similar to][]{Jian18}. We also couple explicit CR transport and losses to an explicit, local stellar feedback model which time-resolves {\em individual} SNe explosions, as well as stellar mass-loss and radiative feedback, which together enable self-consistent generation of galactic winds and a turbulent, multi-phase ISM, critical for understanding CR transport and emission in that same ISM. Specifically, our CR implementation in the code {\small GIZMO} is coupled to the FIRE-2 (Feedback In Realistic Environments 2) algorithm for star formation and stellar feedback \citep{FIRE2,Hopk18sne}.\footnote{\url{http://fire.northwestern.edu/}} Cosmological simulations with these physics (without explicit CR transport) have been shown to successfully reproduce many observed galaxy properties including stellar masses \citep{FIRE2}, galactic winds \citep{Mura15,Angl17,Hafe18}, cored central dark matter profiles \citep{Onor15,Chan15,Wetz16,Fitt17}, the mass-metallicity relation \citep{Ma16} and spatial distribution of gas and metals within galaxies and the CGM \citep{Fauc15,Fauc16,Ma17,Mura17,Hafe17}, typical galaxy star formation rates and histories \citep{Spar17}, the Kennicutt-Schmidt law \citep{Orr18}, and galactic magnetic field structure \citep{Su18}. 

In this, the first paper in a series, we introduce our implementation of the CR propagation model (including isotropic and anisotropic diffusion and streaming), simulate galaxies with several CR propagation models, and focus on constraining the model using the observations of $\sim $GeV  $\gamma$-ray emission from galaxies (and compare our findings with previous theoretical studies). \S~\ref{method} and \S~\ref{simset} discuss numerical methods, simulated physics, and initial conditions. In \S~\ref{CRgalpro}, we investigate how CRs and their propagation influence galactic properties. In \S~\ref{gammaray} we calculate the $\gamma$-ray emission from CRs in our simulations and compare with observational data. In \S~\ref{discussion}, we compare our findings with the previous studies and analyze the relative importances of different CR energy gain and loss processes. We summarize our findings in \S~\ref{conclusion}. 

\section{Method}
\label{method}
\subsection{Simulation code}
\label{sec:simulations}

All the physics and numerical details in this study, except for CRs, follow the FIRE-2 version of the FIRE algorithms presented in detail in \citet{FIRE2}, so we only briefly review them here. Our simulations use the {\small GIZMO}\footnote{\url{http://www.tapir.caltech.edu/\~phopkins/Site/GIZMO}} code \citep{Hopk15} in its mesh-free Lagrangian finite mass (MFM) mode for (magneto)-hydrodynamics; extensive implementation details and tests of the MHD scheme are presented in \citet{Hopk16a,Hopk16MHD}. {\small GIZMO} uses an updated version of the PM+Tree algorithm from Gadget-3 \citep{Spri05} to calculate gravity and adopts fully conservative adaptive gravitational softening for gas \citep{Pric07}. Gas cooling is calculated with tabulated cooling rates from $T=10-10^{10}$\,K, including atomic, metal-line, and molecular cooling. While our simulations are idealized and do not include cosmological environments, we do include the present-day  ultraviolet background, from the \cite{Fauc09} model (accounting for local self-shielding). Stars form in locally self-gravitating, self-shielding, thermally Jeans-unstable gas\footnote{We assume the strong coupling limit between gas and CRs, so the effective sound speed in the virial parameter includes both thermal and CR pressure. See Appendix C in \citealt{FIRE2}.} at densities  $n_{\rm H} \geq 100\, {\rm cm^{-3}}$. Once formed, we calculate the energy, momentum, mass and metal return for each star according to the STARBURST99 stellar population synthesis model \citep{Leit99}, for a \citet{Krou02} IMF, accounting for SNe Types Ia \&\ II, O/B and AGB star mass-loss, and radiation (photo-electric and photo-ionization heating and radiation pressure with a five-band approximate radiation-hydrodynamics treatment). For details see \citet{FIRE2,Hopk18sne,hopkins:fire2.radiation}.

\subsection{Cosmic Rays}

The implementation of CR physics in {\small GIZMO} includes fully-anisotropic cosmic ray transport with streaming and advection/diffusion, CR cooling (hadronic and Compton, adiabatic, and streaming losses), injection in SNe shocks, and CR-gas coupling. The CRs are treated as an ultra-relativistic fluid (adiabatic index $\gamma_{\rm cr}=4/3$) in a ``single bin'' approximation.\footnote{One can think of this as evolving only the CR energy density at the energies $\gtrsim$\,GeV, which dominate the CR pressure, and approximating the CR energy spectrum as having a universal shape at all positions.} Integrating over the CR distribution function and spectrum, the usual ideal-MHD equations solved for gas density $\rho$, velocity ${\bf v}$, magnetic field ${\bf B}$, and specific energy $e$, are extended with the equation for the CR energy density $e_{\rm cr}$ \citep{McKe82}: 
\begin{align}
\label{eqset}
&\frac{\partial \rho}{\partial t}+\nabla\cdot \left ( \rho {\bf v} \right )=0,\nonumber\\
&\frac{\partial \rho{\bf v}}{\partial t}+\nabla\cdot \left ( \rho {\bf v}\otimes{\bf v}+P_T\mathbb{I}-{\bf B}\otimes{\bf B}  \right )=0,\nonumber\\
&\frac{\partial \rho e}{\partial t}+\nabla\cdot \left [ (\rho e+P_T){\bf v}-({\bf v}\cdot{\bf B}){\bf B}  \right ]\nonumber\\
&\ \ \ \ \ \ \ \ \ \ \ =P_{\rm cr}\nabla\cdot{\bf v}+\Gamma_{\rm st}+S_{\rm g}-\Gamma_{\rm g},\nonumber\\
&\frac{\partial {\bf B}}{\partial t}+\nabla\cdot \left ({\bf v}\otimes {\bf B}-{\bf B}\otimes {\bf v}  \right )=0,\nonumber\\
&\frac{\partial e_{\rm cr}}{\partial t} + \nabla \cdot {\bf F}_{\rm cr} = {\bf v}\cdot \nabla P_{\rm cr}- \Gamma_{\rm st}+ S_{\rm cr} - \Gamma_{\rm cr},
\end{align}
where $P_{\rm cr} = (\gamma_{\rm cr}-1)\,e_{\rm cr}$ is the CR pressure; $P_T$ is the total (thermal+magnetic+CR) pressure; $\Gamma_{\rm st}=-{\bf v}_{\rm st} \cdot \nabla P_{\rm cr}$ is the CR ``streaming loss term'' discussed below; $S_{\rm g}$ and $S_{\rm cr}$ are gas and CR source terms (e.g.\ injection); $\Gamma_{\rm g}$ and $\Gamma_{\rm cr}$ are gas and CR sink/loss (or ``cooling'') terms; and ${\bf v}_{\rm st}$ is the CR streaming velocity. ${\bf F}_{\rm cr}$ is the CR energy flux, which can be written ${\bf F}_{\rm cr} = (e_{\rm cr} + P_{\rm cr})({\bf v}+{\bf v}_{\rm st})+ {\bf F}_{\rm di}$  where the first term represents advection and streaming, whereas the second term is a diffusive-like flux (e.g.\ given by ${\bf F}_{\rm di}= -\kappa \hat{\bf B}\otimes\hat{\bf B}\cdot\nabla e_{\rm cr}$ in the ``pure diffusion'' or ``zeroth moment'' approximation, but we explicitly evolve this; see \S\ref{diffM1}). 

For the gas equations-of-motion, note when solving the Riemann problem between neighboring fluid elements, $P_{T}$ includes the CR pressure (i.e.\ we make the local strong-coupling approximation: CRs and gas strongly interact), and the effective sound speed of the two-fluid mixture is modified to $(c_{s}^{2})_{\rm eff} = \partial P/\partial \rho = (c_{s}^{2})_{\rm gas} + \gamma_{\rm cr}\,P_{\rm cr}/\rho$, but no other modifications to the MHD method is required. 

\subsubsection{CR Transport: Advection \&\ Streaming}

In our method, each mesh-generating point (which defines the gas resolution ``elements'') represents a finite-volume domain that moves with the fluid velocity ${\bf v}={\bf v}_{\rm gas}$ in a quasi-Lagrangian fashion. After operator-splitting the source/injection and loss/cooling terms, it is convenient to re-write the advection and streaming terms in the following Lagrangian, finite-volume form (see e.g.\ \citealt{Uhli12}): 
\begin{align}
\label{eqn:finite.volume}\frac{D E_{\rm cr}}{D t} = -\int_{\Omega} d^{3}{\bf x}\, {\Bigl \{} & {P}_{\rm cr}\,  (\nabla \cdot {\bf v}) +\Gamma_{\rm st} +  \nabla \cdot \tilde{\bf F}_{\rm cr} {\Bigr \}}
\end{align}
where $D/Dt = \partial/\partial t + {\bf v} \cdot \nabla$ is the Lagrangian derivative co-moving with the gas, and $E_{\rm cr}^{i} = \int_{\Omega_{i}} e_{\rm cr}\,d^{3}{\bf x}$ is the conserved total CR energy in the finite-volume domain $\Omega_{i}$ belonging to element $i$. Here $\tilde{\bf F}_{\rm cr} \equiv {\bf F}_{\rm cr} - {\bf v}\,(e_{\rm cr}+P_{\rm cr}) = {\bf v}_{\rm st}\,(e_{\rm cr}+P_{\rm cr}) + {\bf F}_{\rm di}$. Pure advection with the gas is automatically handled in this description. In cosmological simulations, the Hubble flow is included in $\nabla \cdot {\bf v}$. 

The ${P}_{\rm cr}(\nabla \cdot {\bf v})$ term represents adiabatic changes to the CR energy via compression/expansion (the ``PdV work''), which exchanges energy with gas. We will refer to this as the ``adiabatic'' term throughout.\footnote{To ensure manifest energy conservation, this is solved when the mesh positions are updated. Calculating the volume changes $\Delta V_{i} = \int dt\,\int_{\Omega_{i}}\,d^{3}{\bf x}\,(\nabla \cdot {\bf v})$ with the kernel-weighted divergence of the {\it fluid} velocity field (which is the {\it exact} discrete change in the domain volume as defined in \citealt{Hopk15})}, we have $\Delta E_{\rm cr} = -P_{\rm cr}\,\Delta V_{i}$. This is removed from the total energy equation after the hydrodynamic Riemann problem is solved to determine the total gas energy update.

The $\Gamma_{\rm st}=-{\bf v}_{\rm st}\cdot \nabla {P}_{\rm cr}$ term represents ``streaming loss'', which transfers energy to gas and is always positive because CRs always stream down the CR pressure gradient (see the next section). As CRs stream, instabilities excite high-frequency Alfven waves (frequency of order the gyro frequency, well below our simulation resolution limits; see e.g. \citealt{Went68,Kuls69}) which are damped and thermalize their energy effectively instantly (compared to our simulation timescales).\footnote{With the streaming velocity defined below, the streaming loss term can be written $D E_{\rm cr} / D t = - E_{\rm cr} / \tau_{\rm st}$ with $\tau_{\rm st}^{-1} = (\gamma_{\rm cr}-1)\,|\hat{\bf B} \cdot \hat{\nabla} e_{\rm cr}|^{2}\,|v_{\rm st}\,\nabla e_{\rm cr}|/e_{\rm cr}$. When this is updated the resulting energy lost $\Delta E_{\rm cr} = \int dt\, \tau_{\rm st}^{-1}\,E_{\rm cr}$ is added to the gas thermal energy.}

Finally, the $\int_{\Omega} d^{3}{\bf x}\,\nabla \cdot \tilde{\bf F}_{\rm cr}$ term does not change the total CR energy, but represents flux of energy between resolution elements, caused by CR streaming and diffusion. This can be transformed via Stokes's law into a surface integral, $\int_{\partial \Omega} d{\bf A} \cdot \tilde{\bf F}_{\rm cr}$, which is then solved via our usual second order-accurate, finite-volume Godunov MFM method (in a manner identical to the hydrodynamic equations, see \citealt{Hopk15} for details). 

We explicitly evolve the conserved quantities $E_{\rm cr}$ and total gas energy $E_{\rm gas}$ which are exchanged (either between gas elements or one another), ensuring manifest total energy conservation.\footnote{\label{adaptiveerr}Because we do not evolve a total energy equation, if we use adaptive timesteps, total energy conservation is formally exact at integration-error level rather than machine-accurate. However we have verified that the errors are typically small (percents-level over hundreds of millions of years evolution, although in the most extreme case we find the cumulative amount over 500 Myr can be $\lesssim 20\%$ of the injection), and negligible compared to physical non-conservative terms (e.g.\ collisional/radiative losses, injection).}

\subsubsection{The Streaming Velocity}
\label{strmethod}
CRs stream at some speed $v_{\rm st}$ down the local CR phase-space density gradient (which is equivalent in our single-bin approximation to CR pressure or energy density gradient), projected along the magnetic field lines, i.e.\ ${\bf v}_{\rm st} \equiv -v_{\rm st}\,\hat{\bf B} \,(\hat{\bf B} \cdot \hat{\nabla} P_{\rm cr})$ where $\hat{\nabla} P_{\rm cr} = \hat{\nabla} e_{\rm cr} = (\nabla P_{\rm cr})/|\nabla P_{\rm cr}| = (\nabla e_{\rm cr})/|\nabla e_{\rm cr}|$ is the direction of the CR pressure/energy density gradient. 

It is generally believed that micro-scale instabilities limit the streaming velocity to Alfven speed $v_{\rm A}\,(=B/\sqrt{4\pi\rho}\,$) in the low-$\beta$ limit (see \citealt{Skil71,Holm79}, or more recently \citealt{Kuls05,Yan08,Enss11}), so we adopt a fiducial value $v_{\rm st} = v_{\rm A}$.

But in the weak-field, plasma $\beta \gg 1$, regime, the streaming velocity can be boosted by significant wave damping \citep[see discussion in][]{Enss11,Wien13,Rusz17}, so we have also tested various streaming speeds in Appendix~\ref{diffstr}. Although the streaming velocity can in principle exceed $v_{A}$ by a large factor, \citet{Wien13} and \citet{Rusz17} argued that the streaming loss $\Gamma_{\rm st}$ should be still limited by $\Gamma_{\rm st}=-{\bf v}_{\rm A} \cdot \nabla P_{\rm cr}$, because this term is mediated by the excitation of Alfven waves. Therefore, regardless of streaming velocities, we set the streaming loss to $-{\bf v}_{\rm A} \cdot \nabla P_{\rm cr}$. When streaming is disabled we simply eliminate terms relevant to streaming.

\subsubsection{Diffusive Transport Terms: Two-moment Method}
\label{diffM1}
It is common in the literature to treat ${\bf F}_{\rm di}$ in the ``zero-th moment'' expansion, i.e.\ approximate it as an anisotropic scalar diffusion with ${\bf F}_{\rm di} = -\kappa\,\hat{\bf B}\,(\hat{\bf B} \cdot \nabla e_{\rm cr})$, where $\kappa$ is the effective diffusion coefficient,  which parameterizes the unresolved CR physics. However at high resolution this is problematic for several reasons: (1) it imposes a quadratic timestep criterion (if evaluated with an explicit scheme: $\Delta t < C_{\rm cour}\,\Delta x^{2}/\kappa$, where $\Delta x$ is the resolution and $C_{\rm cour}$ the Courant factor) which becomes very small; (2) it implies unphysical super-luminal CR transport when the gradient-scale length $ e_{\rm cr}/|\nabla e_{\rm cr}|$ becomes smaller than $\kappa/c \sim 3\,{\rm pc}\,(\kappa / 3\times 10^{29}\,{\rm cm^{2}\,s^{-1}})$ (resolution often reached for simulations in this paper); (3) it cannot smoothly handle the transition between streaming and diffusion-dominated regimes; (4) it will develop spurious numerical oscillations near extrema when handling streaming \citep{Shar10}; and (5) it encounters the usual difficulties with anisotropic diffusion operators in moving-mesh codes described in \citet{Hopk17diff} (including e.g.\ difficulty if CRs are ``trapped'' at local maxima). 

Hence, we adopt a simple two-moment approximation for CR diffusion and streaming, independently developed for this work but similar in concept to the recently-presented implementations in \citet{Jian18} and \citet{Thom19} (although the concept and use in CR dynamics are well-established; see \citealt{Snod06} for examples). Rather than set $\tilde{\bf F}_{\rm cr}=-\kappa\,\nabla e_{\rm cr}$, we explicitly evolve the flux equation:
\begin{align}
\frac{1}{\tilde{c}^{2}}\,\left[ \frac{\partial\tilde{\bf F}_{\rm cr}}{\partial t} + \nabla\cdot\left({\bf  v}\otimes\tilde{\bf F}_{\rm cr} \right) \right] + \nabla_{\|} P_{\rm cr}  &= -\frac{(\gamma_{\rm cr}-1)}{\kappa^{\ast}}\,\tilde{\bf F}_{\rm cr},
\label{Fcr}
\end{align}
where $\tilde{\bf F}_{\rm cr}$ is the CR flux measured in the frame comoving with the fluid, $\nabla_{\|}P_{\rm cr} \equiv \hat{\bf B}\otimes\hat{\bf B}\cdot \nabla P_{\rm cr}$, $\tilde{c}$ is the (reduced) speed of light, and $\kappa^*$ is the composite parallel (magnetic field-aligned) diffusion coefficient in the rest frame of the fluid, 
\begin{align}
\label{kappa.effective}\kappa^*=\kappa+\frac{v_{\rm st}(e_{\rm cr}+P_{\rm cr})}{|\hat{\bf B}\cdot\nabla e_{\rm cr}|},
\end{align}
where the second term includes the CR streaming with the streaming velocity specified above.

For the numerical implementation of CR energy and flux, we follow the treatment of diffusion operators in MFM, outlined in Section 2 in \citet{Hopk17diff} with a few modifications.

We solve a general evolution equation of conserved quantities $\left( V\bf{U} \right )_i$ of cell i (e.g. CR energy) with a volume $V$ by summing over all adjacent cells j:
\begin{align}
\frac{\mathrm{d} }{\mathrm{d} t}\left ( V\bf{U} \right )_i=-\sum_j{\bf F}^*_{\rm diff,ij}\cdot {\bf A}_{i,j},
\end{align}
where V is the cell volume and ${\bf F}^*_{\rm diff,ij}$ is the interface value of the flux and ${\bf A}_{i,j}$ is the effective face area defined in \cite{GIZMO}, in the following steps:
\begin{enumerate}

  \item  We calculate all relevant coefficients, using the standard gradient estimator in GIZMO for MFM to estimate gradients, e.g. $\nabla P_{\rm cr}$ and $\nabla e_{\rm cr}$, as described in \cite{GIZMO};  
  \item  We estimate the values on the left and right sides of the face from the values of cells i and j through a linear reconstruction and use them to solve the Riemann problem; 
  \item We compute the interface face value of the flux ${\bf F}^*_{\rm diff,ij}$ by solving the Riemann problem (RP) through the \cite{Hart83HLL} (HLL) method, using a MINMOD slope limiter, in an operator split manner from the pure MHD.

  \item Finally, the source term in Eq. \ref{Fcr} (not considered in \citealt{Hopk17diff}) is added implicitly to ensure stability.
\end{enumerate}
  
  We differ from \citet{Hopk17diff} since (1) we explicitly evolve the CR flux (instead of calculating it from the CR energy gradient) and (2) in the Riemann solver, we consider the fastest wavespeed to be generally $\tilde{c}$  (since we choose $\tilde{c}$ to be faster than other physical processes).

Unlike \citet{Jian18}, we do not modify the momentum transfer from CRs to gas, i.e. the second line of Eq. \ref{eqset}, since we assume the strong coupling limit between gas and CRs throughout the paper. This, however, will over-estimate the momentum transfer from CRs to gas when the strong coupling assumption breaks down, i.e. in regimes where the CR and gas coupling is weak and the CR mean free paths are long. In all of our simulations, the mean free paths $\sim (1\,{\rm pc})(\kappa/10^{29}{\rm cm^2/s})$ are smaller than the resolved length-scales, so the strong coupling assumption is probably relevant in most of the situations.\footnote{ The formulation in \citet{Jian18} will also over-estimate the momentum transfer in the weakly coupling regime if the ``reduced speed of light'' approximation is introduced (see below and \S~5.2 in \citealt{Jian18}).} 

We stress that while the flux equation can be generically obtained by integrating the first moment of the Vlasov equation (with some model for closure of higher  moments, equation-of-state, and scattering terms), one should not take Eq.~\ref{Fcr} to represent a physical two-moment expansion of the relativistic Vlasov equation for CRs. Doing so requires making a number of additional assumptions about e.g.\ the CR phase space distribution function, ratio of gyro radii to resolved scales, and order of truncation in terms $\mathcal{O}(v/c)$. We discuss some of the subtle differences that can arise in Eq.~\ref{Fcr} as a result, in Appendix~\ref{sec:testanis}, but stress that on large and/or long time scales these vanish, and so they have no effect on our conclusions in this paper. For our purposes here, it is better to think of it as a generic two-moment numerical expansion of the anisotropic diffusion + streaming equation which eliminates all of the numerical pathologies (1)-(5) above. In future work, it will be interesting to explore more detailed physically-derived transport models that include these higher-order terms, and attempt to actually predict the coefficients $\kappa$ and $v_{\rm st}$ on physical grounds \citep[see e.g.][]{Zwei17,Thom19}.

For now, if we ignore streaming, we see that in steady-state and/or when $\tilde{c}$ is large, or $\Delta t \gg \kappa/\tilde{c}^{2}$ (or on spatial scales $\gg \kappa/\tilde{c}$), this equation becomes ${\bf F}_{\rm cr} \approx -\kappa\,\hat{\bf B}\,(\hat{\bf B} \cdot \nabla e_{\rm cr})$, and we recover the usual diffusion equation (see Appendix \ref{reducedSOL} for a comparison between the pure diffusion and two-moment methods). However, the two-moment method smoothly limits the maximum bulk transport velocity of the CRs to $\tilde{c}$, and makes the timestep criterion $\Delta t < C_{\rm cour}\,\Delta x / \tilde{c}$,\footnote{We adopt $C_{\rm cour}=0.4$ throughout, and have validated stability (as expected) for this value. $\Delta x$ in the Courant condition is defined in the same manner as \citet{FIRE2} as the local mean inter-particle separation (i.e.\ the equivalent of the grid spacing in a regular-grid code), $\Delta x \equiv (m/\rho)^{1/3}$.} which is only first-order, instead of quadratic, in $\Delta x$. 

For true micro-physical CR motion, however, $\tilde{c} \approx c$, the speed of light, which still requires a impractically small timestep. Fortunately, for our purposes in these simulations -- where we only capture bulk CR properties in the fluid limit -- it is more convenient to consider $\tilde{c}\ll c$ (namely the ``{\it reduced speed of light}'' approximation), since galaxy properties should still converge, regardless of $\tilde{c}$, provided it is set to some value faster than other relevant physical processes, e.g. the CR cooling or the {\em actual} bulk flow speeds realized in our simulations. We have experimented extensively with this and find that, for the simulations here, values $\tilde{c} \sim 500-2000\,{\rm km\,s^{-1}}$ are sufficient to give converged results, e.g. SFR and $\gamma$-ray emission (see Appendix \ref{reducedSOL}). 

In Appendix \ref{purediff}, we compare the results using the simpler pure-diffusion (zeroth-moment) approximation: we then simply assume ${\bf F}_{\rm di} \rightarrow -\kappa\,\hat{\bf B}\,(\hat{\bf B} \cdot \nabla e_{\rm cr})$ and solve the anisotropic diffusion equation (with the stricter Courant condition) as described in \citet{Hopk17diff}. This is equivalent to adopting $\tilde{c}\rightarrow\infty$, in our Eq.~\ref{Fcr}. For  the same $\kappa_{\ast}$, this gives nearly-identical results to our default Eq.~\ref{Fcr} in our galaxy simulations, demonstrating that the form  of the CR flux  equation is not a significant source of uncertainty here. We also find an excellent agreement between the zeroth- and two-moment methods in a pure diffusion test given a high enough reduced speed of light.

It is worth noting that our CR treatment is akin to the first-moment or “M1”
moment-based method for radiation hydrodynamics (with different closure relations and the scattering terms), with the ``reduced speed of light'' $\tilde{c}$ \citep{Leve84}, while the ``pure diffusion'' approximation is akin to flux-limited diffusion (without the limiter).

\begin{table*}
\centering
    \begin{tabular}{lllllllllllll}
    \hline\hline
    Name & $M_{\rm vir}$&$R_{\rm vir}$& c  &$M_{\rm *,disk}$ & $M_{\rm *,bulge}$ &$M_{\rm g, disk}$ & $M_{\rm g, halo}$ &$d_{\rm *, disk}$& $h_{\rm *, disk}$&  $d_{\rm g, disk}$    & $m_{\rm b}$           \\
         & [$10^{10}\msun$] &${\rm [10\;kpc]}$& &[$10^{10}\msun$]&[$10^{10}\msun$] & [$10^{10}\msun$]& [$10^{10}\msun$] &[kpc]&[kpc]&[kpc]&[$10^{3}\msun$] \\
    \hline
    	\hline
{\bf Dwarf} & 2.9& 63 & 15&0.019 &0.0014 & 0.1 &0.01&1.0&0.2&5 &3.3 \\
 {\bf  Starburst} &   21 &   121     &  11 &0.57  & 0.14& 4.0 & 1.0&1.0&0.2& 15.0  & 20.0\\
{\bf $L\star$ Galaxy} &    150 &   234    & 12 & 4.7 & 1.5&0.9  &0.1 &3.2&0.24&6.4 &2.6 \\
    \hline  
    \end{tabular}
 	\caption{Simulation parameters. $M_{\rm vir}$ is the virial mass; $c$ is the halo concentration;$M_{*,{\rm disk}}$ is the mass of stellar disk; $M_{*,{\rm bulge}}$ is the mass of stellar bulge;  $M_{\rm g, disk}$ is the mass of gas disk; $M_{\rm g, halo}$ is the mass of gas halo; $d_{\rm *, disk}$ is the stellar disk radial scale length; $h_{\rm *, disk}$ is the thickness of stellar disk; $d_{\rm g, disk}$ is the gas disk radial scale length; $m_{\rm b}$ is the gas particle mass.} 
\label{SIC}
\end{table*}

\begin{table*}
\centering
    \begin{tabular}{llllllllllll}
    \hline\hline
    & {\bf {\footnotesize Hydro}} &{\bf {\footnotesize MHD}}&{\bf {\footnotesize Advec-} } & {\footnotesize${\rm \kappa}=3e27$} & {\footnotesize${\rm \kappa=3e28}$} & {\footnotesize${\rm \kappa=3e29}$} &{\bf MHD} &{\bf MHD} &{\bf MHD} \\ 
    & {\bf {\footnotesize no CR}} &{\bf {\footnotesize no CR}}& {\bf {\footnotesize -tion} }&& &&&{\footnotesize${\rm \kappa=3e28}$} &{\footnotesize${\rm \kappa=3e28}$}\\
        &  && && & &{\footnotesize \bf Streaming}& &{\footnotesize \bf Streaming}\\
        \hline
    	\hline
       
MHD&Off&On&Off&Off&Off&Off&On&On&On  \\     
Streaming&Off&Off&Off&Off&Off&Off&On&Off&On\\
$\kappa$ \;\;[${\rm cm^2/s}$]&-&-&0&$3\times 10^{27}$&$3\times 10^{28}$&$3\times 10^{29}$&0&$3\times 10^{28}$&$3\times 10^{28}$  \\
$\tilde{c}$ \;\;[km/s]&-&-&-&500&1000&2000&1000&1000&1000\\
    \hline  
    \end{tabular}

 	\caption{Different propagation models of CRs. Each column gives the name of our simulation models, while rows list the physics/parameters of the propagation model. The ``MHD'' column row indicates whether magnetic fields are included. The ``Streaming'' column indicates whether CR streaming is considered. $\kappa$ gives the isotropic/parallel CR diffusion coefficient (CRs will diffuse isotropically if MHD is off, while CRs will diffuse along magnetic fields if MHD is on). $\tilde{c}$ is the ``reduced speed of light'' in the two-moment method (see \S~ \ref{diffM1}). } 
\label{CRprop}
\end{table*}

\subsubsection{The Diffusion Coefficient}
\label{diffcoeff}
The only remaining unspecified parameter in the CR treatment is the effective diffusion coefficient $\kappa$\footnote{ We do not attempt to calculate the diffusion coefficients from microphysics, but treat them as empirical parameters to be varied/constrained.}. However, there is still substantial uncertainty on its value from a theoretical or observational perspective. In the self confinement picture, it depends on wave damping mechanisms, which are currently not well constrained \citep{Wien13,Zwei13}. In the extrinsic turbulence picture, CRs are scattered through turbulent magnetic fields, but we have limited knowledge of the small scale magnetic fluctuations and the coupling between magnetic field turbulence and CRs \citep{Enss03,Enss07}.

Fortunately, there are some empirical constraints on the effective diffusion coefficients, i.e., the diffusion coefficients that broadly reproduce observations of cosmic rays in the Milky Way (even though it is possible that the microscopic model of diffusion is not the correct one for cosmic-ray transport).

For example, \citet{Trot11} constrained the {\em isotropically-averaged} diffusivity $\kappa$ to be $\sim 6\times 10^{28}{\rm cm^2/s}$ to within a factor of a few, at $\sim$GeV energy with {\small GALPROP}, using the measured energy spectra and abundances of nuclei species in CRs, and adopting a flat halo diffusion model (\citealt{Ginz76}; see Introduction for a brief description). Implicitly, these abundances depend on the residence time of CRs in the Galaxy, so there is a degeneracy between $\kappa$ and the CR halo height $z_{\rm h}$ (typically 1-10 kpc), out of which CRs can freely propagate (see Figure 3 in \citealt{Trot11} or Figure 10 in \citealt{Lind10}; this issue was also discussed in \citealt{Ginz76}). Even in this model, it is possible to match the observational data with a significantly larger $\kappa$ (up to factors of several) if a larger halo size is adopted.\footnote{For example, Fig. 10 in \citet{Lind10} shows isotropically-averaged $\kappa\sim 3\times 10^{28}{\rm cm^2/s}$ with $z_{\rm h}\sim 3\;{\rm kpc}$ but $\kappa\sim 10^{29}{\rm cm^2/s}$ with a larger $z_{\rm h}\sim 5\;{\rm kpc})$.}

There are other substantial uncertainties in the estimates of $\kappa$, as these empirical constraints usually neglect e.g. local variations in $\kappa$ or magnetic field structure, the role of advection, halo density profiles (in addition to sizes), small-scale gas density variations (``clumpiness''), and the complicated spatial and temporal distributions of CR sources. The value of $\kappa$ is even more poorly constrained outside the MW. 

Given these uncertainties, we do not attempt a self-consistent calculation of the diffusion coefficient. Instead, we simply assume a constant $\kappa$, which is a common approach in the literature \citep[e.g.][]{Boot13,Sale14,Pakm16,Pfro17gamma,Wien17}, and test a wide range of $\kappa$.

Unlike a flat halo or ``leaky box''-type diffusion model, where CRs simply freely escape after crossing the boundary of the halo, we assume CR diffusion with constant $\kappa$ {\it everywhere}, even at large heights above the disk. It is therefore likely that our simulations will require a larger $\kappa$ than the value from a flat halo model with a small halo size.

We will also consider anisotropic CR diffusion with a constant parallel diffusivity. Because the above estimate is isotropically-averaged, if magnetic fields are tangled or toroidal, the equivalent anistropic diffusion coefficient $\kappa$ would be factor $\gtrsim 3$ larger.

\subsubsection{Sources \&\ Injection} 
\label{sectionsource}
We assume CR injection from SNe (including Type Ia and Type II), with a fixed fraction $\epsilon_{\rm cr}$ ($=0.1$, as our default value) of the initial ejecta energy ($\Delta E_{\rm cr} = \epsilon_{\rm cr}\,E_{\rm SNe}$ with $E_{\rm SNe} \approx 10^{51}\,{\rm erg}$) of every SNe explosion going into CRs. SNe explosions inject thermal and kinetic energy into neighboring gas resolution elements according to the algorithm described in detail in \citet{Hopk18sne}; we therefore reduce the coupled energy by $1-\epsilon_{\rm cr}$ and inject the remaining $\epsilon_{\rm cr}$ energy alongside the metals, mass, and thermal+kinetic energy using the same relative ``weights'' to determine the CR energy assigned to each neighbor. Likewise the CR flux is updated assuming the CRs free-stream at injection (${\bf F}_{\rm cr} \rightarrow {\bf F}_{\rm cr} + \Delta {\bf F}_{\rm cr}$ with $\Delta {\bf F}_{\rm cr} = \Delta e_{\rm cr}\,\tilde{c}\, \hat{\bf r}$ from the source, where $\hat{\bf r}$ is a unit vector pointing outwards from the source). The injection is therefore operator-split and solved discretely (associated with each SNe).

\subsubsection{Hadronic \&\ Coulomb Losses (``Cooling'')}
\label{cooling}
We adopt the estimate for combined hadronic ($\tilde{\Lambda}_{\rm cr,had}$) plus Coulomb ($\tilde{\Lambda}_{\rm cr,\,Cou}$) losses, $\Gamma_{\rm cr}$, from \citet{Volk96} and \citet{Enss97} as synthesized and updated in \citet{Guo08}: 
\begin{align}
\label{crcooling}
&\Gamma_{\rm cr} =\tilde{\Lambda}_{\rm cr}\,e_{\rm cr}\,n_{\rm n} = (\tilde{\Lambda}_{\rm cr,had} + \tilde{\Lambda}_{\rm cr,\,Cou})\,e_{\rm cr}\,n_{\rm n} \\ 
\nonumber =&\, 5.8\times 10^{-16}\,(1 + 0.28\,x_{e})\left(\,\frac{e_{\rm cr}}{\rm erg\, cm^{-3}}\right)\left(\frac{\,n_{\rm n}}{\rm cm^{-3}}\right)\,{\rm erg \,cm^{-3}s^{-1}}\,
\end{align}
where $n_{\rm n}$ is the number density of nucleons and $x_{e}$ is the number of free electrons per nucleon. Following \citet{Guo08} we assume $\sim 1/6$ of the hadronic losses and all Coulomb losses are thermalized, adding a volumetric gas heating term 
\begin{align}
S_{\rm gas} =&\; 0.98\times10^{-16}\,(1 + 1.7\,x_{e})\left(\,\frac{e_{\rm cr}}{\rm erg\,cm^{-3}}\right)\times\nonumber\\
&\left(\frac{\,n_{\rm n}}{\rm cm^{-3}}\right)\,{\rm erg \,cm^{-3}s^{-1}}
\end{align}
The remaining CR losses are assumed to escape in the form of $\gamma$-rays and other products to which the gas is optically thin.  

Due to the hadronic and Coulomb losses, we have to consider the Boltzmann equation with a weak collision term, instead of the Vlasov equation. Since the collision term affects both CR energy density and flux, in the two moment method, we also update the CR flux as ${\bf F}_{\rm cr} \rightarrow {\bf F}_{\rm cr} (1- \tilde{\Lambda}_{\rm cr}\,n_{\rm n}\Delta t)$.

The loss and heating terms are operator-split and solved together with all other gas heating/cooling terms with our usual fully-implicit cooling scheme described in \citet{FIRE2}.

\subsubsection{``Isotropic'' Runs}

By default, we solve the CR equations coupled to the ideal MHD equations, and treat the CR transport (streaming and advection/diffusion) fully anisotropically. However in many of the tests below we consider isotropic CR diffusion without MHD and streaming, so we simply solve the hydrodynamic equations, remove the terms relevant to streaming, and replace $\hat{\bf B}$ wherever it appears above (representing projection of motion along field lines) with $\hat{\nabla} P_{\rm cr}$.

\section{Simulation setup}
\label{simset}
\subsection{Initial conditions}

We study the impact of CRs on three characteristic types of galaxies, dwarf ({\bf Dwarf}), sub-$L\star$ starburst ({\bf  Starburst}) and $L\star$ ({\bf $L\star$ Galaxy}) galaxies, whose details are listed in Table \ref{SIC}. 

All of the runs have exponential stellar and gas disks with scale radii $d_{\rm *, disk}$ and $d_{\rm g, disk}$ respectively. We also include small stellar bulges with Hernquist profiles \citep{Hern90} and gas halos with beta profiles (beta=2). The latter enable CRs to diffuse far from the galaxies, since CRs cannot diffuse without the presence of neighboring gas particles in our numerical scheme. 

Halo spin parameters (which determine the rotation of the halo gas and dark matter) are set to be 0.033, close to the median value of simulated halos in \cite{Bull01}, and the initial Toomre Q is set to one uniformly in the gas and stellar disks.  We set the metallicity of all star and gas particles in our initial conditions (ICs) in {\bf Dwarf}, {\bf Starburst} and $L\star$ {\bf Galaxy} to be 0.1, 0.3, and 1.0 $Z_\odot$ respectively. Ages of stars present in our ICs are set to $>$ 10 Gyr to avoid excessive SNe from old stellar populations when the simulation begins, which could significantly affect the early evolution of our simulations.

In all {\bf $L\star$ Galaxy} and {\bf Dwarf} runs, we delay turning on the CRs because of initial instabilities from settling of the ICs and to allow magnetic fields to first amplify to a steady-state strength. We enable CRs after initial evolution of 150\,Myr in {\bf $L\star$ Galaxy} and 300\,Myr in {\bf Dwarf}. In the runs with magnetic fields we start with a seed magnetic field with $10^{-2}\mu$G uniformly (over all gas particles) pointing along the direction of disk angular momentum. The magnetic fields rapidly amplify to $\sim \mu$G in dense gas and develop toroidal morphologies with significant turbulent structure, by around a hundred Myr (see \citealt{Su17,Su18}). In the following, we define $t=0$ at the time when CRs are turned on.

{\bf  Starburst} is designed to mimic dwarf galaxies with high gas surface density ($\sim 0.1\; {\rm g/cm^2}$) and SFR ($\sim 5\; \msun/{\rm yr}$) (e.g. M82 or NGC253). We set up a massive gas reservoir with the extended disk and halo such that gas can continuously accrete to the galaxy and trigger intense star formation for an extended period of time. In {\bf  Starburst} runs, we inject CRs immediately at the beginning of the run, since we want to study the transient phenomena (namely, the starburst).

For a subset of our runs we have performed resolution studies and show (see Appendix \ref{resostudy}) that global quantities of interest are robust at our default resolution indicated in Table \ref{SIC}, and that main qualitative effects of CRs on galaxies can be captured at this resolution. 
\begin{figure*}
\begin{centering}
 \includegraphics[width={0.95\textwidth}]{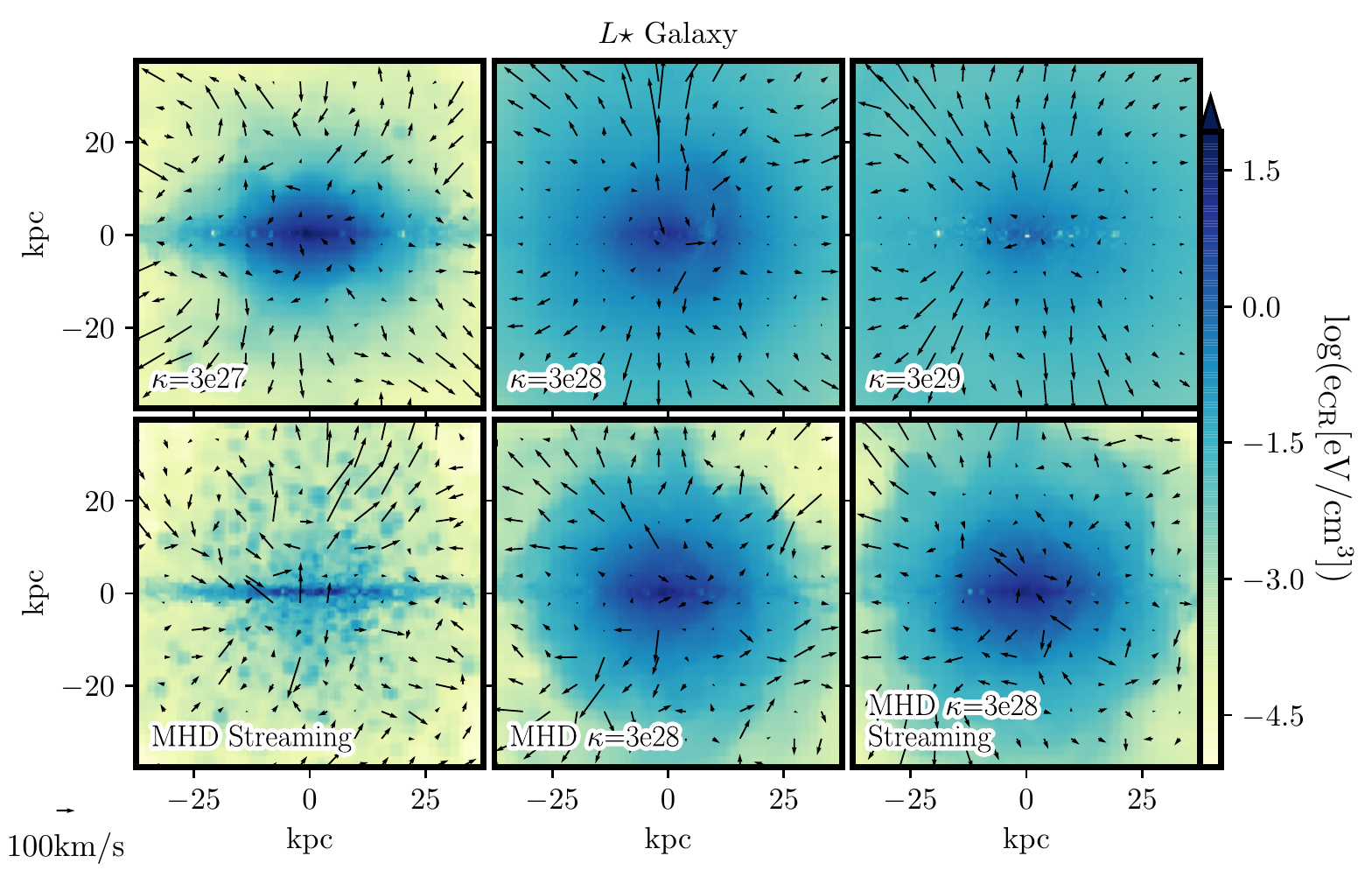}
 \end{centering}
 \caption{Slice plots of CR energy density $e_{\rm cr}$ (in a plane perpendicular to the galactic disk), in runs (of our $L{\star}$ galaxy model) with different CR transport assumptions (Table~\ref{CRprop}), after $500\,$Myr of evolution (in quasi-steady-state). Arrows show gas velocities parallel to the slices. CR halos are more extended with larger $\kappa$, somewhat smaller with magnetic fields included (owing to suppression of perpendicular diffusion), and somewhat larger again with streaming also included.}
\label{MWcrden}
\end{figure*}

\begin{figure*}
\begin{centering}
 \includegraphics[width={0.95\textwidth}]{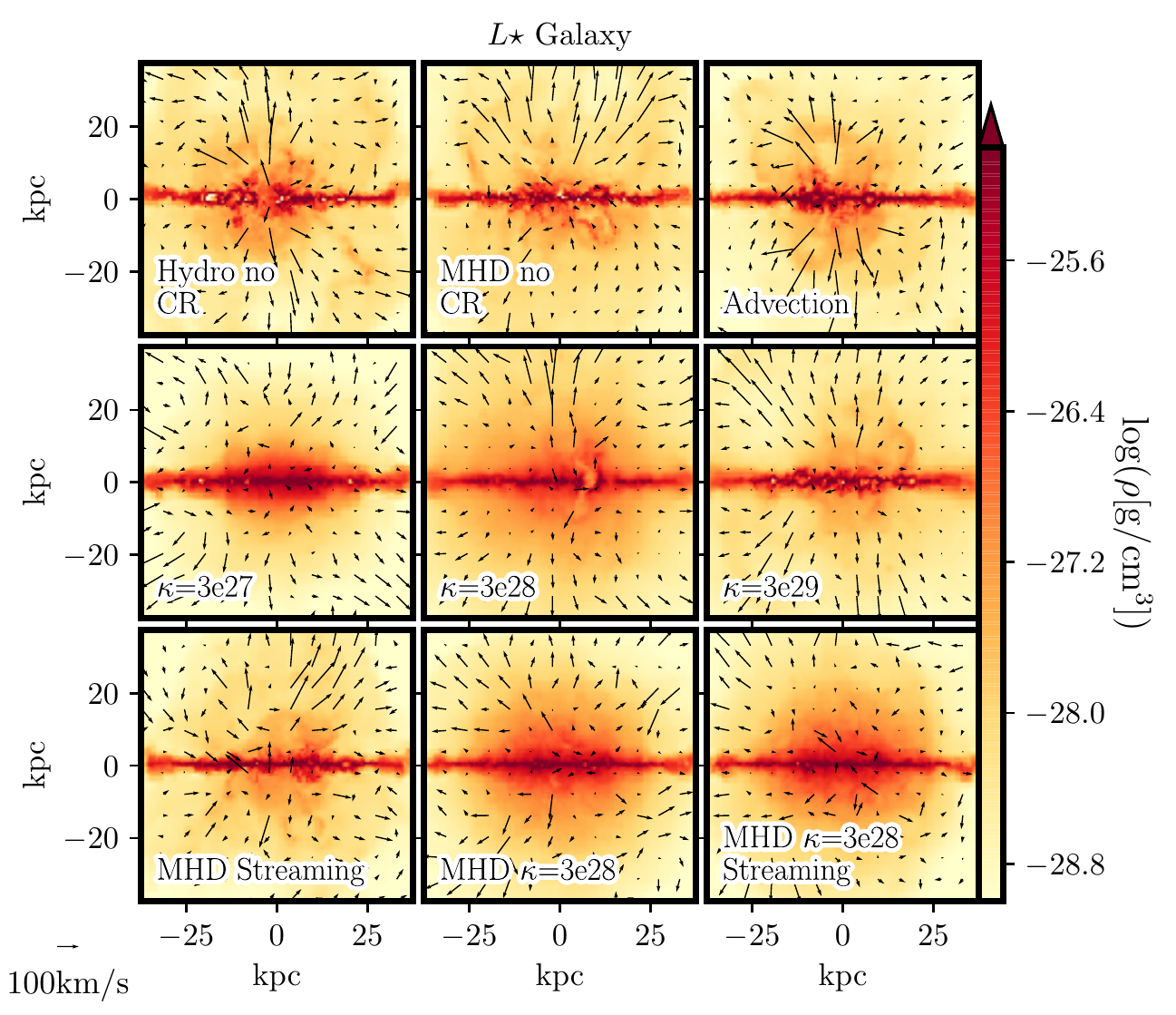}
 \end{centering}
 \caption{Slice plots of gas density and velocity, for the same runs (and in the same style as) Fig.~\ref{MWcrden}. The gaseous halo responds more weakly to changes in CR assumptions: gas disks are thicker with CRs at low diffusivity (because CRs are trapped), but outflows more ordered at large scales with high diffusivity.}
\label{MWden}
\end{figure*}

\begin{figure*}
\begin{centering}
  \includegraphics[width={0.45\textwidth}]{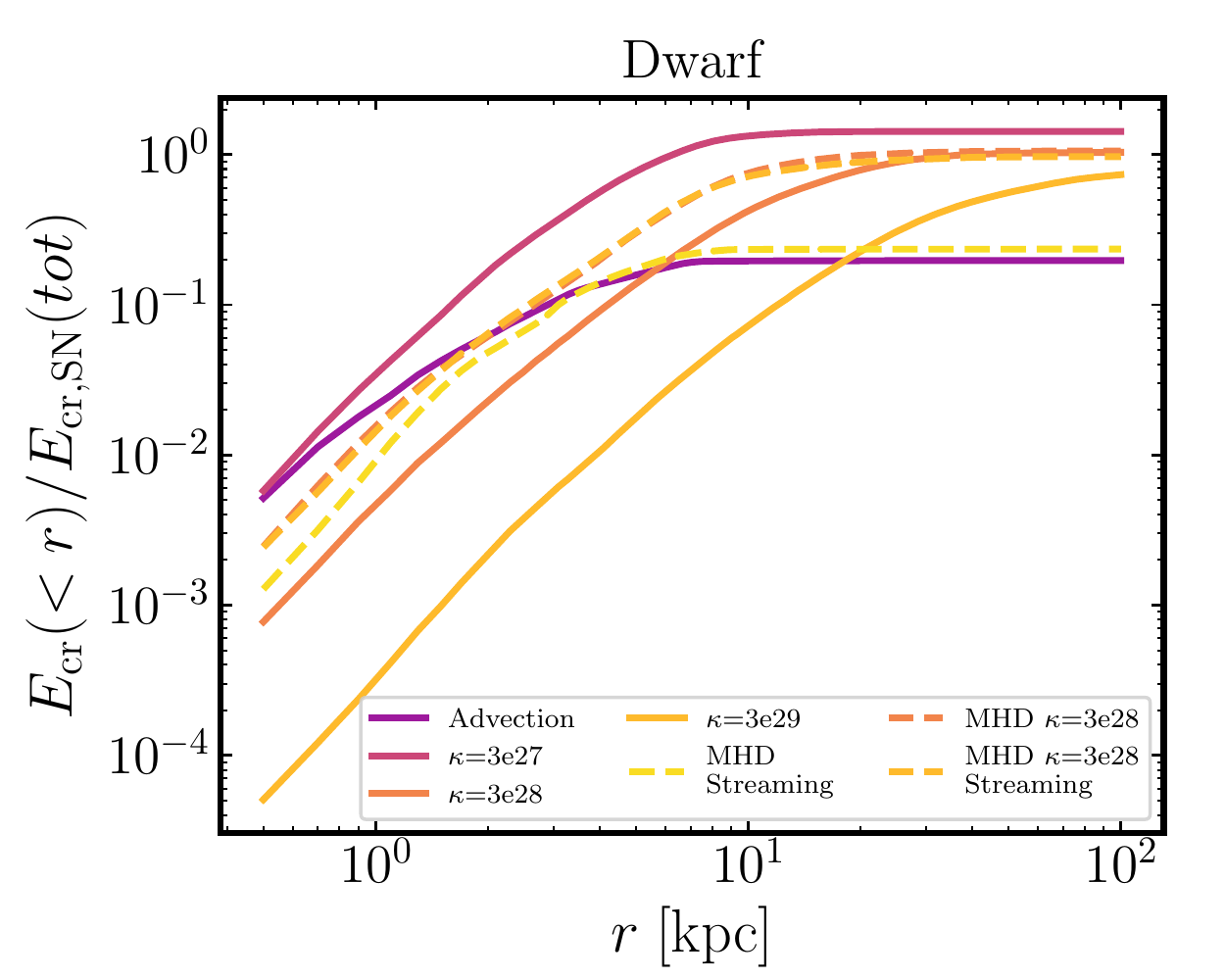}
  \includegraphics[width={0.45\textwidth}]{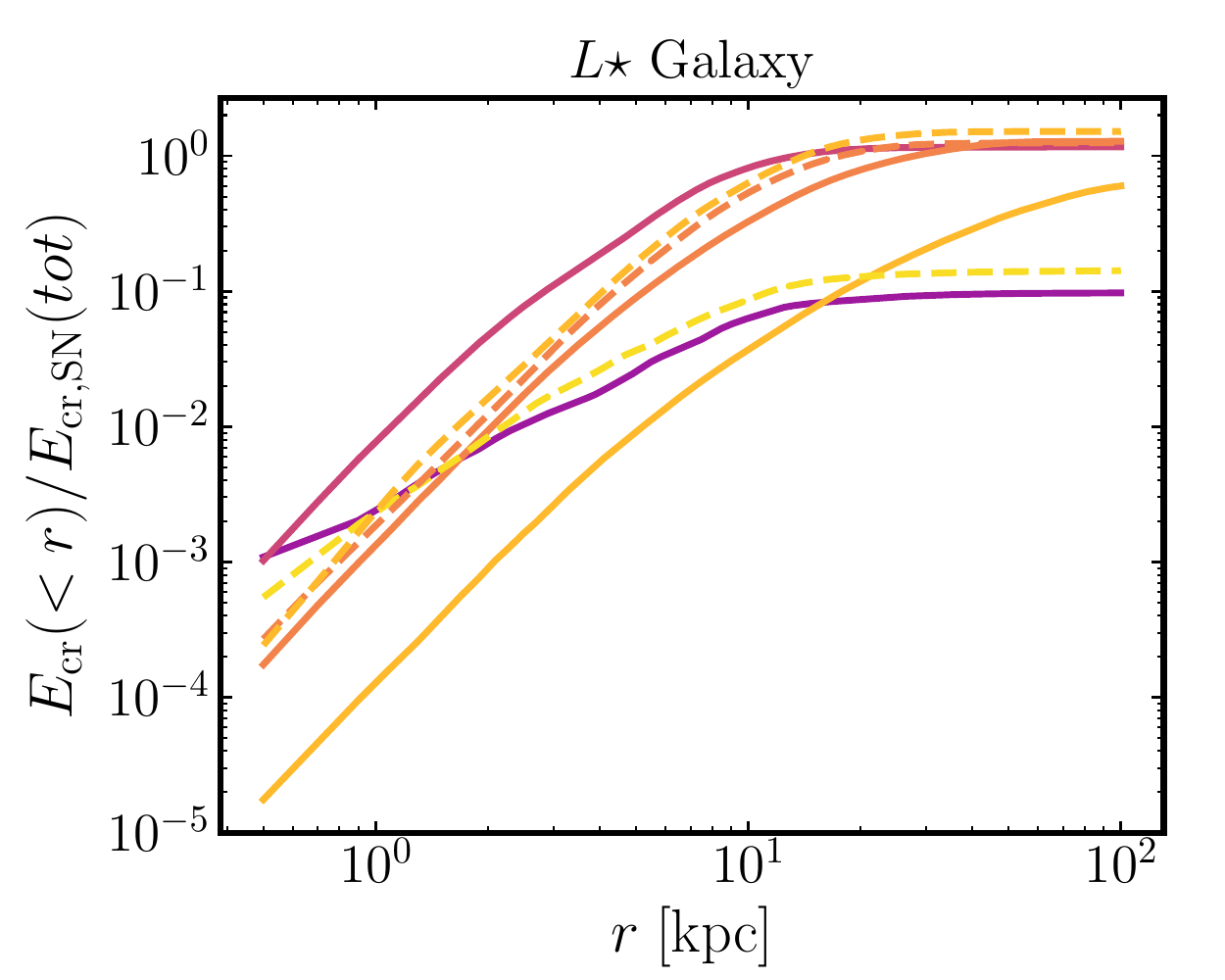}
  \end{centering}
\caption{Cumulative CR energy as a function of radius from the galaxy center (at $t=500\,$Myr), normalized by the total CR energy injected by SNe since $t=0$, in our {\bf Dwarf} ({\em left}) and $L\star$ galaxy models ({\em right}), from Table~\ref{SIC}, with different CR propagation models (Table \ref{CRprop}). Higher-$\kappa$ leads to larger CR scale radii and lower CR densities at a given radius, as expected.}
\label{CRgasz}
\end{figure*}

\subsection{Cosmic Ray Propagation Models}

We consider several different CR propagation models, and a range of diffusion coefficients. All models are listed in Table \ref{CRprop}. In particular, we consider diffusion coefficients up to ten times higher and lower than the common inferred isotropically-averaged MW values,  sampling a range $\kappa = 3 \times 10^{27} {\rm cm^2/s} - 3 \times 10^{29}{\rm cm^2/s}$.

The most complete (and potentially the most realistic) CR propagation model we test includes fully anisotropic diffusion with MHD and streaming. However, given the uncertainties in the magnetic field configuration on small scales as well as uncertainties in the streaming parameters, we evolve a range of simulations with isotropic diffusion without streaming. This model also enables straightforward comparison with other work as it is the most prevalent propagation model in the literature \citep[see e.g.][]{Stro98,Jube08,Lack10}.

We apply the newly developed two-moment method (\S~\ref{diffM1}) to both streaming and diffusion with a reduced speed of light, $\tilde{c}$. In Appendix \ref{reducedSOL} we test different choices for this parameter and demonstrate that physical properties, e.g. SFR or $\gamma$-ray emission, are not affected by the choice of $\tilde{c}$ as long as it is equal to or larger than the values listed in Table \ref{CRprop}.

\section{Results}
\subsection{Distribution of cosmic rays  and the effects on galactic properties}
\subsubsection{Dwarf and $L\star$ galaxies}
\label{CRgalpro}

\begin{figure*}
\begin{centering}
 \includegraphics[width={0.45\textwidth}]{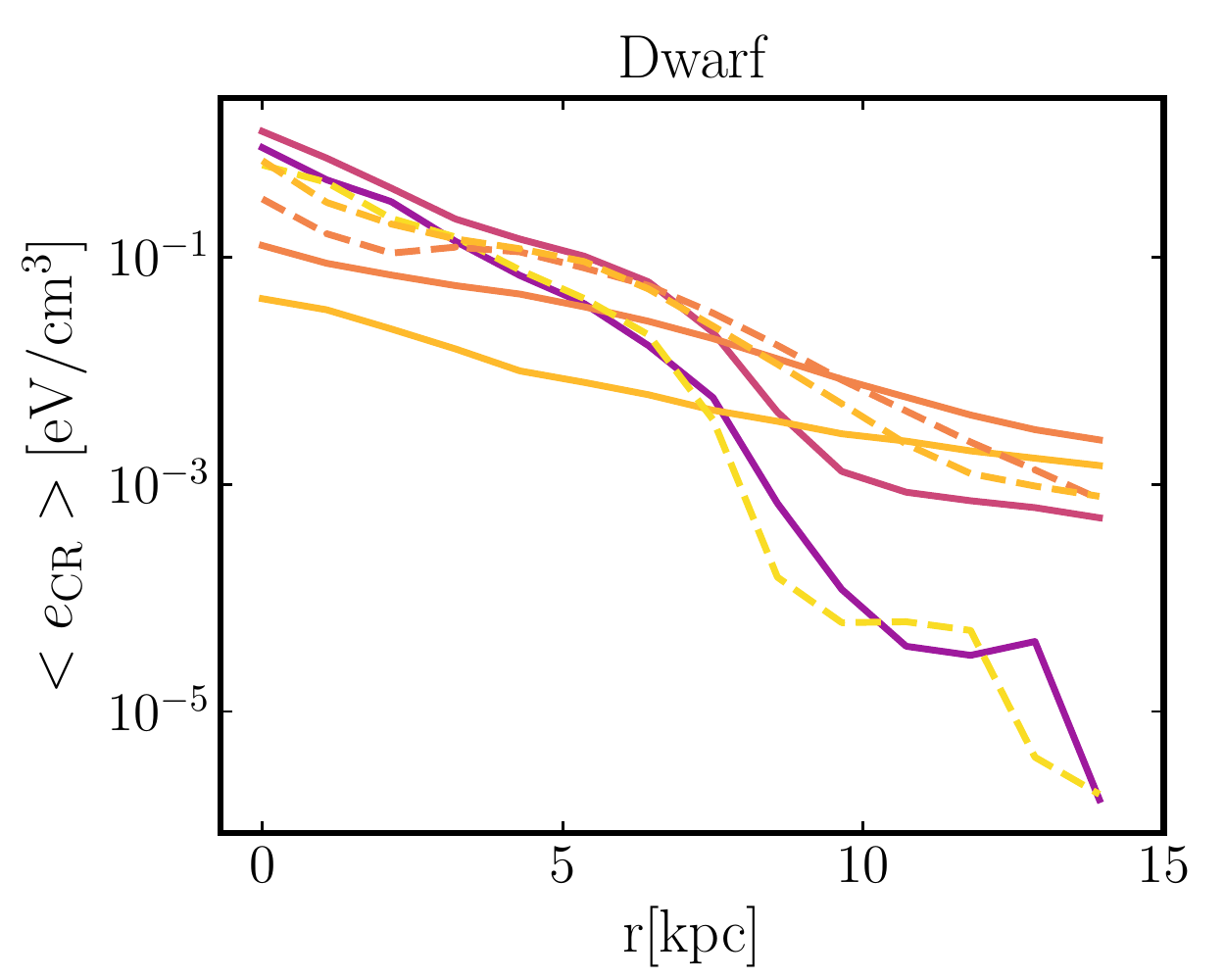}
 \includegraphics[width={0.45\textwidth}]{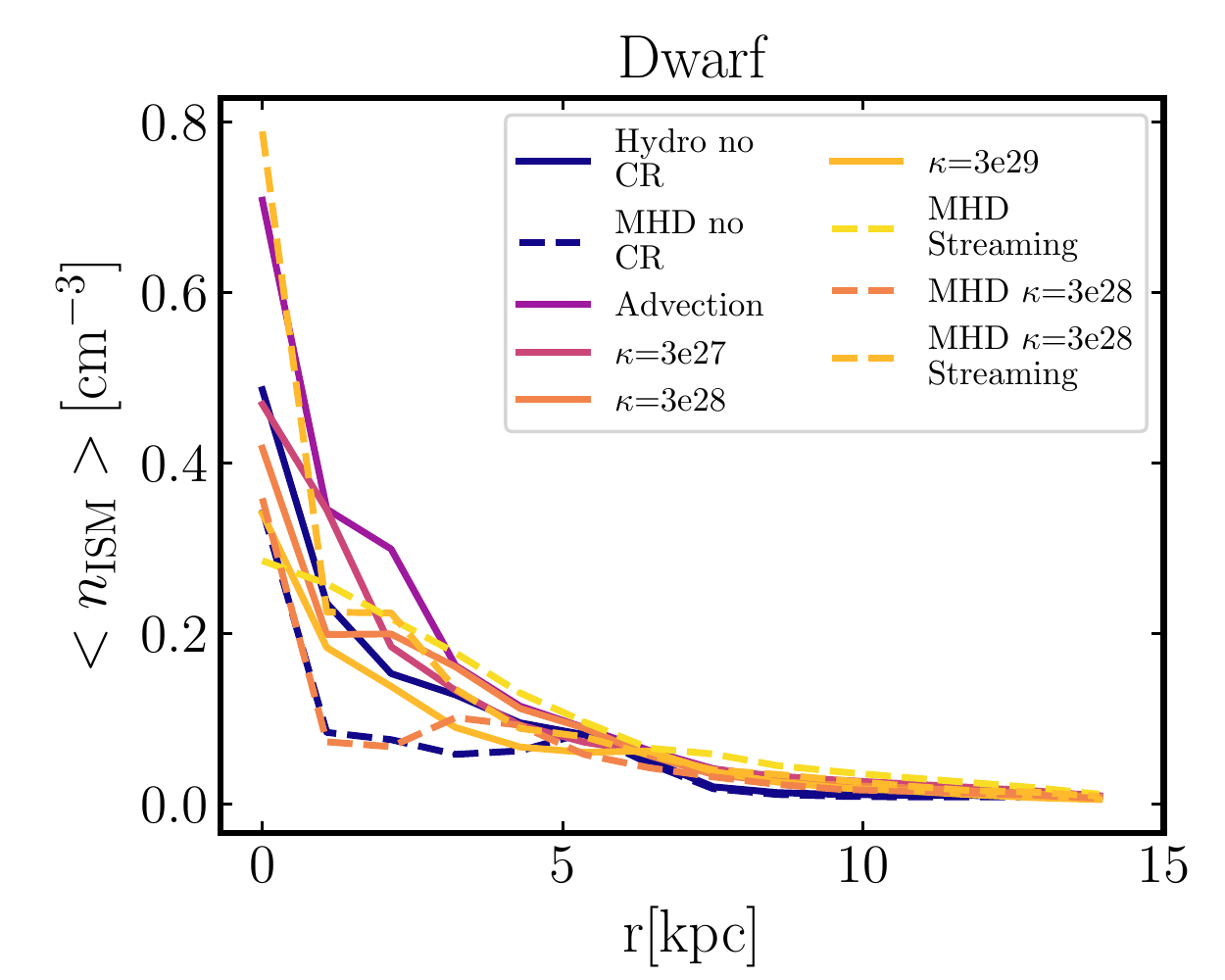}
   \includegraphics[width={0.45\textwidth}]{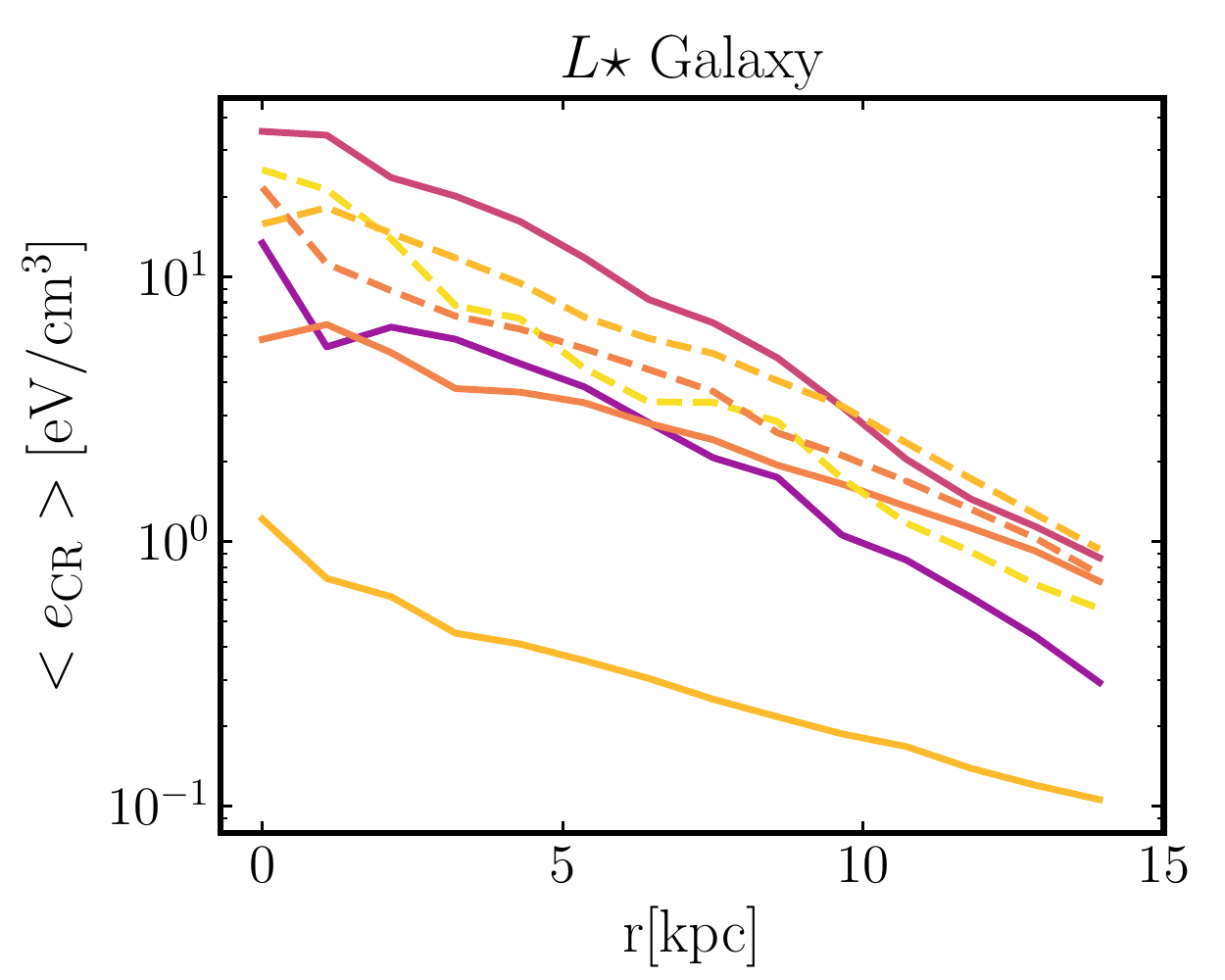}
  \includegraphics[width={0.45\textwidth}]{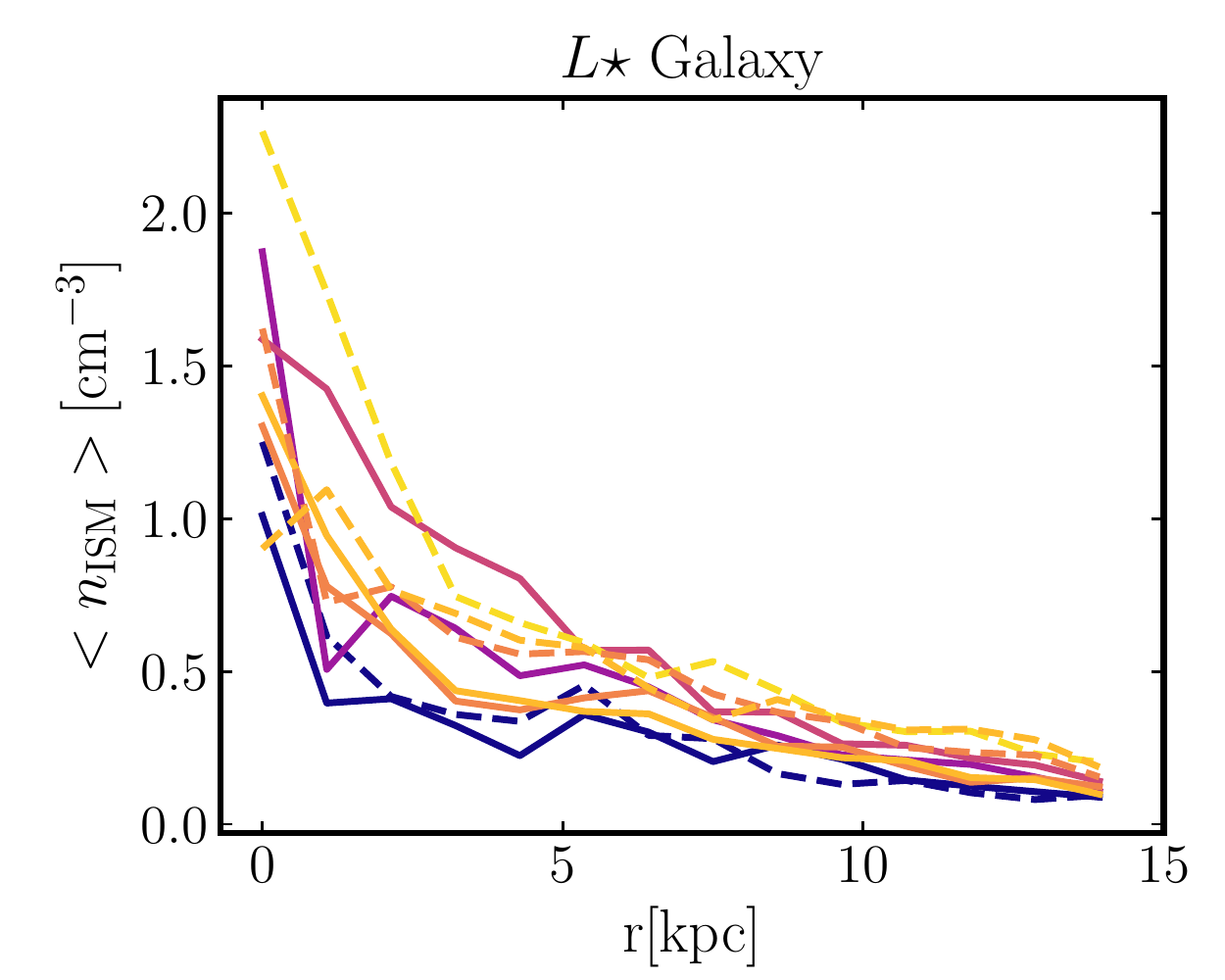}
  \end{centering}
\caption{{\it Left:} Mid-plane value of CR energy density (averaged in a $200\,$pc-thick slab) at $500\,$Myr, for {\bf Dwarf} and $L\star$ galaxies. {\em Right:} Mid-plane gas density. The gas density does not have an obvious dependence on the CR propagation models, as the latter influences both the midplane pressure and the gas flows from/onto the gas disk.}
\label{CRgasmid}
\end{figure*}

\begin{figure}
\begin{centering}
  \includegraphics[width={0.95\columnwidth}]{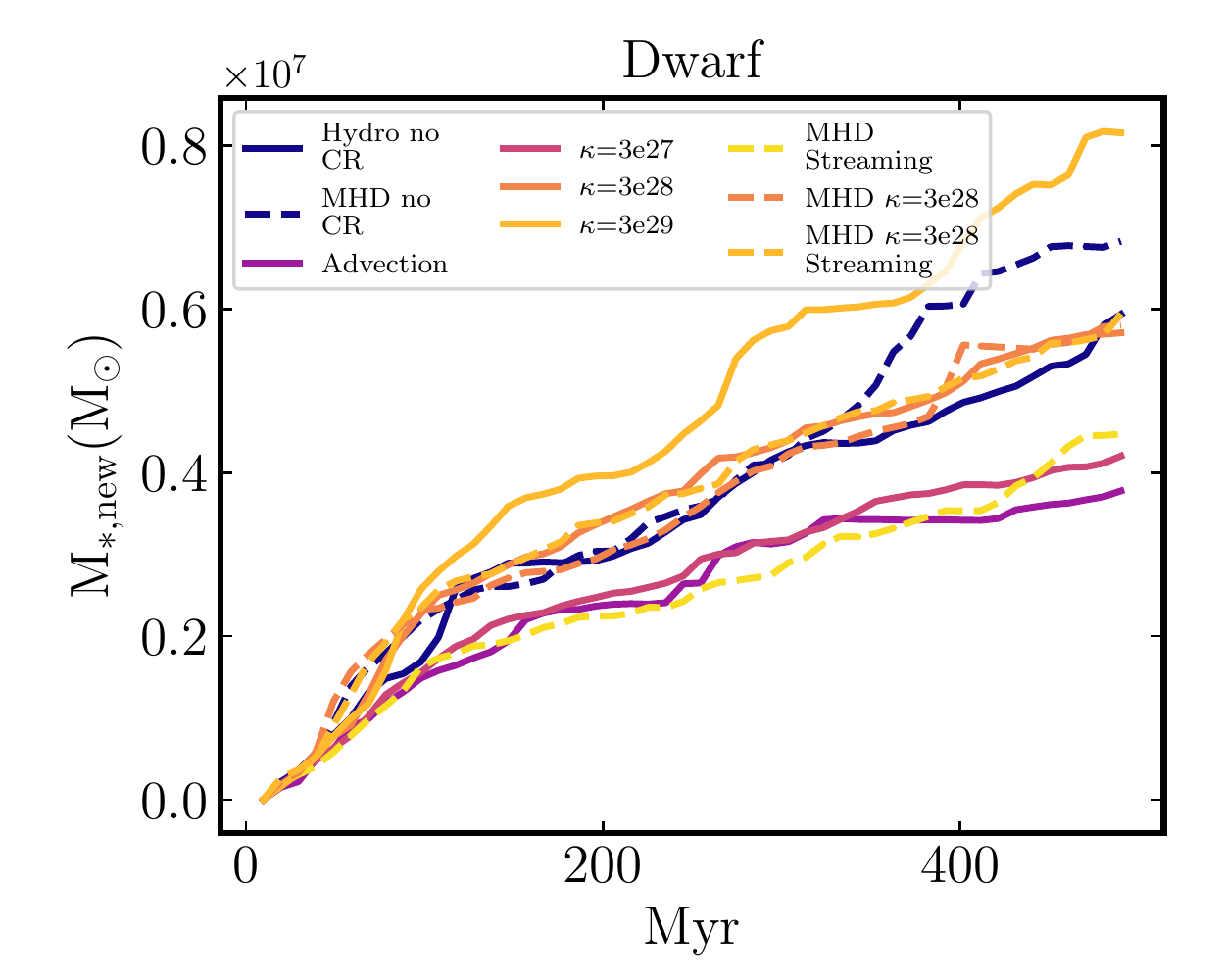}
\includegraphics[width={0.95\columnwidth}]{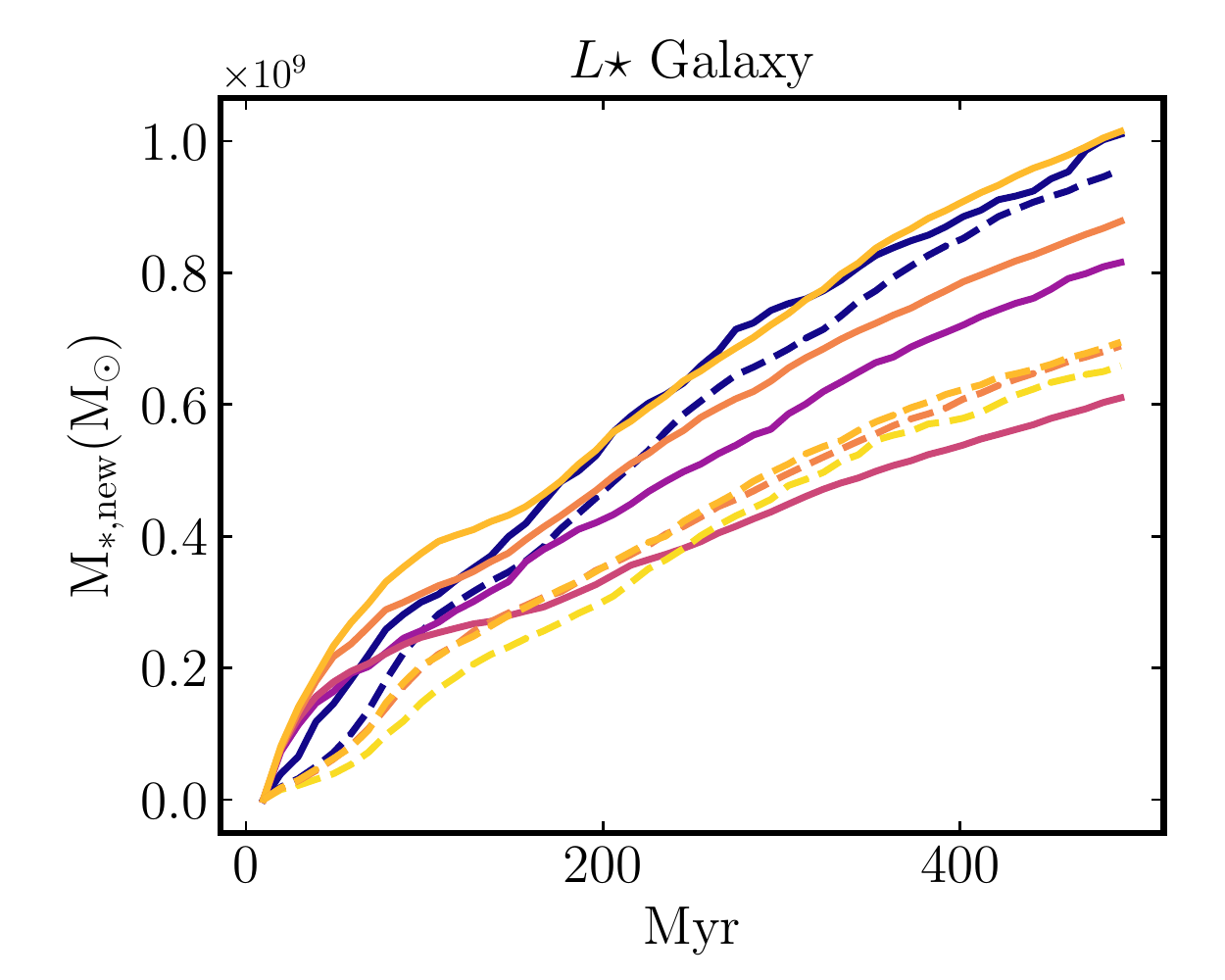}
  \includegraphics[width={0.95\columnwidth}]{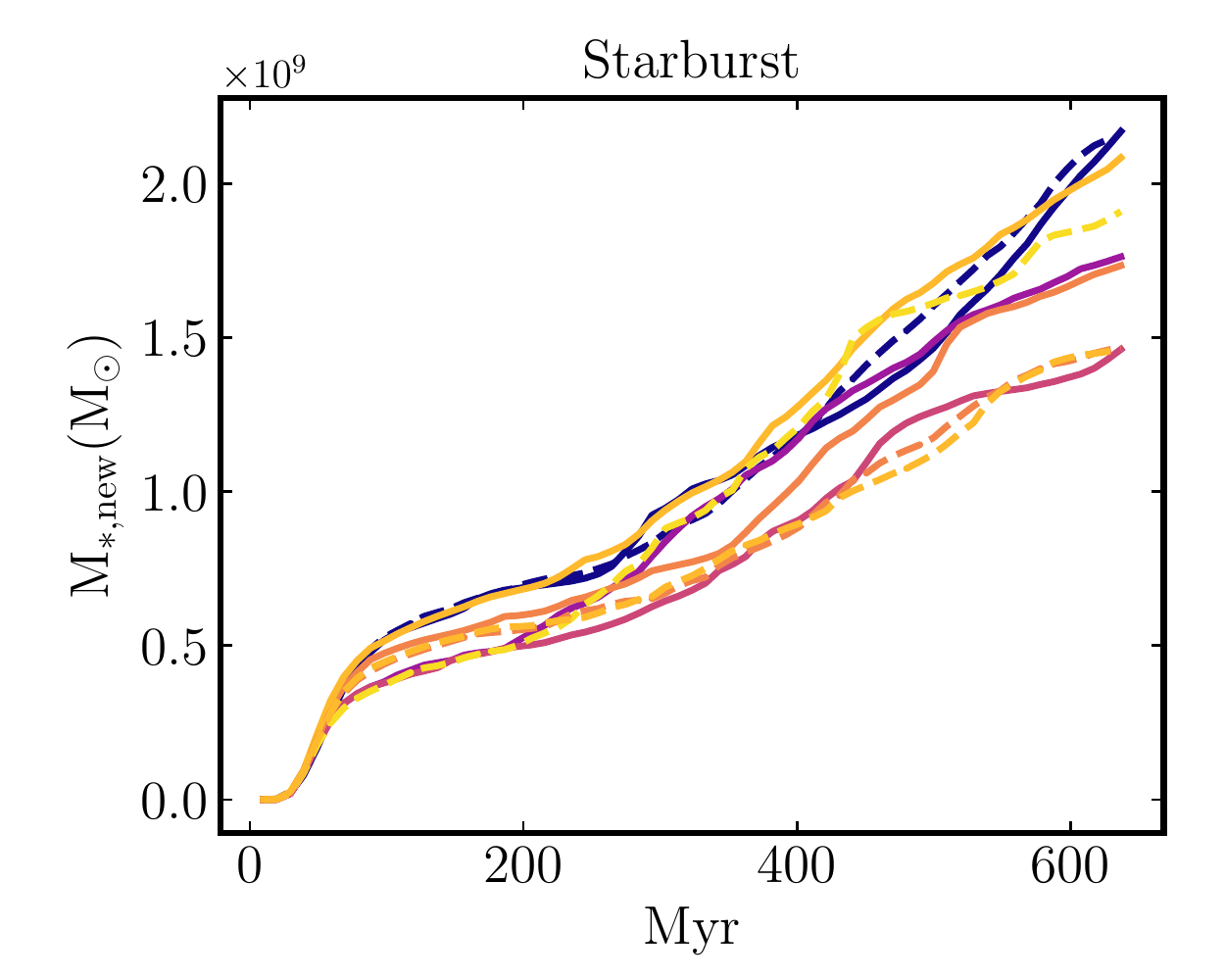}
  \end{centering}
  \vspace{-0.25cm}
\caption{Cumulative mass of stars formed (after CRs are ``turned on''), in different galaxies (labeled) with different CR propagation models (Table \ref{CRprop}). CRs with very low diffusivity (e.g.\ $\kappa=3{\rm e}27$, ``MHD Streaming'' or ``Advection'' models) can modestly suppress star formation (by factors $\sim 1.5$ relative to models without CRs), while models with larger diffusivity have no effect (or even slightly enhance SFRs).}
  \vspace{-0.25cm}
\label{sfr}
\end{figure}

We begin with a brief overview of the evolution of the gas, CRs and magnetic fields in our {\bf $L\star$ Galaxy} simulations with different CR propagation models. Once initial transients are damped away, the galaxy has a relativity steady, low SFR with weak galactic winds driven by SNe and other stellar feedback processes. The magnetic fields amplify and develop irregular yet roughly toroidal morphology through dynamo action \citep{Su18}. After 150 Myr when we turn on CRs, the galaxy is in approximate steady state. SNe inject a fraction of energy into the surrounding gas as CRs, which is transported via advection, diffusion, or streaming. The timescale for CR hadronic and Coulomb losses is long enough that steady-state CR pressure support can arise within/around the galaxy. The total CR energy at any time roughly follows the CR injection from SNe, which is proportional to the total stellar mass formed (see Figure \ref{sfr} and the related text). However, there are also other CR energy gain and loss processes, which we will investigate in \S~ \ref{comparisongamma}. But in all runs, the CR source distributions are much more concentrated than CR densities, as CRs move from their ``birthplace''. 

Figure \ref{MWcrden} shows the distribution of the CR energy density in a 60kpc$\times$60kpc slice centered on our simulated {\bf $L\star$ Galaxy}. Runs with higher diffusion coefficients result in lower CR energy density at the galaxy center but develop stronger CR pressure away from the disks. The strong CR pressure gradients continue accelerating gas out to a large radius in the radial direction, although stellar feedback without CRs can also drive winds. In Figure \ref{MWden}, we also show that galaxies with CR diffusion in general have the smoother CGM structure, and outflowing gas is present further from the galactic centers. 
The study of CR driven outflows, including a comparison with thermally-driven outflows and their effect on the CGM, will be presented in a companion paper (Chan et al., in preparation).  Simulations with only streaming ($\kappa=0$ but $v_{\rm st}\ne 0$) are similar to cases with very low diffusivity ($\kappa \lesssim 3\times10^{27}\,{\rm cm^{2}\,s^{-1}}$), where CRs are largely confined to the galaxy. From Figure \ref{MWcrden} it is clear that combining streaming with diffusion lowers the concentration of CRs in the disk plane and spreads them to larger distances. In almost all of our runs with non-negligible diffusion ($\kappa \gtrsim 10^{27}\,{\rm cm^{2}\,s^{-1}}$), diffusion makes the CR energy distribution approximately spherical, as opposed to flattened (only the streaming-only and advection-only runs show strongly flattened $e_{\rm cr}$, as the CRs do not efficiently escape the star-forming disk).

These scalings are easy to qualitatively understand. In the ISM, bulk transport speed for streaming is typically, $v_{\rm st}\sim 10\,{\rm km\,s^{-1}}(B/\mu {\rm G})(n_{\rm ISM}/0.1\,{\rm cm^{-3}})^{-1/2}$ (see \citealt{Su18}), giving a transport time through a gas halo with a radius $\ell$ of $t_{\rm st} \sim 100\,\,{\rm Myr\;}(\ell/{\rm kpc})(v_{\rm st}/10\,{\rm km\,s^{-1}})^{-1}$, while the corresponding diffusive transport velocity/time is $v_{\rm di} \sim 330\,{\rm km\,s^{-1}}(\kappa/10^{29}{\rm cm^{2}\,s^{-1}})(\ell/{\rm kpc})^{-1}$ and $t_{\rm di} \sim 3\,{\rm Myr}\,(\ell/{\rm kpc})^2(\kappa/10^{29}{\rm cm^{2}\,s^{-1}})^{-1}$ . Thus even for quite low $\kappa$, the diffusive flux dominates transport on sub-kpc scales.

But because the diffusion time scales as $\sim \ell^{2}$, if the CRs establish a smooth profile with scale length $\gtrsim 1-10\,$kpc, then on the larger scales the diffusion time eventually could become larger than the streaming transport time, i.e. outside a scale $\ell \sim 30\,{\rm kpc}\,(\kappa/10^{29}{\rm cm^{2}\,s^{-1}})(v_{\rm st}/{\rm 10\,km\,s^{-1}})^{-1}$.

We quantify the above observations with Figure \ref{CRgasz}, which shows the cumulative distribution of CR energy in the {\bf Dwarf} and {\bf $L\star$ Galaxy} runs. CR energy density is most extended vertically in simulations with the largest diffusion coefficients and it is most concentrated in {\bf Advection} simulations. We define the (3D) CR scale radius $r_{\rm cr, 1/2}$ such that the sphere with $r_{\rm cr, 1/2}$ encloses one half of the total CR energy. In the {\bf $L\star$ Galaxy}, we find $r_{\rm cr, 1/2}$ is around 3 kpc in run ``${\rm \kappa=3e27}$'', but it increases to around 10 kpc in run ``${\rm \kappa=3e28}$'' and 30 kpc in run ``${\rm \kappa=3e29}$.'' Similarly, the scale-height of the CR energy distribution also increases with increasing $\kappa$. Trends of the CR scale radius with $\kappa$ can be understood with a diffusion model where the CR injection time ($E_{\rm cr}/\dot{E}_{\rm inj}$) is comparable to the CR diffusion escape time ($t_{\rm di}\sim r_{\rm cr,1/2}^2/\kappa$) where $\dot{E}_{\rm inj}$ is the CR energy injection rate. Assuming a similar injection time, we find that $r_{\rm cr,1/2}$ is roughly proportional to $\sqrt{\kappa}$, so a faster diffusion leads to a more extended CR distribution.

We show the CR and gas mid-plane densities for our {\bf $L\star$ Galaxy} and {\bf Dwarf} runs in Figure \ref{CRgasmid}. For both galaxy types, CR density profiles are significantly ``flatter'' (more extended and less centrally-concentrated) with higher $\kappa$. Consequently, in runs with fast diffusion, CRs have smaller impact on the central region of galaxies, providing less pressure support to the central gas, but they can be more important in the CGM. Interestingly, the ``{\bf Advection}'' runs have lower CR central densities than ``${\rm \kappa=3e27}$'' because of the smaller adiabatic energy gain (a highly non-linear effect), which we will discuss in the next section and Figure \ref{CRenergies}.
The gas midplane density depends rather weakly on CRs. Low-$\kappa$ (or streaming/advection-only) runs have slightly higher midplane densities while higher-$\kappa$ runs have midplane densities similar to ``no CR'' runs. 
  
This is likely caused by additional pressure support from CRs trapped in the midplane in low-$\kappa$ runs, which allows gas to reach higher densities before fragmenting and forming stars.

It is interesting to compare the CR energy density in our $L\star$ galaxy model with that observed near the solar circle ($e_{\rm cr}\sim 1\;{\rm eV/cm^3}$, see e.g.\ \citealt{Gren15}), but we must recall that the $L\star$ model was {\em not} constructed to be an exact MW analogue. For example, it has a more steeply-rising central gas density, without the gas deficiency that appears in the center of the MW (i.e. it does not have a ``star-forming ring'' and corresponding ``hole'' in the central few kpc), and the gas densities at $\sim 8\,$kpc from its center are lower than the $\sim 1\,{\rm cm^{-3}}$ in the solar neighborhood \citep{Mosk02,Cox05}.

Nevertheless, the model has a stellar mass, gas mass, and SFR similar to the MW. Our runs with isotropic $\kappa \le 3\times 10^{27}\,{\rm cm^{2}\,s^{-1}}$ produce a mid-plane $e_{\rm cr}$ at $\sim 8\,$kpc from the galaxy center which is high relative to the observed value, those with isotropic  $\kappa \ge 3\times 10^{29}\,{\rm cm^{2}\,s^{-1}}$ are lower, while those in-between are reasonably consistent. Turning on MHD (making the diffusion anisotropic) increases $e_{\rm cr}$ by factors $\sim 2-3$, consistent with the isotropically-averaged $\kappa$ being lower by a similar factor (as expected), so values $\kappa \sim 1-3\times10^{29}\,{\rm cm^{2}\,s^{-1}}$ are marginally more favored. Given the lack of a detailed match between our models and the MW, stronger constraints on CR propagation come from $\gamma$-ray emission in \S~\ref{gammaray}. 

Fig.~\ref{sfr} shows cumulative SF histories: akin to the disk midplane-pressure effects above, CR runs with very low $\kappa$ suppress SF by modest factors $\sim 1.5-2$, an effect which vanishes at higher $\kappa$. Smaller variations ($\sim 10\%$-level) are generally dominated by stochastic run-to-run variations. Runs with MHD generally show slightly higher SFRs (all else equal), an effect discussed in detail in \citet{Su17,Su18}, but the stochasticity during the early evolution can wash out such effect.

\begin{figure}
\begin{centering}
 \includegraphics[width={0.95\columnwidth}]{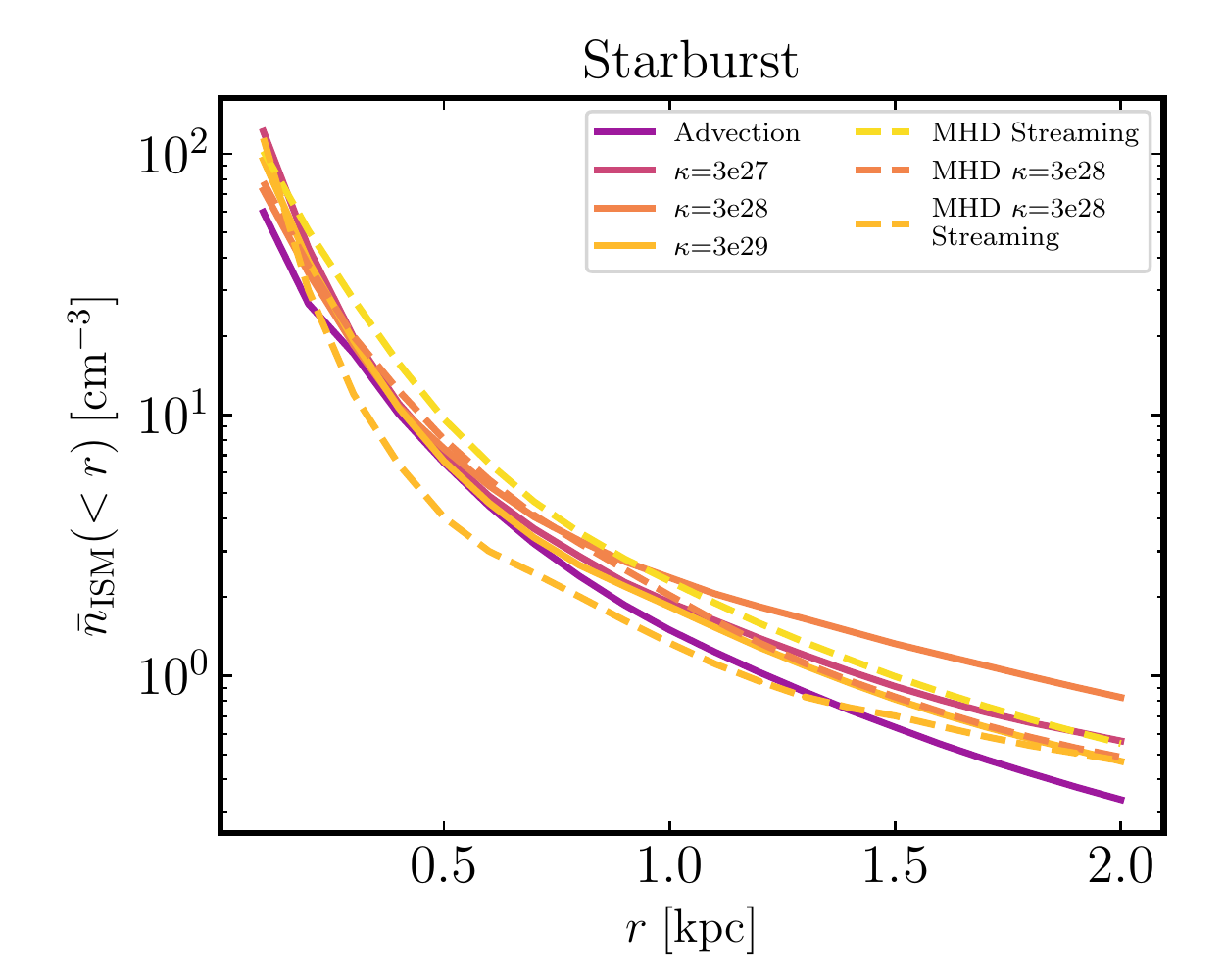}
  \includegraphics[width={0.95\columnwidth}]{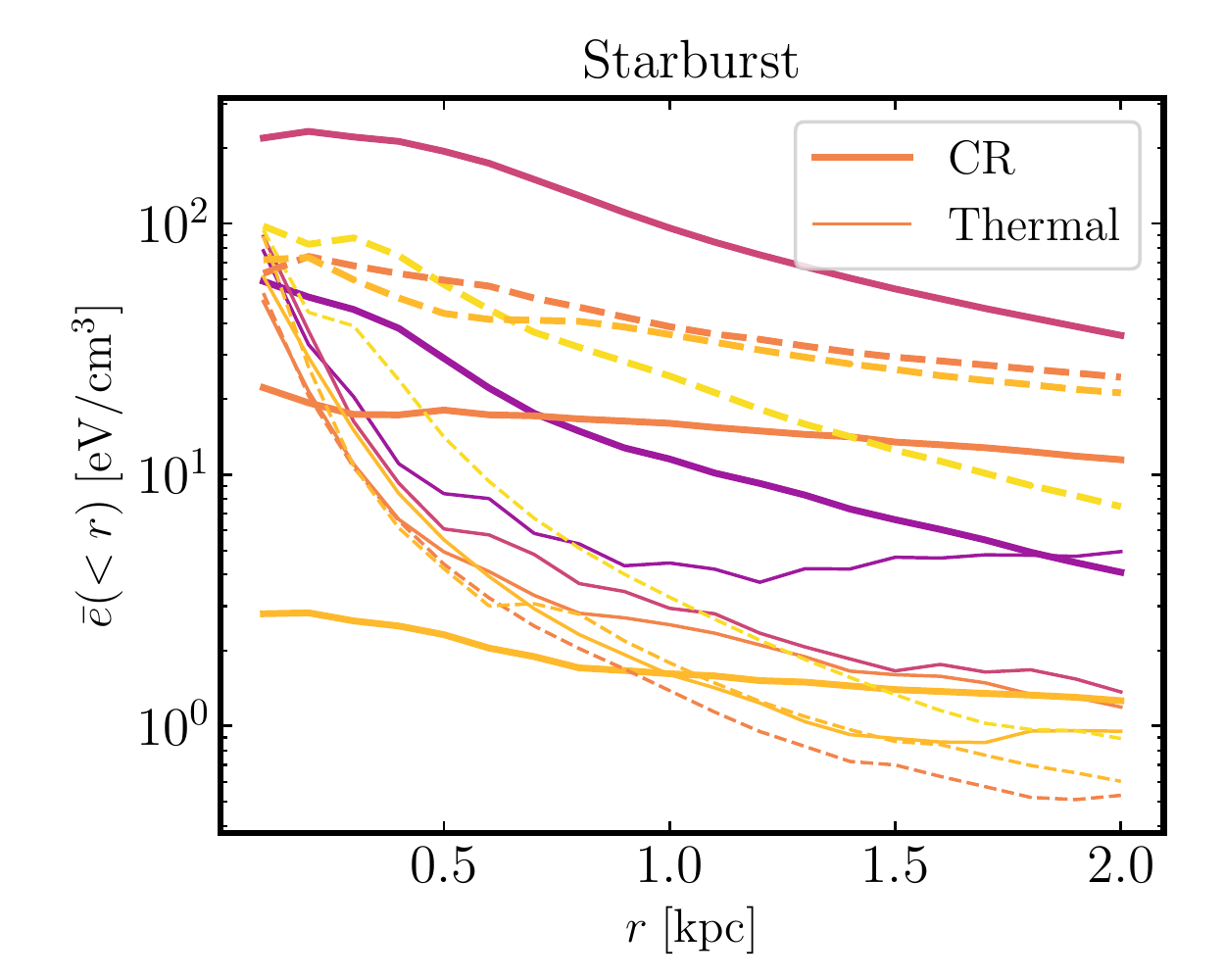}
  \end{centering}
\caption{{\it Upper}: Gas densities averaged within a spherical radius $r$, in {\bf  Starburst}, in the snapshots when the central gas densities are the highest in each run. The high gas density is similar to that of observed starburst galaxies (to which we compare below).
{\it Lower}: CR ({\it thick}) and thermal ({\it thin}) energy densities at the same times. CR energy densities are higher than thermal when the gas densities peak (but still generally less than turbulent energy densities). }
\label{avestarburst}
\end{figure}

\subsubsection{Starburst galaxy}
The {\bf Starburst} model is designed to reach high SFRs and gas densities, which are transient phenomena since strong stellar feedback after the starburst will disrupt the galaxy and reduce the gas density. Thus, our {\bf Starburst} run reaches SFR peaks of $\sim 5\,\msun\,{\rm yr^{-1}}$ with highest central gas densities $\sim 100\,{\rm cm^{-3}}$ or (edge-on) surface densities $\sim 0.1\,{\rm g\,cm^{-2}}$ ($\sim 500\,{\rm \msun\,pc^{-2}})$ (compare Fig.~\ref{avestarburst} and \citealt{Weis01}), which last $\sim 10\,$Myr. Between starbursts the galaxy has lower SFR and gas densities, with correspondingly longer hadronic loss times, so CRs can escape more easily.

Fig.~\ref{avestarburst} shows that CR energy densities during starburst phases are around 100 ${\rm eV/cm^3}$ with slow transport, similar to the value inferred from modeling the observed $\gamma$-ray spectra of e.g.\ M82 (\citetalias{Lack11} and \citealt{Yoas16}). Although these are high relative to the MW, and a factor of several higher than the thermal (or magnetic) energy densities, they are lower than the pressure required for hydrostatic balance ($\pi G \Sigma_{\rm g} \Sigma_{\rm g}\sim 10^{3-4} {\rm eV/cm^3}$), which is primarily comparable to the kinetic energy density in these galaxies (with turbulent velocity dispersions similar to those observed, $\sim 50-100\,{\rm km\,s^{-1}}$). Our findings are therefore consistent with earlier claims by \cite{Lack10}, \citetalias{Lack11}, and others who showed that CRs are dynamically unimportant at least in the cores of the starbursts, but they might be more important away from the central dense region.

\subsection{Pionic $\gamma$-ray emission as a measure of CR propagation}

\label{gammaray}
\begin{figure*}
\begin{centering}
\includegraphics[width={0.95\textwidth}]{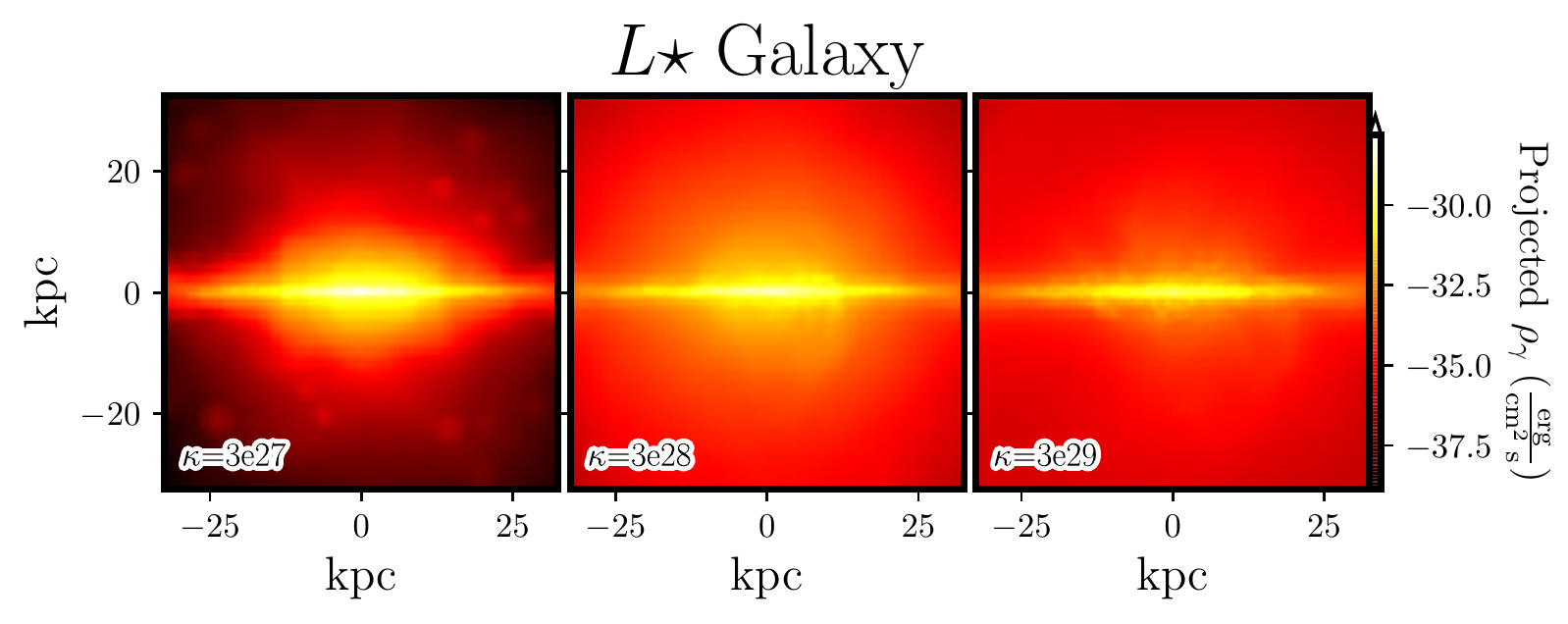}  
\end{centering}
\caption{Projected pionic $\gamma$-ray surface brightness ($E_{\gamma}$ $>1$\,GeV) with different isotropic diffusion coefficients in {\bf $L\star$ Galaxy} at $t=500$\,Myr. $\gamma$-ray emission is stronger and more compact for lower $\kappa$.}
\label{MWgamma}
\end{figure*}

\begin{figure}
\begin{centering}
\includegraphics[width={0.9\columnwidth}]{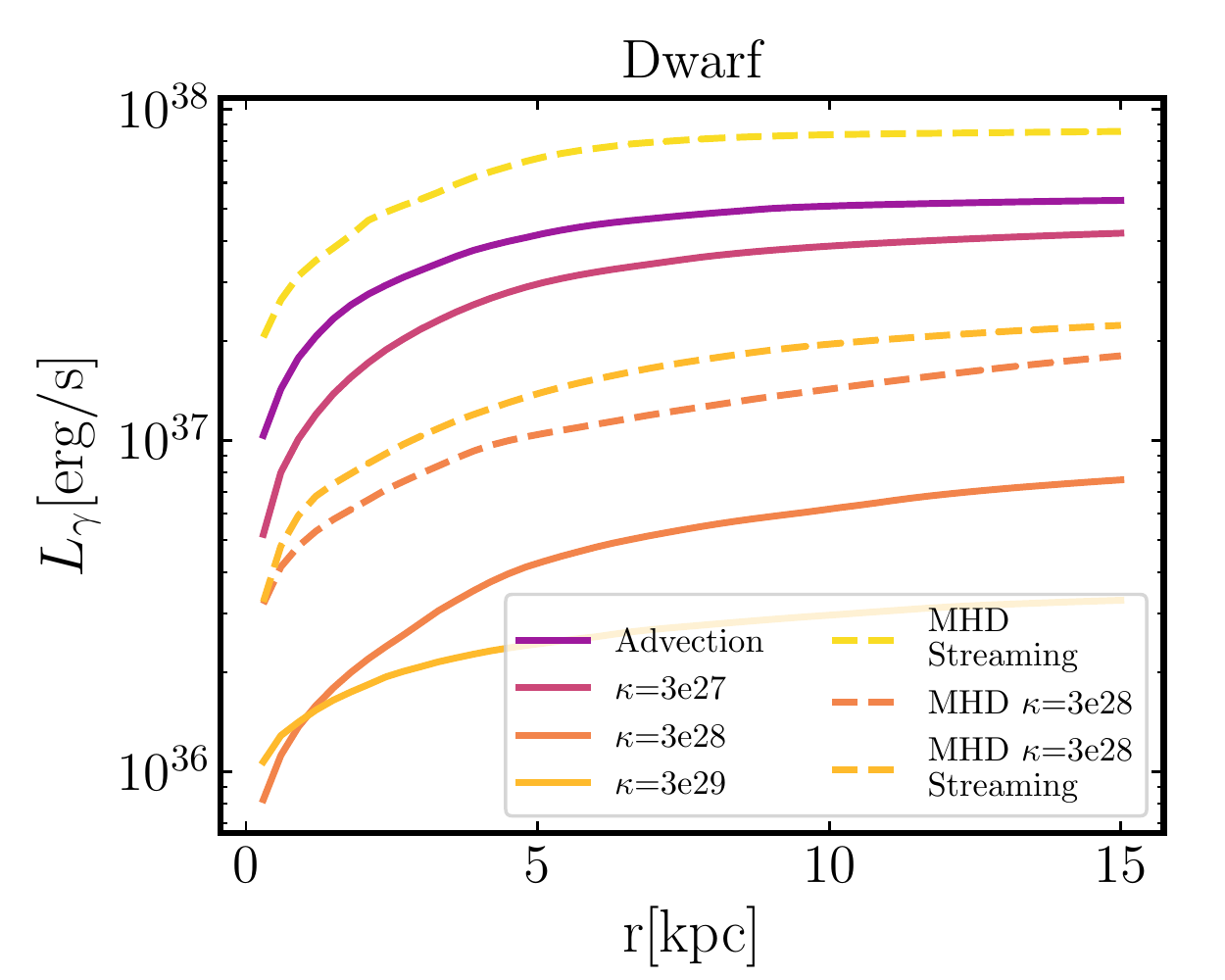}
 \includegraphics[width={0.9\columnwidth}]{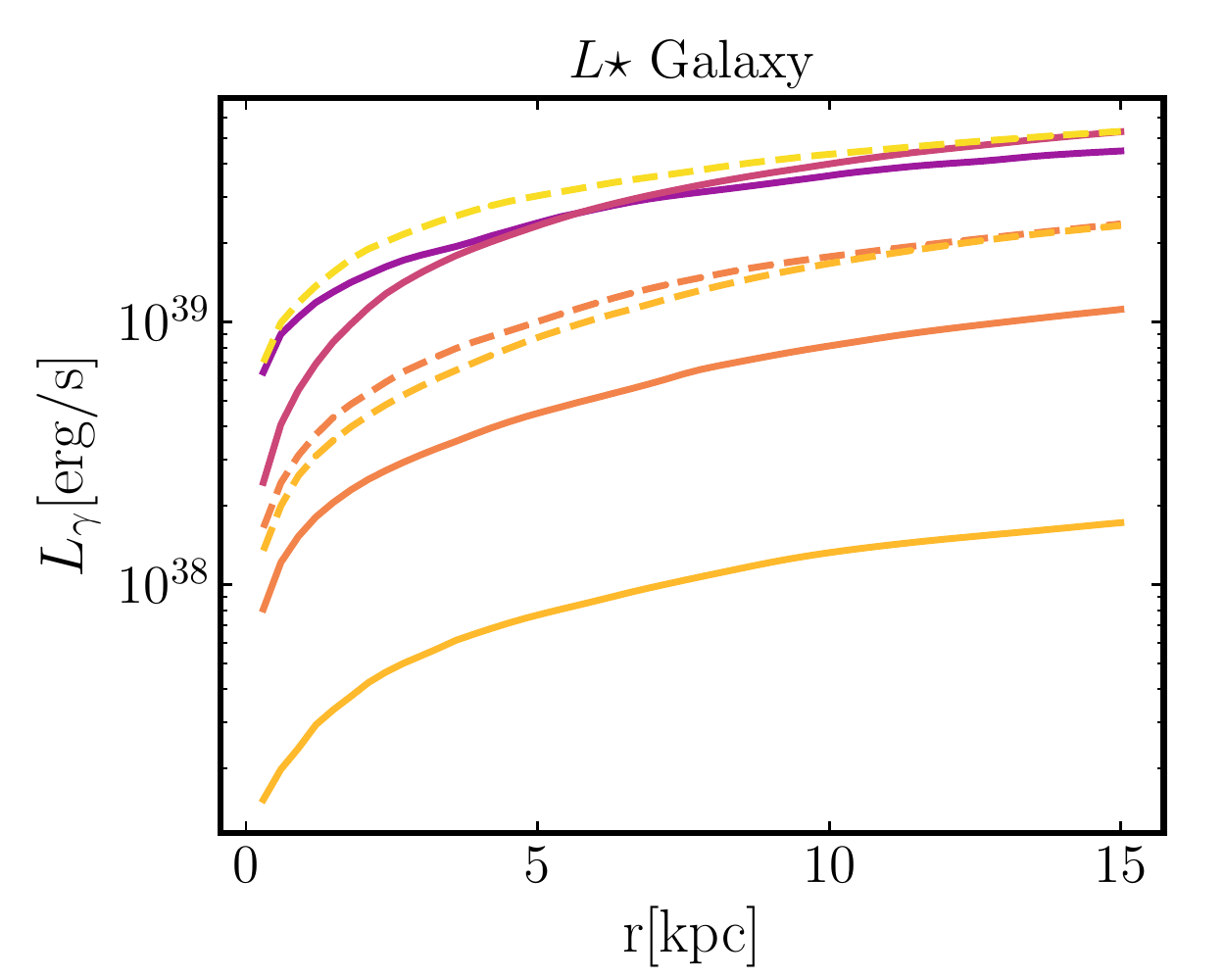}
   \includegraphics[width={0.9\columnwidth}]{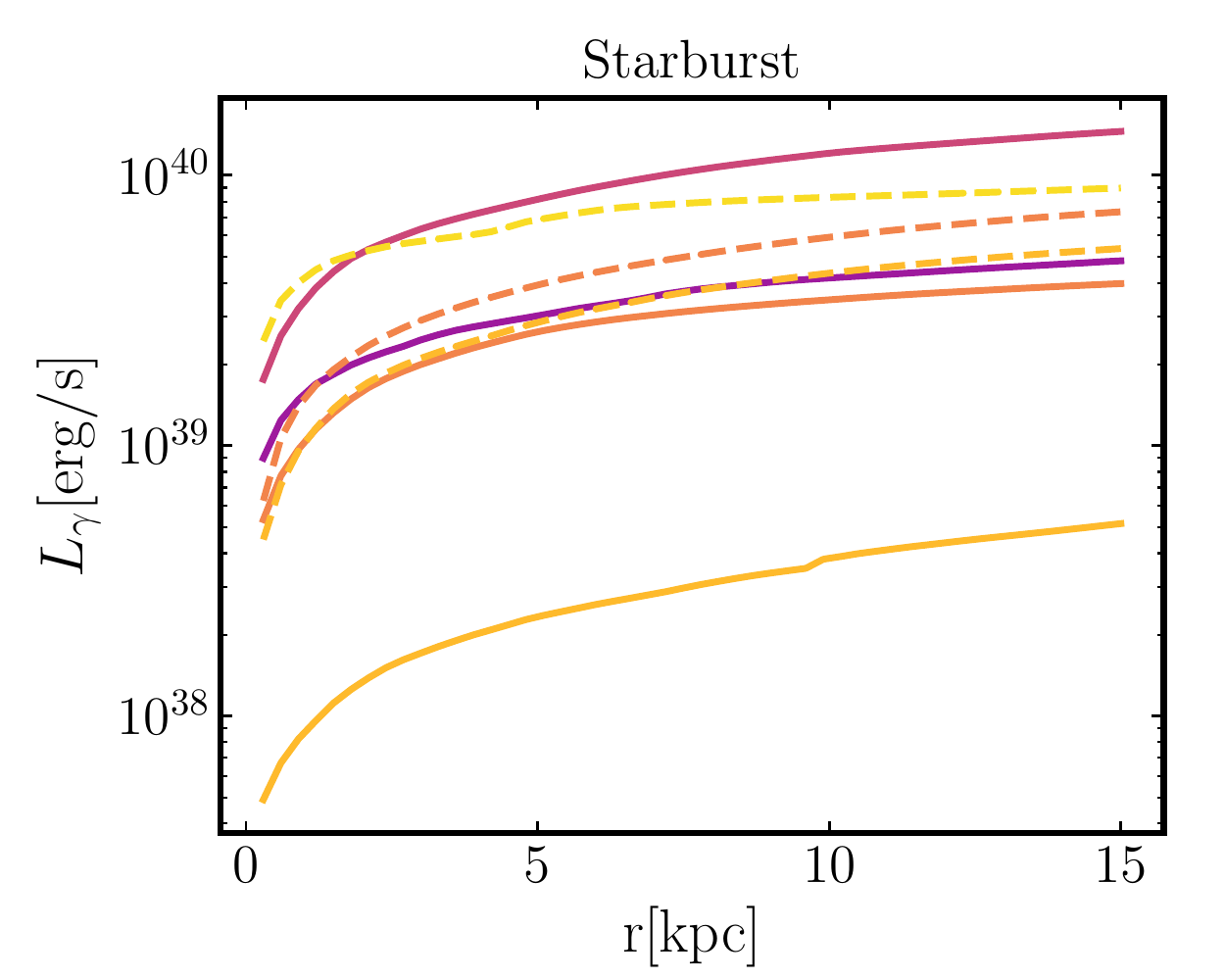}
   \end{centering}
\caption{Cumulative pionic $\gamma$-ray luminosity ($E_{\gamma}$ $>1$\,GeV) as a function of spherical radius (averaged over t = 400 - 500 Myr) in {\bf Dwarf} ({\it upper}), {\bf $L\star$ Galaxy} ({\it middle}) and {\bf  Starburst} ({\it lower}). The $\gamma$-ray luminosity has a spatial extent of a few kpc in dwarf galaxies and more than 10\,kpc in $L_\star$ galaxies.}
\label{gammasph}
\end{figure}

\begin{figure}
\begin{centering}
 \includegraphics[width={0.9\columnwidth}]{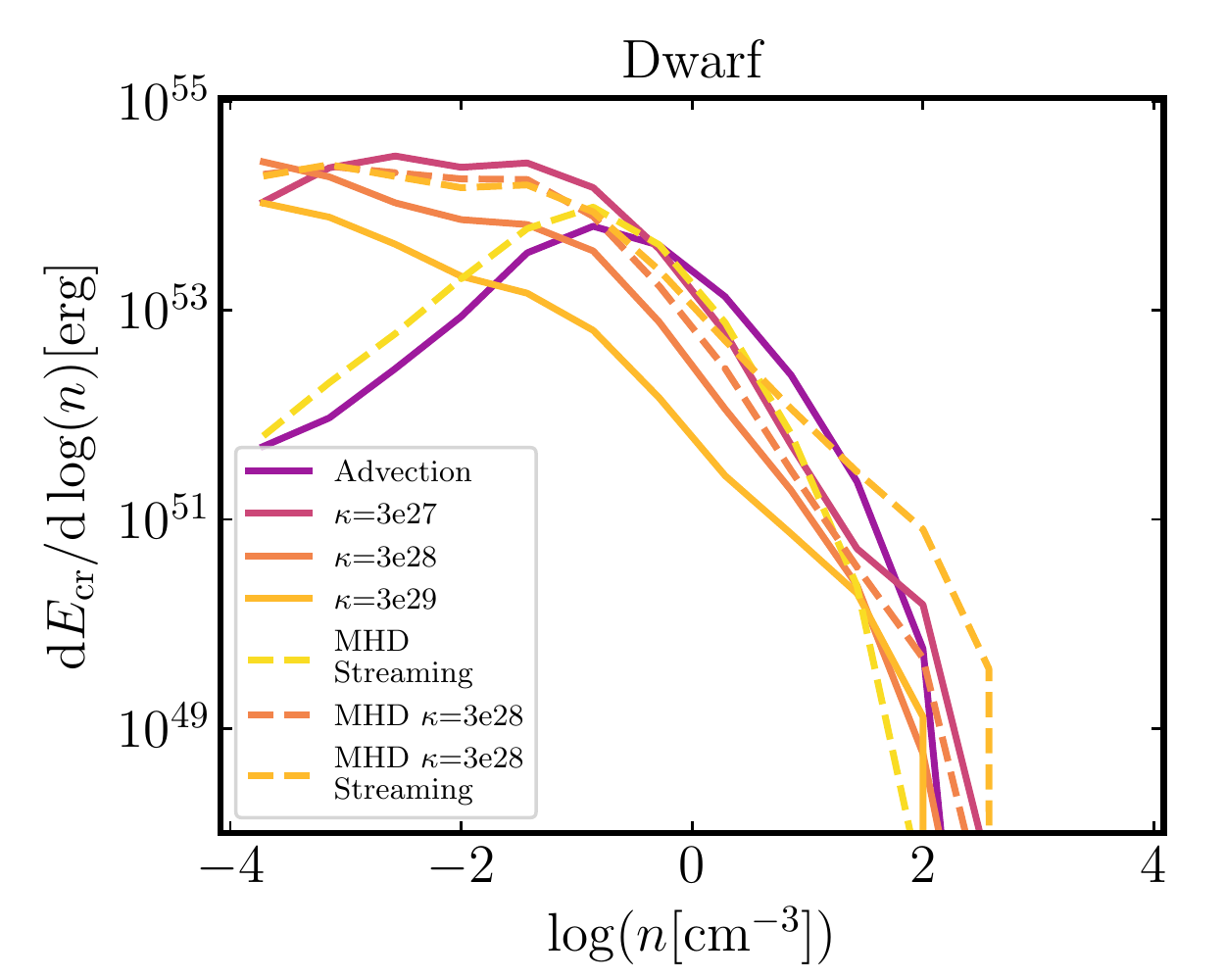}
 \includegraphics[width={0.9\columnwidth}]{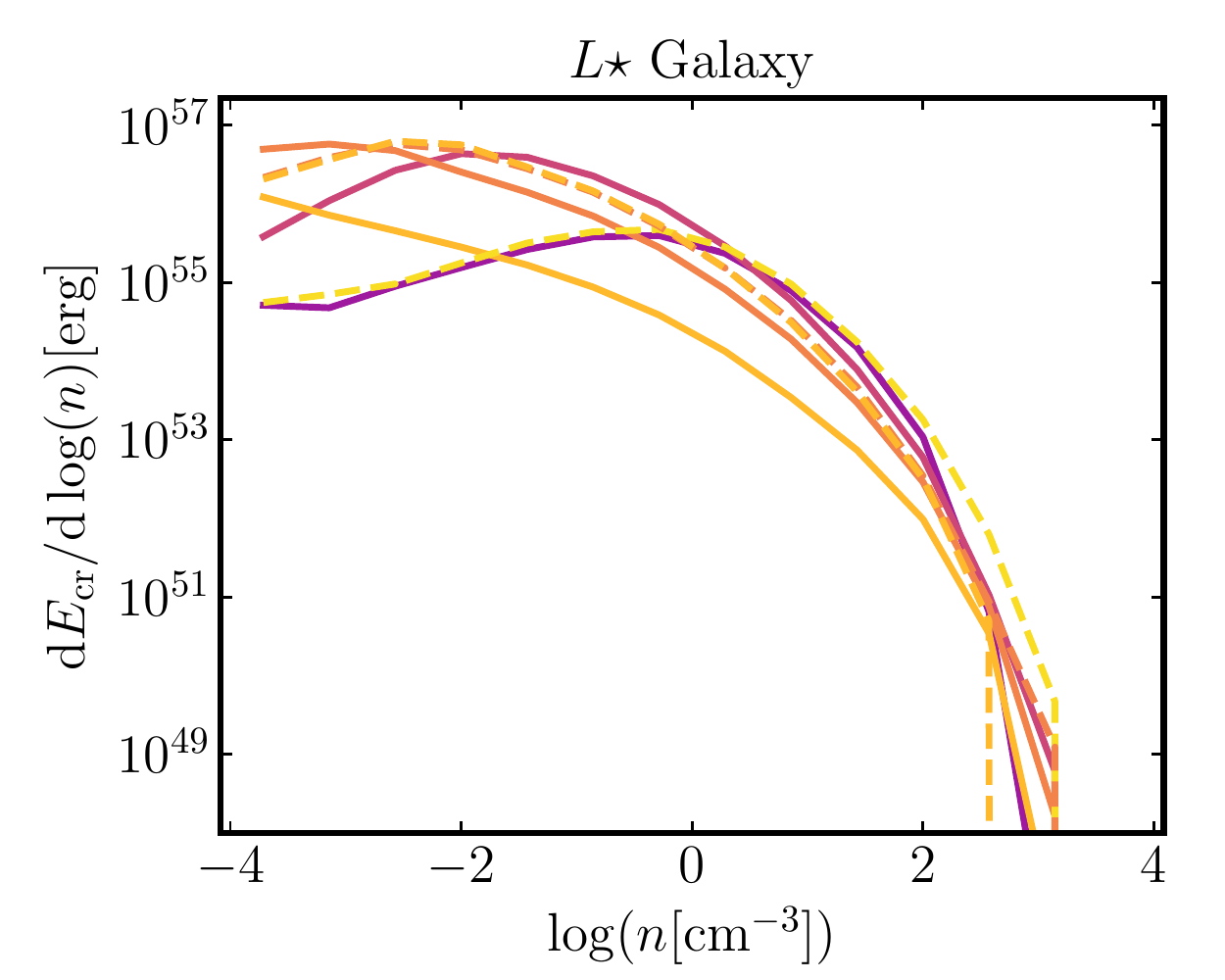}
   \includegraphics[width={0.9\columnwidth}]{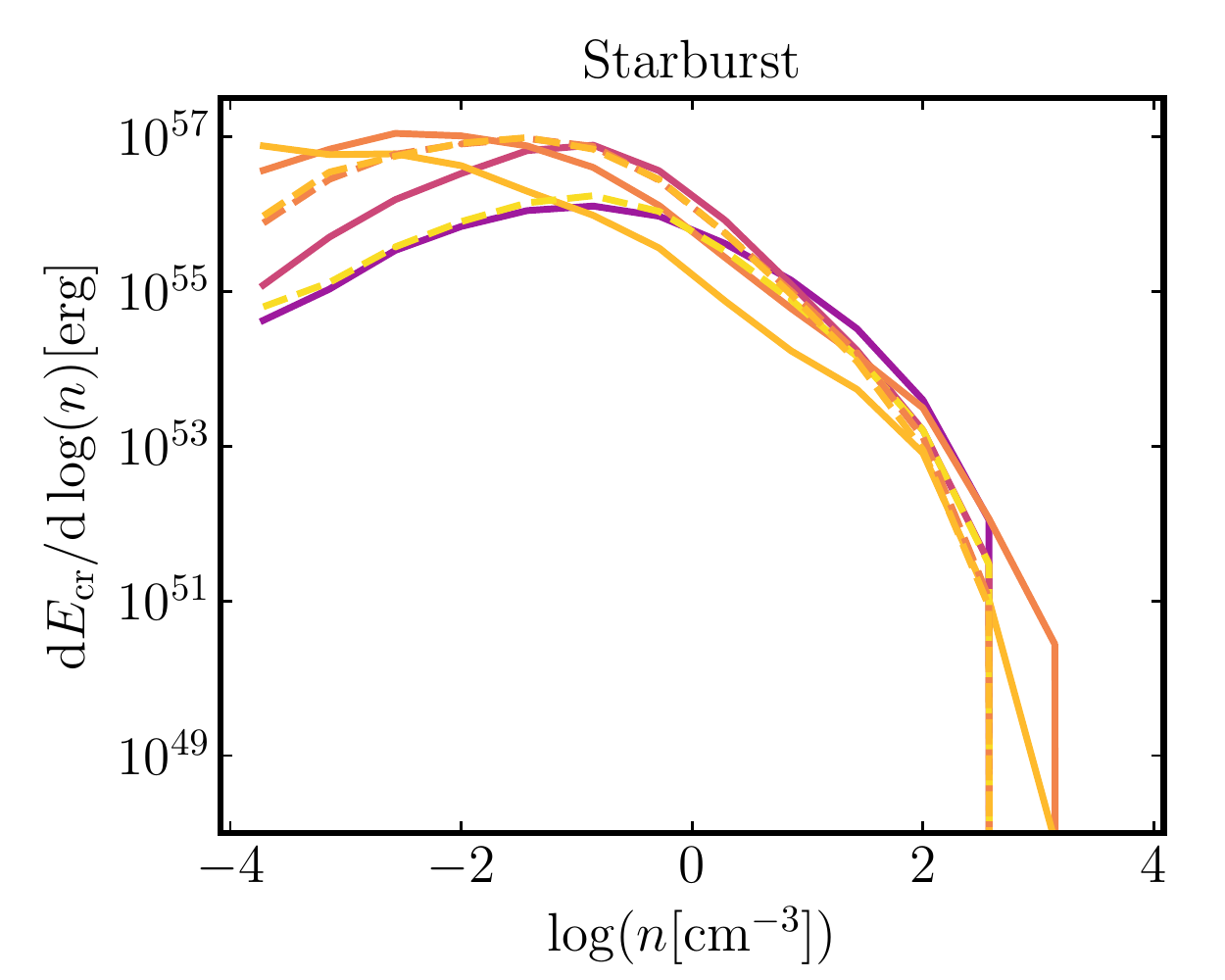}
   \end{centering}
\caption{The CR energy distribution as a function of local ISM or CGM density (at $t = 500$\,Myr), in different galaxies and CR propagation models (Table \ref{CRprop}). CR energy is less concentrated at high densities (e.g.\ within the thin disk, and in dense clouds where SNe explode) when diffusivities are larger, as expected. This reduces the $\gamma$-ray luminosity.}
\label{nismcr}
\end{figure}

\begin{figure}
\begin{centering}
 \includegraphics[width={0.95\columnwidth}]{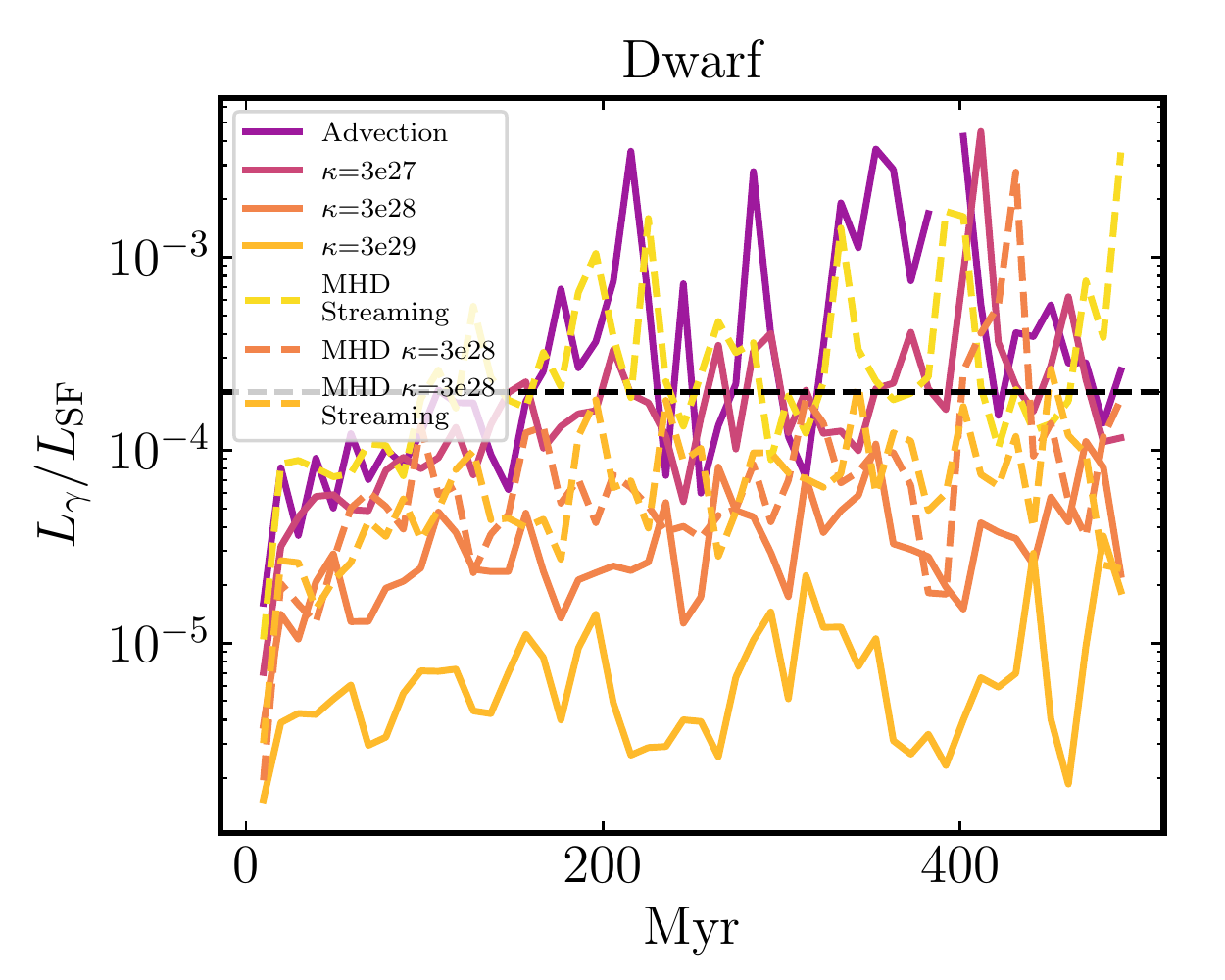}
 \includegraphics[width={0.95\columnwidth}]{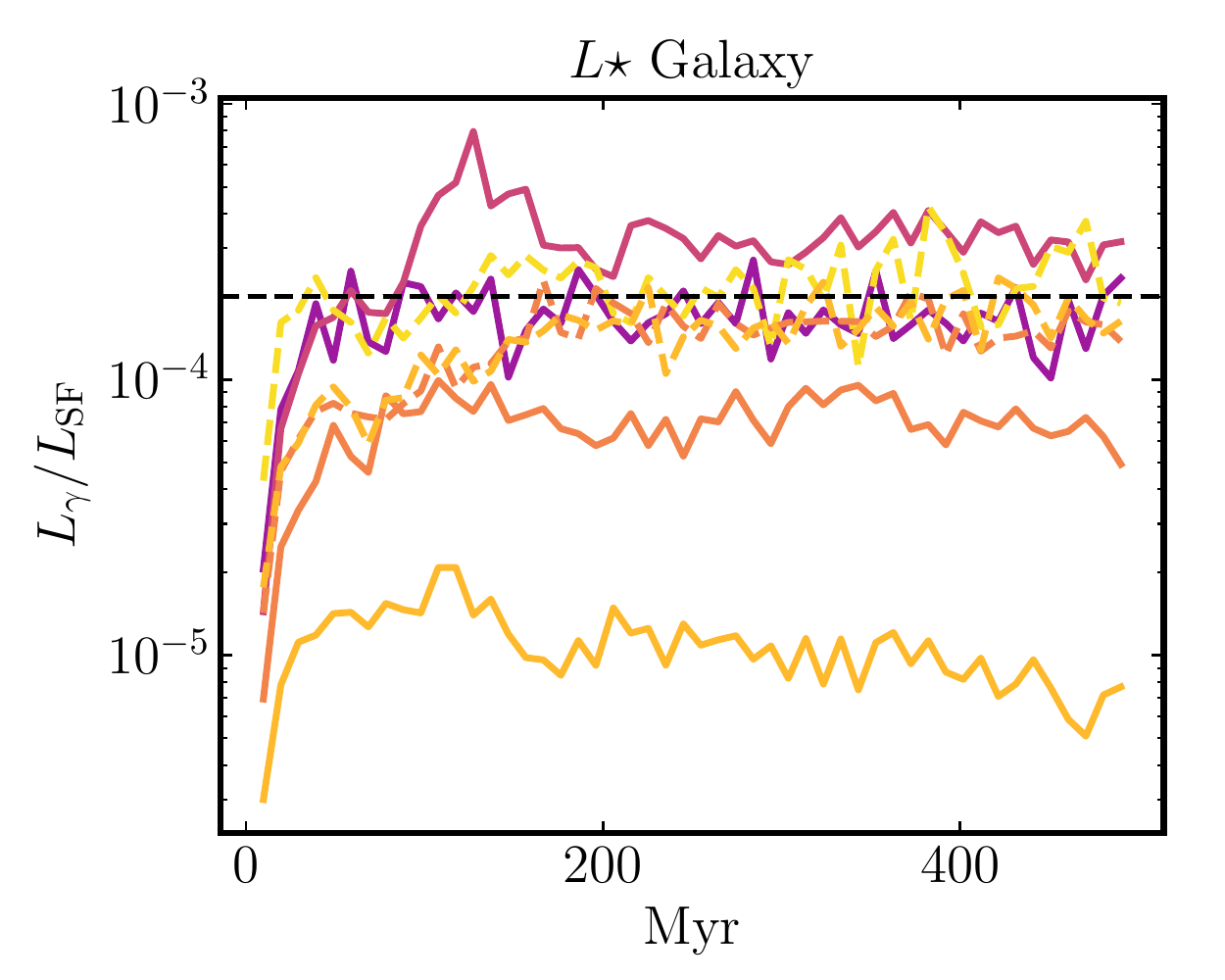}
   \includegraphics[width={0.95\columnwidth}]{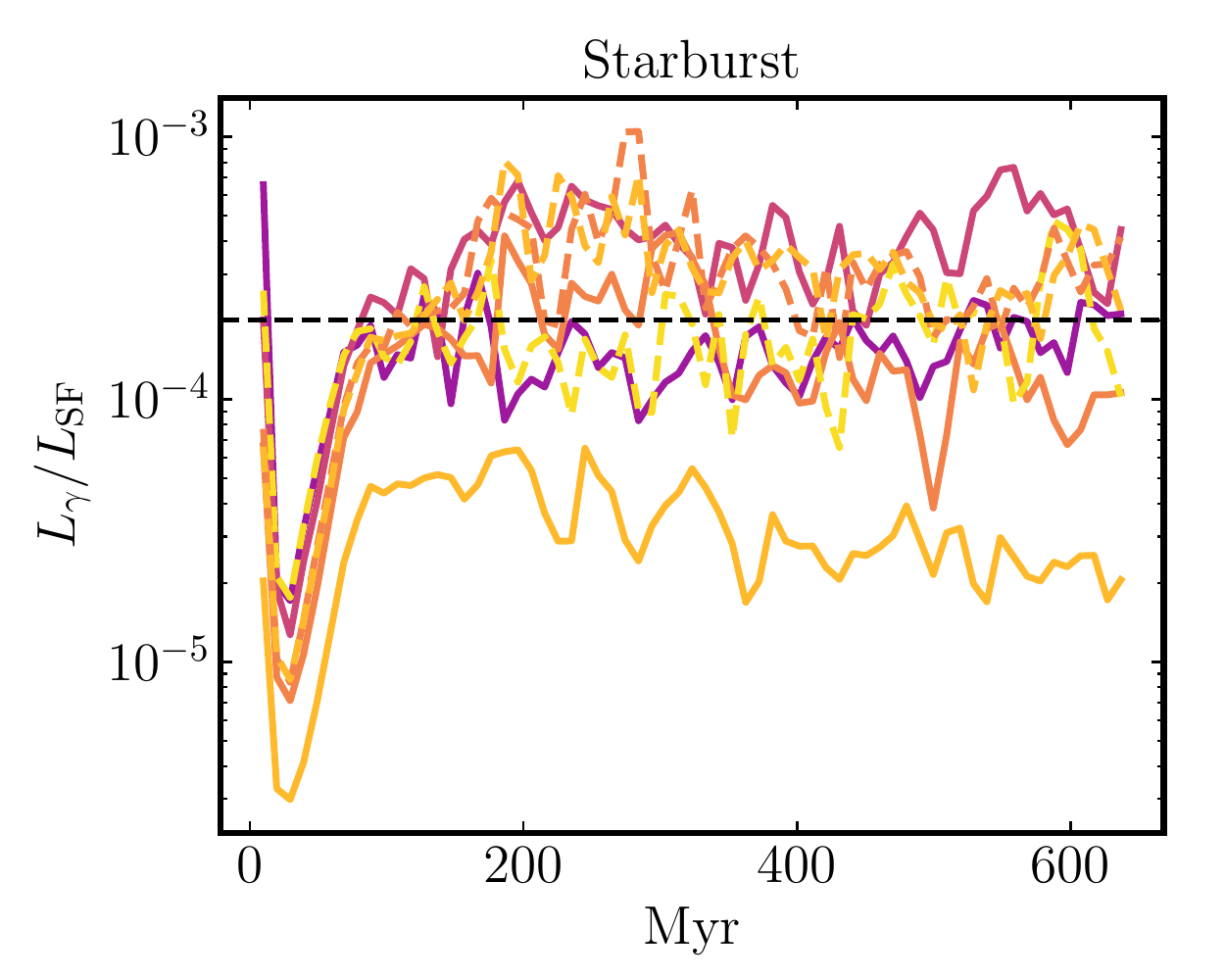}
   \end{centering}
\caption{Ratios between pionic $\gamma$-ray luminosity, $L_{\gamma}(E_{\gamma} >1\,{\rm GeV})$, integrated within $<0.1\,R_{\rm vir}$, and total star formation luminosity (estimated from the SFR averaged over the last $10\,$Myr at each time). Dashed horizontal lines show the calorimetric limit, i.e. CRs cannot escape galaxies and are lost immediately to collisions without gains (see the caveats in the main text). The ratio reaches a steady value after $\sim100$\,Myr and is lower with higher CR diffusion coefficients.}
\label{Fgammasfr}
\end{figure}

Owing to the lack of direct measurements of primary and secondary CRs at low ($\sim$\,GeV) energies from extra galactic sources, pionic $\gamma$-ray emission is one of the few observables that constrain CR propagation outside of the MW. CRs interact with nuclei and produce pions that decay into pionic $\gamma$-rays. While there is a substantial amount of pionic $\gamma$-ray emission with energy < 1 GeV, it is difficult to isolate it observationally owing to contamination by leptonic emission. For $\gamma$-rays with energies $> 1$ GeV, the leptonic emission is less than one tenth of the pionic emission (for CRs with a spectrum consistent with our default model assumptions; see calculations by \citealt{Pfro17gamma}). We will ignore additional potential channels of  $> 1$ GeV $\gamma$-ray production, e.g. pulsars or dark matter annihilation. Hence, in the following, we assume all $> 1$ GeV $\gamma$-rays are pionic (if there is substantial pulsar contamination, the pionic $\gamma$-ray emission is lower, and higher diffusivities $\kappa$ are required).

The $> 1$ GeV $\gamma$-ray luminosity for $\gamma$-rays $L_\gamma(\geq {\rm GeV})$ can be calculated as:
\begin{align}
\label{eqlackigamma}
L_\gamma(\geq {\rm GeV})\approx \sum_i \frac{1}{3}\beta_\pi \tilde{\Lambda}_{\rm cr,had}\,e_{\rm cr}\,n_{\rm n} \Delta V_i,
\end{align}
where we sum over gas particle i with volume $\Delta V_i$. First, the most of the hadronic loss ($\tilde{\Lambda}_{\rm cr,had}\,e_{\rm cr}\,n_{\rm n}$ in Eq. \ref{crcooling}) is responsible for the pion production. Second, only one third of the pions ($\pi^0$) produce $\gamma$-rays. Third, $\beta_\pi(\approx 0.7)$ is the fraction of the pionic $\gamma$-rays with energy above GeV\citep{Lack11}, which is calculated with the GALPROP pionic cross sections \citep{Mosk98,Stro98,Stro00} built on \cite{Derm86}, assuming the CR spectrum (between 1 GeV and 1 PeV) follows $E^{-p}$, where $E$ is the CR proton energy and $p\;(= 2.2)$ is the spectral index. 

If CRs can propagate fast enough that a significant fraction can leave galaxies without interacting with ISM, the $\gamma$-ray emission will be relatively weak, compared to the expectations from the CR injection. We follow \citet{TQW} and \citetalias{Lack11} and quantify this by comparing the pionic $\gamma$-ray luminosity $L_\gamma$ (above $>1$\,GeV) with the bolometric ``star formation luminosity'' $L_{\rm SF}$ (UV/optical/IR luminosity ultimately contributed by stellar radiation from massive stars, estimated assuming a time-constant SFR and the same stellar IMF as in our simulations), since the CR injection is proportional to SN injection rate and thus to the SFR. If the SFR is constant and we are in the ``proton calorimetric limit'' (all CR protons instantly lose their energy to collisional processes, without any other processes influencing their energies or spatial distribution, assuming the same time-constant SFR), then the ratio $L_\gamma/L_{\rm SF}$ is approximately constant. 

The value of $L_{\gamma}$ in the calorimetric limit is derived as follows. If the SFR is constant, the SNe rate is dominated by Type-II events, and the CR injection rate is: 
\begin{equation}
\label{CRSNegy}
\frac{\dot{E}_{\rm cr,SN}}{{\rm [erg/s]}} =\epsilon_{\rm cr}  \frac{E_{\rm SN}}{{\rm [erg]}}\frac{\rm [\msun]}{<m_*>}\xi(m_*>8\msun)\frac{\rm SFR}{\rm 3.2\times 10^7[\msun/yr]},
\end{equation}
where $\epsilon_{\rm cr}(=0.1)$ is the fraction of SNe energy going into CRs (constant by assumption in our simulations), $E_{\rm SN}(=10^{51}\;{\rm erg})$ is the energy from one supernova (also constant by assumption), $\xi(m_*>8\msun)(=0.0037)$ is fraction of stars that end as supernovae, and $<m_*>(=0.4\;\msun)$ is the mean stellar mass, both calculated for the same \citep{Krou02} IMF used in our simulations. If this injection rate of CRs is balanced by collisional losses without any other energy gain/loss processes, i.e. $\dot{E}_{\rm cr,SN}=e_{\rm cr}\tilde{\Lambda}_{\rm cr}$, then the pionic $\gamma$-ray luminosity is 
\begin{align}
L_\gamma(\geq {\rm GeV})^{\rm calor} \approx 6.7\times 10^{39}\frac{\rm SFR}{\rm [\msun/yr]}\;{\rm erg/s }.
\end{align}

The corresponding ``star formation luminosity'' $L_{\rm SF}$ for a constant SFR assuming again the same \citet{Krou02} IMF adopted in our simulations is
\begin{align}
\label{LSF}
L_{\rm SF}^{\rm constant} \approx 3.5\times 10^{43}\frac{\rm SFR}{\rm [\msun/yr]} {\rm erg/s},
\end{align}
where the prefactor is calculated with STARBURST99\footnote{\url{http://www.stsci.edu/science/starburst99/docs/default.htm}} \citep{Leit99,Vazq05,Leit10,Leit14}.\footnote{\citetalias{Lack11} adopted a different conversion factor because they assumed a Salpeter IMF \citep{Salp55} (following \citealt{Kenn98}). But the ratio $L_\gamma/L_{\rm sf}$ in \citetalias{Lack11} is only higher by $\sim 10\%$  since the SNe rate is also adjusted accordingly.} So in the constant-SFR, calorimetric limit, we would expect $L_{\gamma}/L_{\rm SF} \approx 2\times 10^{-4}$ \citep{TQW}.

\begin{figure*}
\begin{centering}
 \includegraphics[width={0.95\textwidth}]{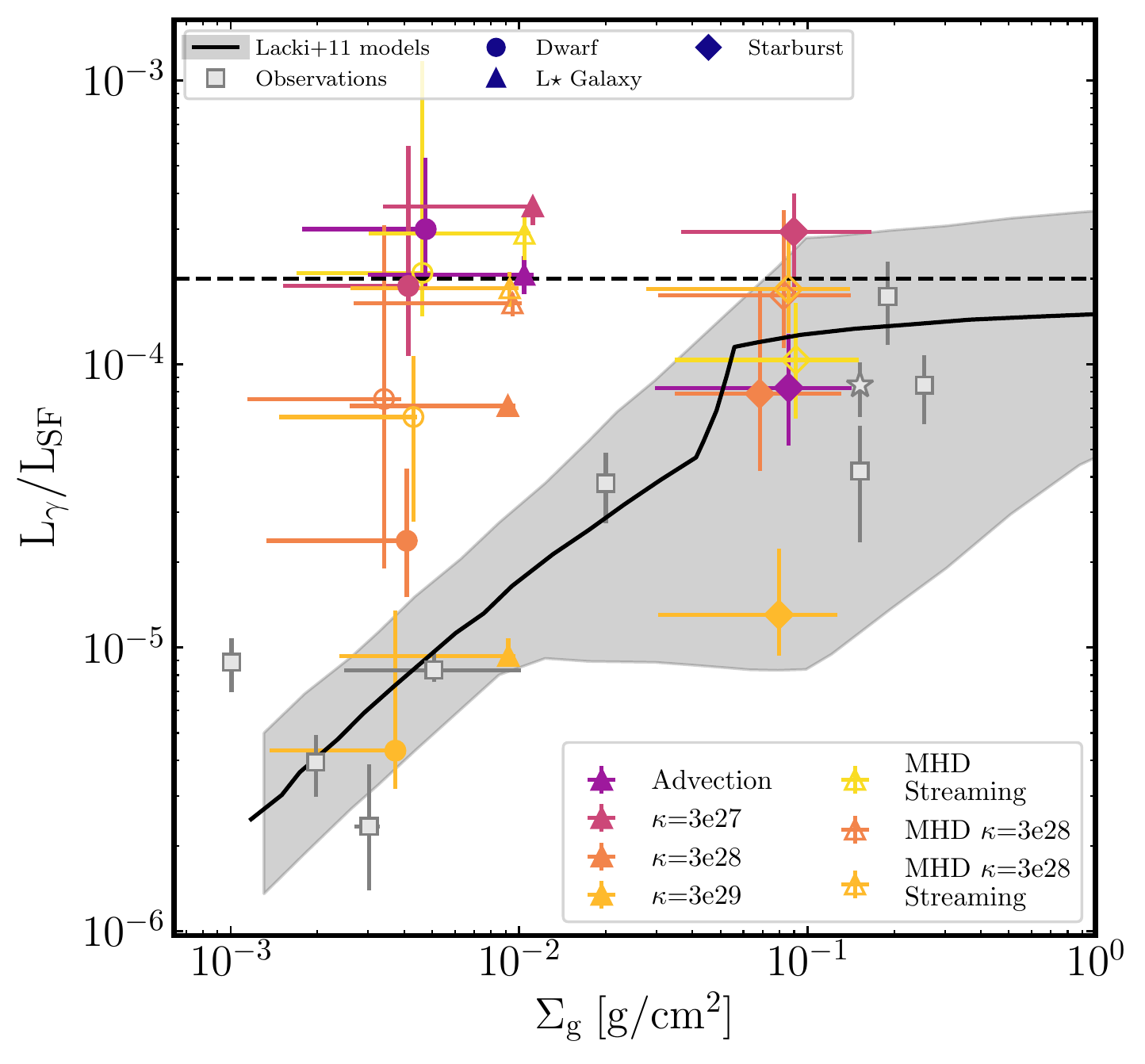}
 \end{centering}
 \vspace{-0.35cm}
\caption{Ratio of pionic $\gamma$-ray luminosity ($L_{\gamma}$; $E_{\gamma}$ $>1$\,GeV) to SF luminosity ($L_{\rm SF}$) as a function of gas surface density ($\Sigma_{\rm g}$, averaged over inclination). We compare our {\bf Dwarf} (circle), {\bf $L\star$} (triangle), and {\bf Starburst} (diamond) galaxy models, with different CR transport models (colors; Table \ref{CRprop}). Dashed line is calorimetric (Fig.~\ref{Fgammasfr}). Points+error bars indicate median and $\pm 1\sigma$ range of values over the time range $\sim 400-500\,$Myr (smoothed on 10Myr timescales). In order to compare with ``active'' starbursts, in our {\bf Starburst} runs we only consider snapshots that reach $\Sigma_{\rm g} > 0.08\,{\rm g\,cm^{-2}}$ for at least one inclination during an extended t=250-650 Myr interval ($L_{\rm SF}$ and $L_\gamma$ are averaged on 5 Myr timescales). Grey squares show observed values compiled in \citetalias{Lack11} (left-to-right: M31, LMC, SMC, MW, NGC1068, NGC253, NGC4945, M82; star is the NGC253 core). Solid line and shaded range shows the range of ``successful'' models considered in \citetalias{Lack11} which simultaneously fit the available observational constraints on CR $\gamma$-ray emission, spectra, and Milky Way constraints. The simulations of low surface density galaxies are consistent with observations for $\kappa \sim 3\times10^{29}\,{\rm cm^{2}\,s^{-1}}$, while lower effective $\kappa$ might be preferred in {\bf Starburst} runs (but note that typical gas densities in {\bf Starburst} model are lower than in observed starbursts, so here we use only a handful of snapshots that reach highest central gas densities). 
Lower gas densities $\Sigma_{\rm g}$, or higher diffusivity $\kappa$, produce lower $L_{\gamma}/L_{\rm SF}$.
 \vspace{-0.25cm}}
\label{gassurFgFsf}
\end{figure*}

\begin{figure}
\begin{centering}
\includegraphics[width={0.95\columnwidth}]{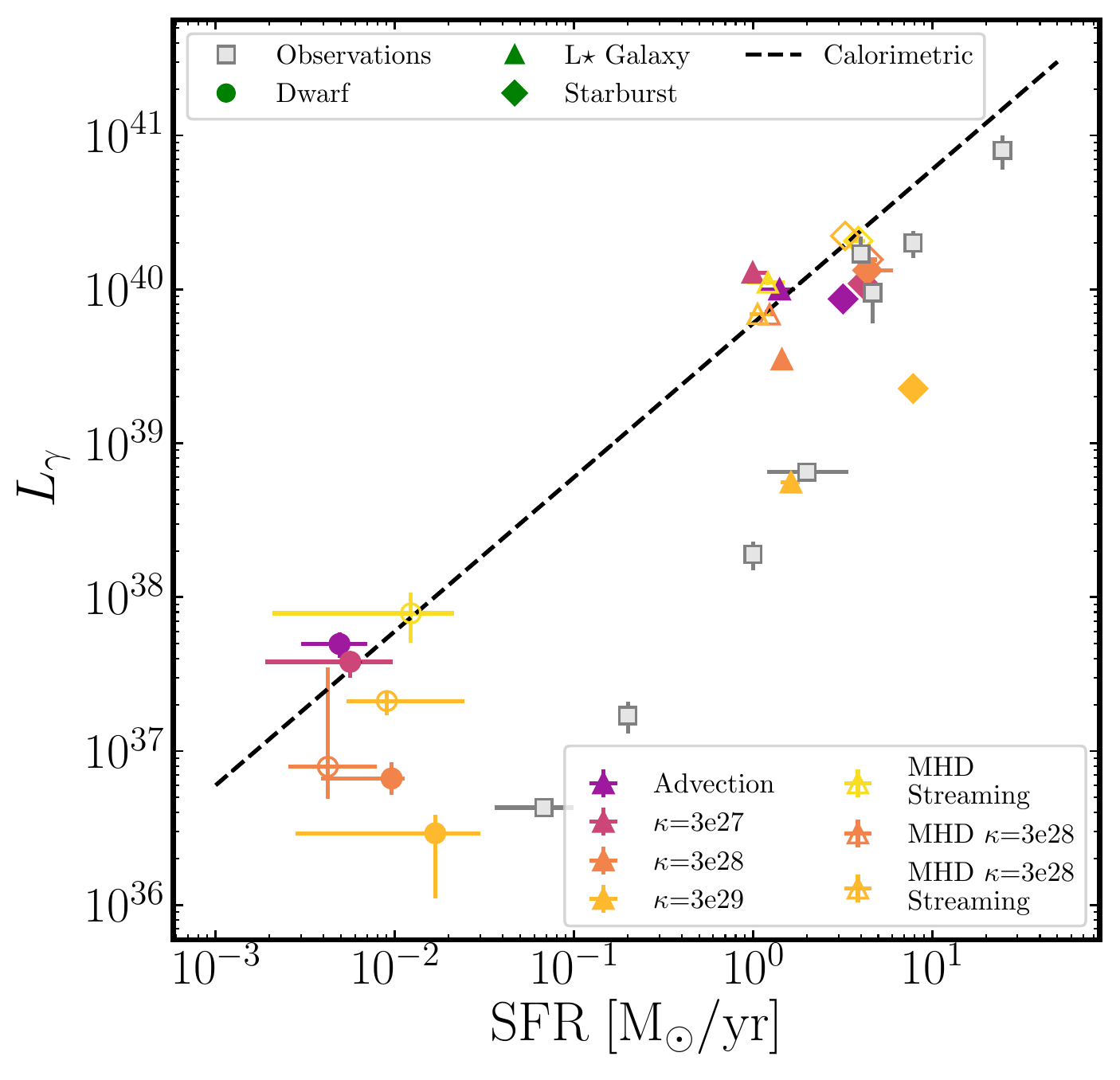}
\end{centering}
\caption{Pionic $\gamma$-ray luminosity ($E_{\gamma}$ $>1$\,GeV) vs SFR (averaged over $\sim 10\,$Myr) from our simulations and observations (as in Fig.~\ref{gassurFgFsf}). For the {\bf Starburst} models we restrict to times ``during starburst'' (SFR $>3\,\msun\,{\rm yr^{-1}}$) and take 5-Myr averaged SFR. Trends are similar to Fig.~\ref{gassurFgFsf}: high-SFR galaxies have $L_{\gamma}$ close to calorimetric (dashed), while low-SFR galaxies have much lower $L_{\gamma}$, indicating efficient CR escape. Again $\kappa\sim 3\times10^{29}\,{\rm cm^{2}\,s^{-1}}$ matches the observations in low SFR galaxies. }
\label{SFRLg}
\end{figure}

\subsubsection{Pionic $\gamma$-ray emission}
\label{sec:pionic.emission.overview}

Figure \ref{MWgamma} shows the projected pionic $\gamma$-ray surface brightness of the {\bf $L\star$ Galaxy} run, for different values of $\kappa$. $\gamma$-rays mostly originate from galactic disk, i.e. where gas and CR densities are the highest. The $\gamma$-ray surface brightness drops by over an order of magnitude a few kpc away from the disk plane. For higher-$\kappa$, the emission is dimmer but more spatially extended (reflecting the CR energy distribution). 

Fig.~\ref{gammasph} quantifies the distribution of the $\gamma$-ray luminosity within spheres of increasing radii for all of our runs with cosmic rays. Consistent with the discussion above, $\gamma$-ray emission is much weaker for large $\kappa$.\footnote{As with the CR energy density, we note that sometimes the runs with finite-but-low $\kappa$ exhibit slightly higher $L_{\gamma}$ even than the ``{\bf Advection}'' runs owing to non-linear effects discussed in \S~\ref{sectionCRegy}.} For our {\bf $L\star$ Galaxy}, half of the $\gamma$-ray luminosity originates from the inner 5-7\;kpc. The half-luminosity radius is smaller for our dwarf galaxy, as expected, since the galaxy itself (gas and stellar) is smaller.

Fig.~\ref{nismcr} breaks down the distribution of CR energy as a function of local gas density, which (since $L_{\gamma} \propto e_{\rm cr}\,n_{\rm gas}$) effectively determines $L_{\gamma}$. With low-$\kappa$ (or with advection/streaming only), CRs reside longer in the high-density regions where they are injected. If the density (on some scale $\ell$ of the cloud or disk) is larger than $\sim 10\,{\rm cm^{-3}}\,(\ell/{\rm kpc})^{-2}\,(\kappa/10^{29}\,{\rm cm^{2}\,s^{-1}})$, then the collisional loss time of CRs becomes shorter than the diffusion time, so the CRs decay close to their injection. This means $L_{\gamma}$ is lower at higher $\kappa$, even when the SFR (injection rate) is higher, because the bulk of the CR energy is at lower $n_{\rm n}<0.1\,{\rm cm^{-3}}$.

\subsubsection{$L_{\gamma}/L_{\rm SF}$ ratio and scalings}
\label{implication}

Fig.~\ref{Fgammasfr} shows the time evolution of $L_{\gamma}/L_{\rm SF}$. With a lower $\kappa$ (``{\bf Advection}'' or ``${\rm \kappa=3e27}$''), the galaxies are closer to the calorimetric limit (as expected).\footnote{Note there are periods where $L_{\gamma}/L_{\rm SF}$ exceeds calorimetric -- this is perfectly allowed. Usually it occurs because of short-timescale drops in the SFR and hence $L_{\rm SF}$, where the CRs take somewhat longer to decay so $L_{\gamma}$ is $\sim$\,constant. But it can also owe to adiabatic compression increasing CR energy, or the contribution of SNe Ia's, or smoothly declining SF histories, all of which violate the assumptions used to derive the calorimetric limit.} But the ``${\rm \kappa=3e29}$'' runs are lower than the calorimetric limit by more than an order of magnitude in our {\bf Dwarf} and $L\star$ runs.

Turning around our argument from \S~\ref{sec:pionic.emission.overview} above, if CRs are injected in a structure of size $\ell$ and gas density $n$ with an isotropically-averaged diffusivity $\kappa \lesssim 10^{28}\,{\rm cm^{2}\,s^{-1}}\,(\ell/{\rm kpc})^{2}\,(n/{\rm cm^{-3}})$, the collisional loss time becomes shorter than the escape time, so we expect near-calorimetric $L_{\gamma}/L_{\rm SF}$. On the other hand, at larger $\kappa \gg 10^{28}\,{\rm cm^{2}\,s^{-1}}$ in the limit where CRs do escape, if we assume the galaxy gas structure is otherwise similar, then the ratio of escape time to loss time (proportional to the fraction of CR energy lost in escaping, and therefore $L_{\gamma}/L_{\rm SF}$) should scale $\propto \kappa^{-1}$.

In \S~\ref{sectionCRegy}, we will show that adiabatic processes are of secondary importance relative to diffusion in reducing $L_{\gamma}$ in runs with high-$\kappa$, but in runs with low-$\kappa$ (e.g.\ ``${\rm \kappa=3e27}$''), they boost $L_{\gamma}$ considerably.

As expected, anisotropic diffusion tends to suppress the isotropically-averaged $\kappa$ by factors $\sim 1.5-3$, and correspondingly increase $L_{\gamma}/L_{\rm SF}$. Streaming slightly increases the escape and dissipates CR energy through streaming loss, so decreases $L_{\gamma}/L_{\rm SF}$, but the effect is very small (and streaming alone produces near-calorimetric results). This is  because (as discussed in \S~\ref{CRgalpro} above) the streaming escape time is much longer than the diffusive escape time, even for relatively low $\kappa$, but with the caveat that we do not consider the decoupling between CRs and gas in the cold ISM due to the low ionization fraction and ion-neutral damping \citep{Farb18}, which could significantly reduce $\gamma$ ray emission from dense gas.

For the same CR propagation model, galaxies with higher gas densities and larger sizes (effectively larger column densities of dense gas with which CRs must interact to escape) have a larger $L_{\gamma}/L_{\rm SF}$ , which can be seen in Fig.~\ref{Fgammasfr}.

\subsubsection{Comparison to observations}
\label{comobs}
We now compare the simulations to observational estimates of $L_{\gamma}/L_{\rm sf}$ as a function of either central gas surface density in galaxies ($\Sigma_{\rm g}$) or SFR, as compiled in \citetalias{Lack11}. Most of the observed data is described in \citetalias{Lack11}, but we also include the SFR of the SMC (0.036 ${\rm \msun/yr}$ from \citealt{Wilk04}). We add two extra starburst galaxies in Figs.~\ref{gassurFgFsf}-\ref{SFRLg} (NGC 1068 and NGC 4945; which are listed in Table 2 of \citealt{Lack11} but not in their figures). The SFRs of starburst galaxies (NGC1068, NGC253, NGC4945, M82) are obtained with the \citet{Kenn98} IR to SFR conversion formula, assuming the Kroupa IMF (Their IR luminosities are also listed in Table 2 of \citealt{Lack11}). 

$\gamma$-ray observations of nearby galaxies are limited in spatial extent due to energy resolution and contamination from the diffuse backgrounds and foregrounds \citep{Abdo10SMC,Abdo10LMC,Abdo10M31}. Hence, we only consider $\gamma$-ray emission within 3 kpc for {\bf Dwarf}, matched approximately to that used for the SMC. This choice reduces the $\gamma$-ray luminosity by a factor of two compared to using an infinitely large aperture. For {\bf  Starburst} and {\bf $L\star$ Galaxy}, we take 10 kpc apertures (which only reduces $L_{\gamma}$ by tens of percent compared to an infinitely-large aperture), matched to those used for e.g.\ M31, NGC1068, and M82. See Fig.~\ref{gammasph} for how this scales with size.

We measure the gas surface densities $\Sigma_{\rm g}$ (averaged over viewing angles) of {\bf Starburst} within 250 pc, {\bf $L\star$ Galaxy} with 4 kpc, and {\bf Dwarf} within 2 kpc -- chosen to be twice the sizes of the active star-forming region in {\bf Dwarf} and $L\star$\footnote{This choice is smaller than the optical radii that \cite{Kenn98} and \citetalias{Lack11} used, but the gas surface densities are similar in both choices.} and about equal in {\bf Starburst} (similar to the choice in \citealt{Kenn98} and \citetalias{Lack11}).

Figs.~\ref{gassurFgFsf}-\ref{SFRLg} compare our simulations with the observations (compare to Figure 2 of \citetalias{Lack11}).  As expected based on our discussion above, $L_{\gamma}/L_{\rm SF}$ is high and close to the calorimetric limit for {\bf Dwarf}  and {\bf $L\star$ Galaxy} with slow CR transport, i.e. for ``{\bf Advection}'', ``${\rm \kappa=3e27}$'', and ``{\bf MHD Streaming}''. These values are clearly well above the observationally inferred $L_{\gamma}/L_{\rm SF}$.

With larger diffusion coefficients, $L_{\gamma}/L_{\rm SF}$ decreases as expected. For isotropic diffusion, the observations in dwarf and $L\star$ galaxies appear to require $\kappa_{\rm isotropic} \sim 10^{29}\,{\rm cm^{2}\,s^{-1}}$. Out of the options tested $\kappa = 3\times 10^{29}\,{\rm cm^{2}\,s^{-1}}$ provides the best match but the range of data allows for a slightly lower value as well. 
 
For anisotropic diffusion, $L_{\gamma}$ is somewhat larger owing to suppressed isotropic-averaged diffusivity, as discussed above, so values of the {\em parallel} diffusivity $\kappa_{\|} > 10^{29}\,{\rm cm^{2}\,s^{-1}}$ are favored. 

For galaxies with high gas surface densities and SFRs, i.e. {\bf Starburst}, we found $\kappa$ has to be less than $3\times 10^{29}{\rm cm^2/s}$. On the face value, this implies that CR transport is effectively slower in high gas surface density regions or during starburst. However, in our {\bf Starburst} runs, for the highest diffusion coefficient tested, we did not include MHD and anisotropic diffusion that, depending on the magnetic field configuration, could slow down the transport of CRs out of high density regions. The models also have gas configurations and typical gas densities that are not an excellent match to observed starburst galaxies, and our analysis only includes brief time intervals when central gas density reaches the values similar to observed starbursts.

Many observations, e.g. \cite{Acke12} and \cite{Roja16}, considered $L_{\gamma}$ with $0.1 {\rm GeV}<E_{\gamma}<100{\rm GeV}$, instead of $E_{\gamma}>1{\rm GeV}$. Therefore, in Appendix \ref{Fermicom}, we compare $L_{\rm 0.1-100\;GeV}$ from simulations with observations, and we find that the same high diffusion simulations provide the best match to the observed $\gamma$ ray emission from galaxies.

It is interesting to compare the results from the simple leaky-box model of \citetalias{Lack11}, as well as more detailed models of CR transport in the MW, with our findings. \citetalias{Lack11} predicts $L_{\gamma}/L_{\rm SF}$ as a function of gas surface density by assuming the Kennicutt-Schmidt law and a one-zone leaky box model with CR diffusion (see \citealt{Lack10} for details), with an isotropically-averaged $\kappa = 3\times10^{28}\,{\rm cm^{2}\,s^{-1}}$. The broad contours of their prediction for $L_{\gamma}/L_{\rm SF}$ as a function of $\Sigma_{\rm g}$ or SFR are similar to our simulations, suggesting -- as they argued -- that CR escape is required to reproduce the observed trend of $L_{\gamma}/L_{\rm SF}$. 

In the MW, much more detailed propagation models have been tested (see e.g. \citealt{Trot11} and reference in \S~\ref{introduction}). We again caution that our ``$L\star$'' model is not an exact MW analogue, since it has higher gas surface density and lacks a central gas deficiency like the MW \citep{Mosk02}, both of which could affect $L_{\gamma}$. 

Note that at ``face value'', both MW and \citetalias{Lack11} constraints might appear to favor slightly-lower $\kappa\sim 3-6\times10^{28}\,{\rm cm^{2}\,s^{-1}}$ compared to the best-match here ($\kappa\sim 3\times10^{29}\,{\rm cm^{2}\,s^{-1}}$), but this is a relatively small offset and completely expected if we account for the points below. (1) The MW observations and \citetalias{Lack11} models assume relatively small halos out of which the CRs escape instantly, while we assume a constant $\kappa$ everywhere, meaning that our effective halo size is large ($\sim 10-30\,$kpc). Recall (\S~\ref{introduction} and \S~\ref{diffcoeff}), the inferred $\kappa$ in the observations increases with the halo height. (2) The gas in the simulations is clumpy where CRs are injected, slightly increasing $L_{\gamma}$  \citep{Boet13}, compared to the smooth mass profiles assumed in those studies (requiring larger $\kappa$ by a factor $\sim 1.5-2$). (3) \citetalias{Lack11} did not consider galactic winds and adiabatic losses/gains in their fiducial models; the MW constraints did not account for galactic winds in a self consistent manner (i.e. they do not consider CR-driven winds and the radial/temporal variations of the winds). In our anisotropic runs, we also find that the isotropically-averaged $\kappa$ (what is nominally constrained by the \citetalias{Lack11} study, for example) is a factor $\sim 2-3$ lower than the parallel $\kappa$. Accounting for all of these facts, our favored coefficients appear to be consistent with other state-of-the-art constraints on CR propagation in the MW from e.g.\ \citealt{Trot11}, and references in \S~\ref{introduction}.

\section{Discussion}
\label{discussion}
\subsection{Comparisons to previous studies}
\label{comparison}
\subsubsection{Suppression of star formation by cosmic rays}

In our idealized non-cosmological simulations, we find that SF can be suppressed by CR feedback in simulations with either advection or streaming only, or very low diffusivities $\kappa \lesssim 10^{28}\,{\rm cm^{2}\,s^{-1}}$, consistent with many previous findings, e.g. \cite{Boot13,Sale14,Pfro17cr}. However, such slow transport severely violates constraints from observed $\gamma$-ray emission, and at best results in modest SFR suppression (factor $\sim 1.5-2$). For larger transport speeds required to reproduce the observed $\gamma$-ray emission, CRs have only a weak effect on SF. 

Interestingly, \cite{Jube08} found that while CRs reduce SFRs in dwarf galaxies, they have almost no effect in MW mass galaxies. Their conclusion was likely due to their ``local equilibrium'' assumption, namely that CR injection ($\propto {\rm SFR} \propto \rho^{1.5}$ in their model) is balanced by collisional losses ($\propto \rho$) locally (like in the calorimetric limit), in an isothermal-like ISM, so in their models the CR energy density is proportional to $\rho^{1/2}$ while thermal energy densities are proportional to $\rho$: as a result, CR energy was always sub-dominant to thermal energy in their models at gas densities $n>0.2\,{\rm cm^{-3}}$. In contrast, in our simulations, CRs can propagate far from their injection sites, 
so local equilibrium is not valid and we find that the ratio of CR pressure to gas thermal or turbulent pressure for low $\kappa$ can be significant even at moderate ISM densities, providing mild suppression of the star formation
(similar arguments were presented in 
\citealt{Socr08,Boot13}). For our favorite, large $\kappa$ values CRs escape from the ISM, resulting in practically no effect on the star formation in both our {\bf Dwarf} and {\bf $L\star$ Galaxy} simulations.

However, we caution that because our simulations are non-cosmological, they do not account for the effect of CRs on the CGM and IGM (the source of fuel for galaxies). As CRs escape the {\em galaxies} more efficiently with the favored larger $\kappa$, we have shown they have proportionally much higher energy density/pressure in the CGM, which means they could (in principle) be important for the long-term cosmological evolution and accretion onto galaxies. This is likely most important in more massive galaxies that build quasi-hydrostatic halos whose late-time cooling influences galaxy growth. We will explore this in cosmological simulations in future work\citep{Hopk19cr}.

\subsubsection{$\gamma$-ray emission}
\label{comparisongamma}

Our results are in line with \citetalias{Lack11}: when matching the observed $\gamma$-ray emission, starburst galaxies (with effective isotropic diffusivities $\kappa < 3\times 10^{29}{\rm cm^2/s}$) are nearly proton calorimeters, while galaxies with lower gas surface density or SFRs (with $\kappa\sim 3\times10^{28-29}{\rm cm^2/s}$) are not proton calorimeters (most CR protons escape).

\cite{Sale16} also studied hadronic $\gamma$-ray emission with simulations of MW-mass galaxies and argued for isotropically-averaged coefficients $\kappa\sim 3\times 10^{28}\,{\rm cm^{2}\,s^{-1}}$; but they only considered the $\gamma$-ray emission in the CGM and they did not include hadronic/collisional losses in the simulations, which led to some unphysical results. For example, their predicted pionic $\gamma$-ray luminosity significantly exceeded the CR injection rate at lower $\kappa$. Moreover, as noted by \citet{Jaco18}, neglecting collisional CR losses allows CRs to build up in dense gas or the disk midplane without being rapidly lost (as they should), which artificially enhances the strength of CR-driven winds. Nevertheless, we broadly agree on the preference for a relatively high $\kappa$.

Recently, \cite{Pfro17gamma} also investigated $\gamma$-ray emission with idealized galaxy simulations, assuming CR transport via either advection-only or advection+anisotropic diffusion with $\kappa=10^{28}\,{\rm cm^{2}\,s^{-1}}$. They argued they could (a) reproduce the FIR-$\gamma$-ray correlation and (b) explain the low $L_{\gamma}$ in non-starburst galaxies primarily by adiabatic losses. 

But there are several caveats: 

(1) Their favored model still over-predicted $L_{\gamma}/L_{\rm SF}$ by a factor of a few or more in non starburst galaxies, e.g. dwarf and MW-mass galaxies. For their actual simulated points (see their Fig.~3) without diffusion, the predicted $L_{\gamma}/L_{\rm SF}$ is larger than the SMC, LMC, MW, and M33 (not shown therein, but see \citetalias{Lack11}).

They claimed to match the observed FIR-$L_{\gamma}$ correlation, only if an empirical FIR-SFR conversion relation \citep{Kenn98} is assumed. However, as they acknowledged, this conversion relation over-predicts $L_{\rm FIR}$ in dwarfs, due to much lower dust opacity/absorption/reddening. Their FIR-$L_{\gamma}$ relation might deviate from observations after taking this correction into account.\footnote{Because both the $L_{\gamma}/L_{\rm SF}$ and $L_{\rm FIR}/{\rm SFR}$ ratios drop in dwarfs, a roughly-linear $L_{\rm FIR}-L_{\gamma}$ relation can still maintain in our simulations (directly related to the ``conspiracy'' which maintains the FIR-radio correlation; for discussion see e.g.\ \citealt{Bell03,Lack10}).}

(2) We do not consider the same CR models and the same range of $\gamma$ ray energy. They consider $L_{\gamma}(0.1-100\,{\rm GeV})$, i.e.\ including all CRs from $0.1-100\,$GeV, instead of the choice here and in \citetalias{Lack11}, which is restricted to $L_{\gamma}(>1\,{\rm GeV})$. They also assumed a shallower CR spectrum ($\propto E^{2.05}$, as compared to $\propto E^{2.2}$ here and in \citetalias{Lack11}). Together with this, our $L_{\gamma}$ can differ from theirs by a factor of $\sim 2-3$. However, even if these differences are considered, their $L_{\gamma}/L_{\rm SF}$ are still greater than the observed dwarfs.

They suggested their over-prediction of $L_{\gamma}/L_{\rm SF}$ might be reconciled with simulations that could resolve the multi-phase ISM, since CRs may preferentially spend time in low density regions, which dominate the volume. We do have the multi-phase ISM here, but predict similar results in our advection-only or low-$\kappa$ runs. A possible explanation for the discrepancy is that the observed low $\gamma$-ray luminosities require high diffusion coefficients $\kappa \sim 3\times10^{28-29}\,{\rm cm^{2}\,s^{-1}}$ as favored by our study here and the modern MW constraints \citep{Trot11}.

(3) We will show immediately below that when $\kappa$ is in the favored range, adiabatic processes are less important than CR transport in reducing $L_{\gamma}$, although if $\kappa$ is small, adiabatic processes tend to increase $L_{\gamma}$.

\begin{figure*}
\begin{centering}
 \includegraphics[width={0.95\columnwidth}]{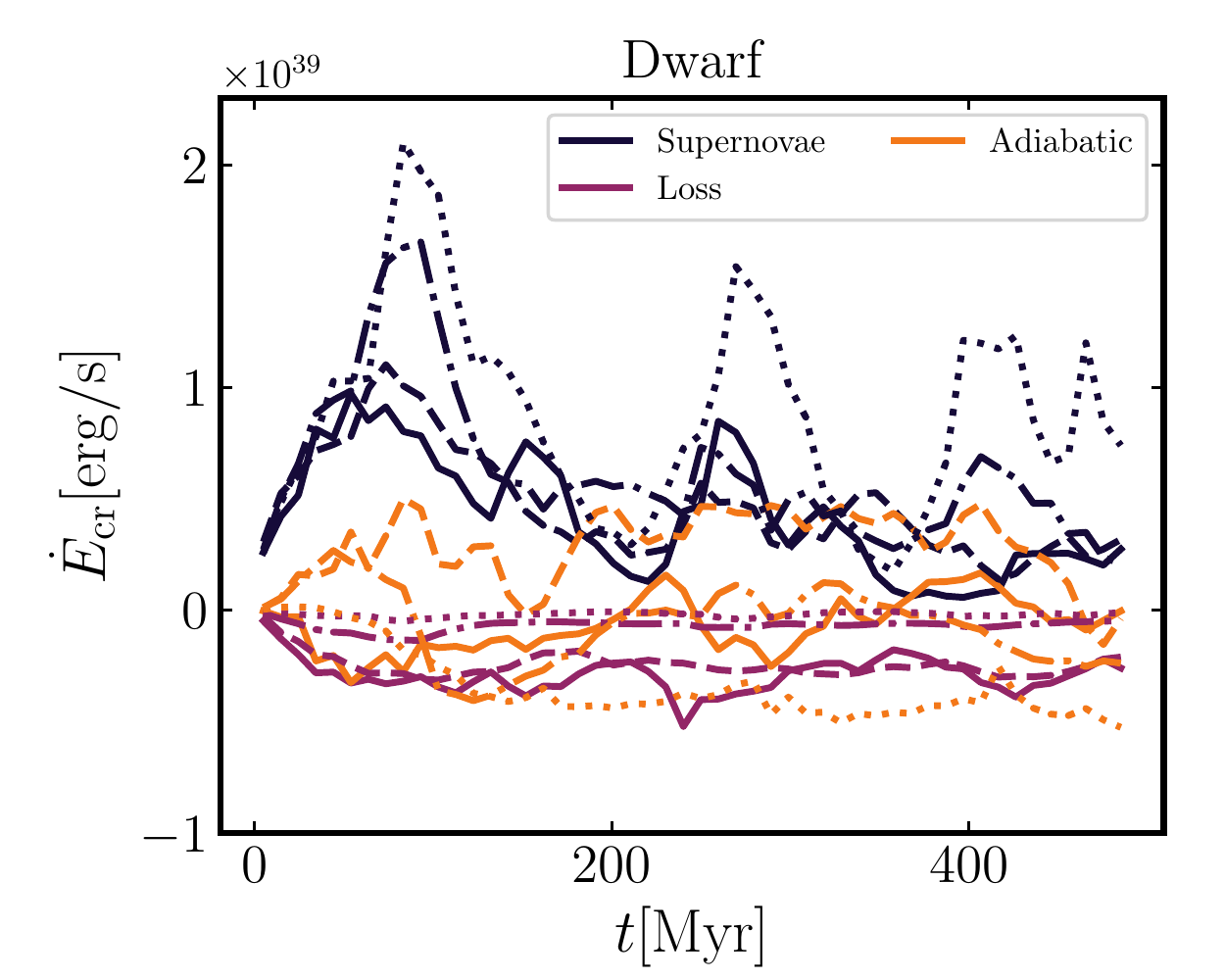}
  \includegraphics[width={0.95\columnwidth}]{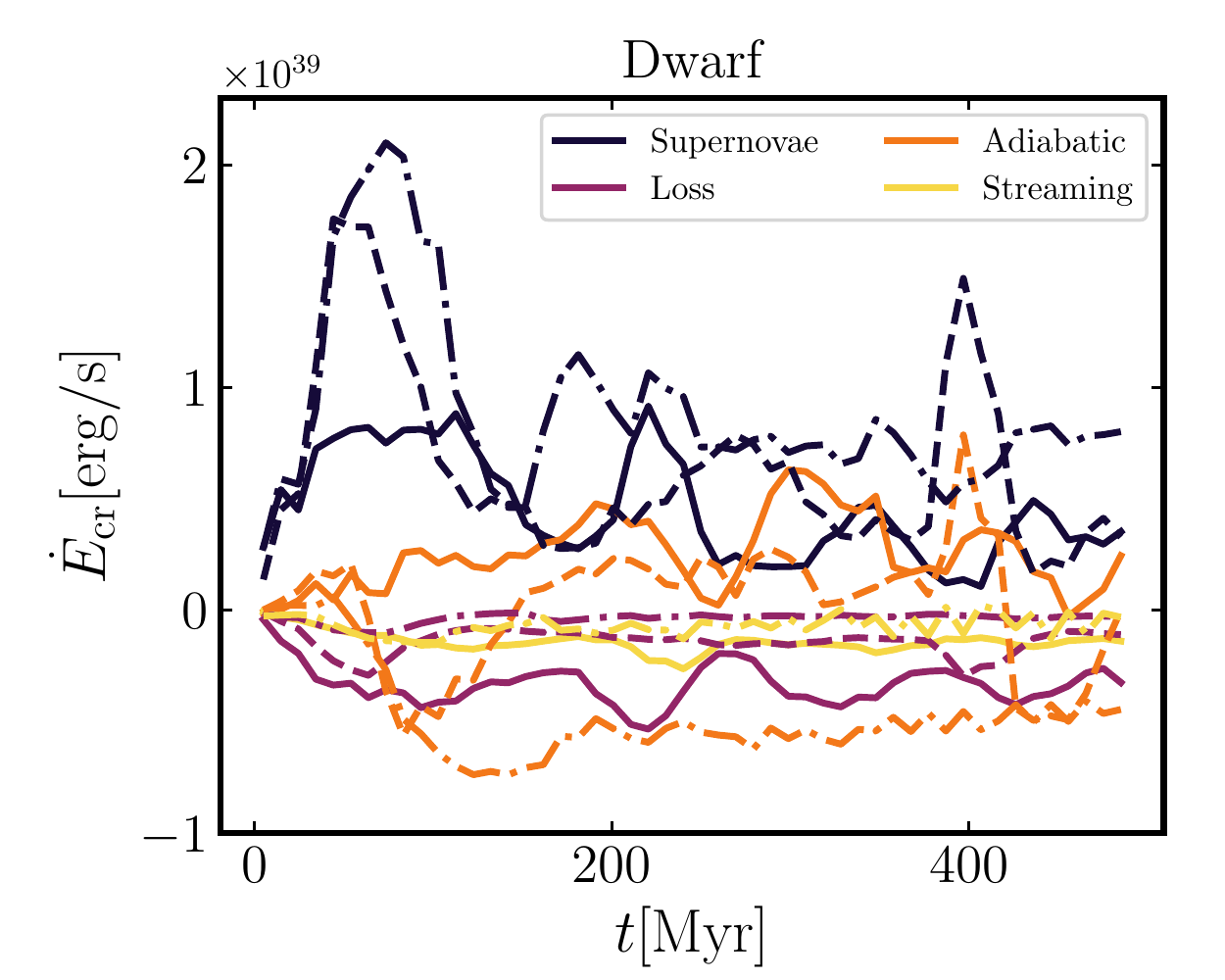}\\
 \includegraphics[width={0.95\columnwidth}]{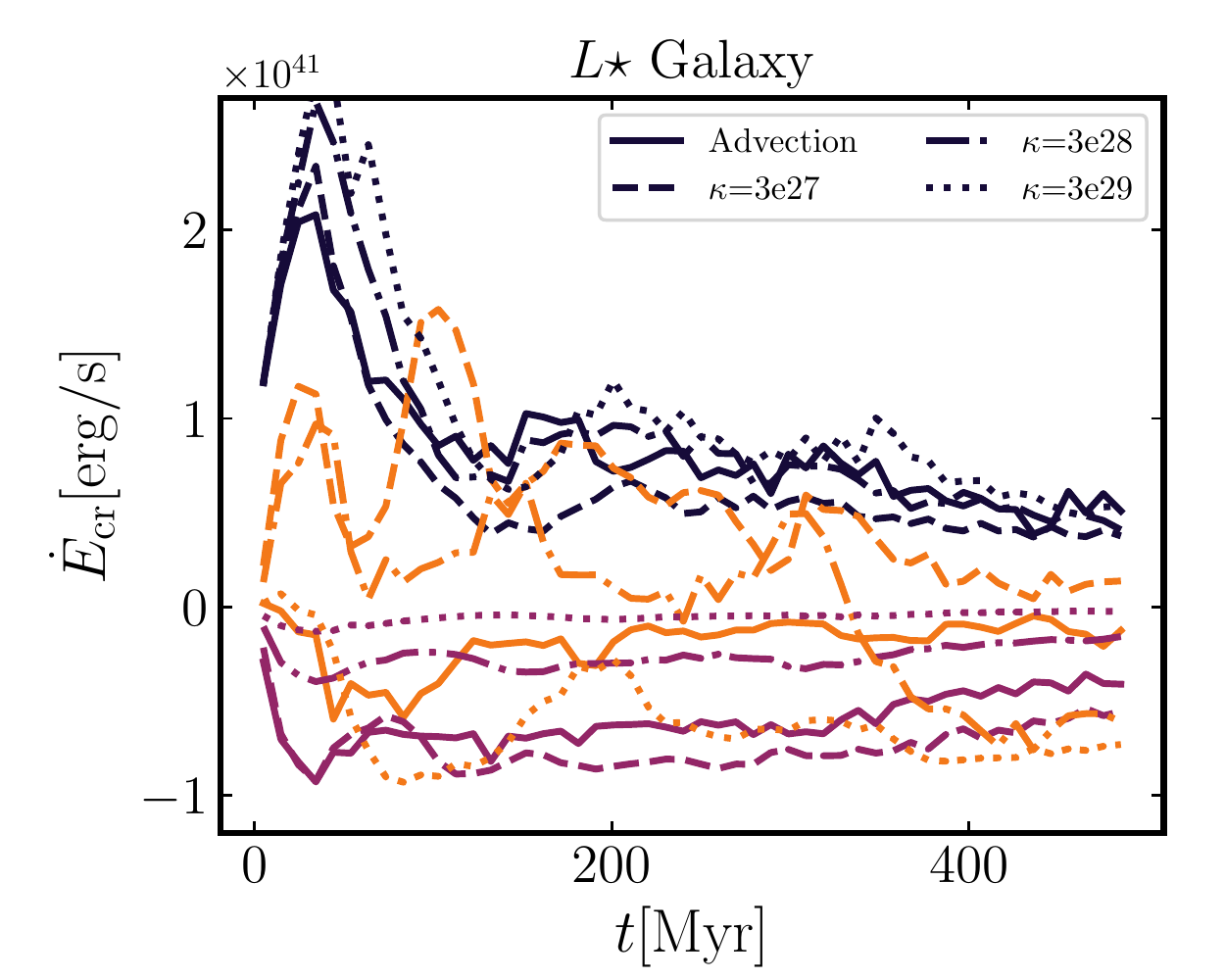}
 \includegraphics[width={0.95\columnwidth}]{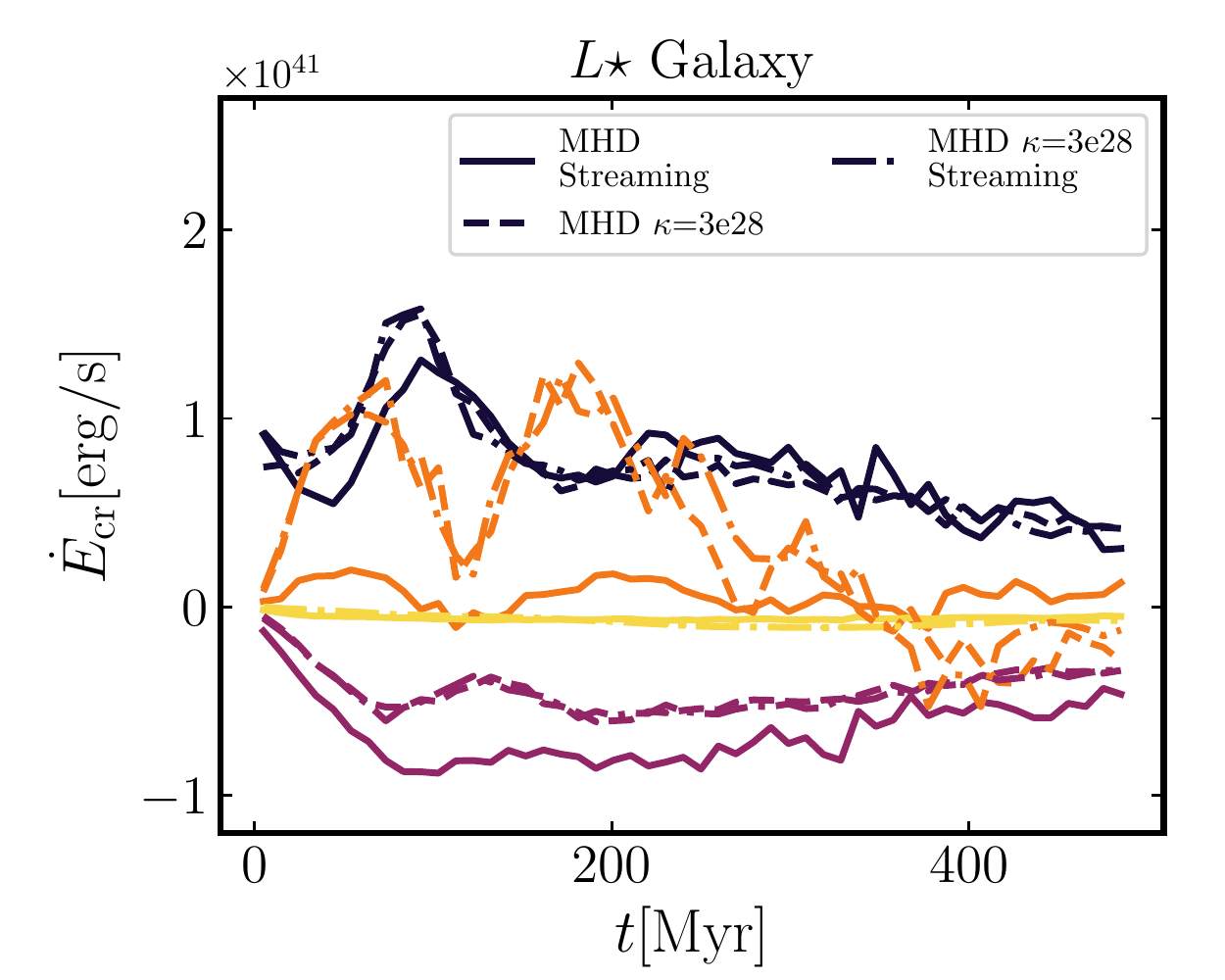}\\
  \includegraphics[width={0.95\columnwidth}]{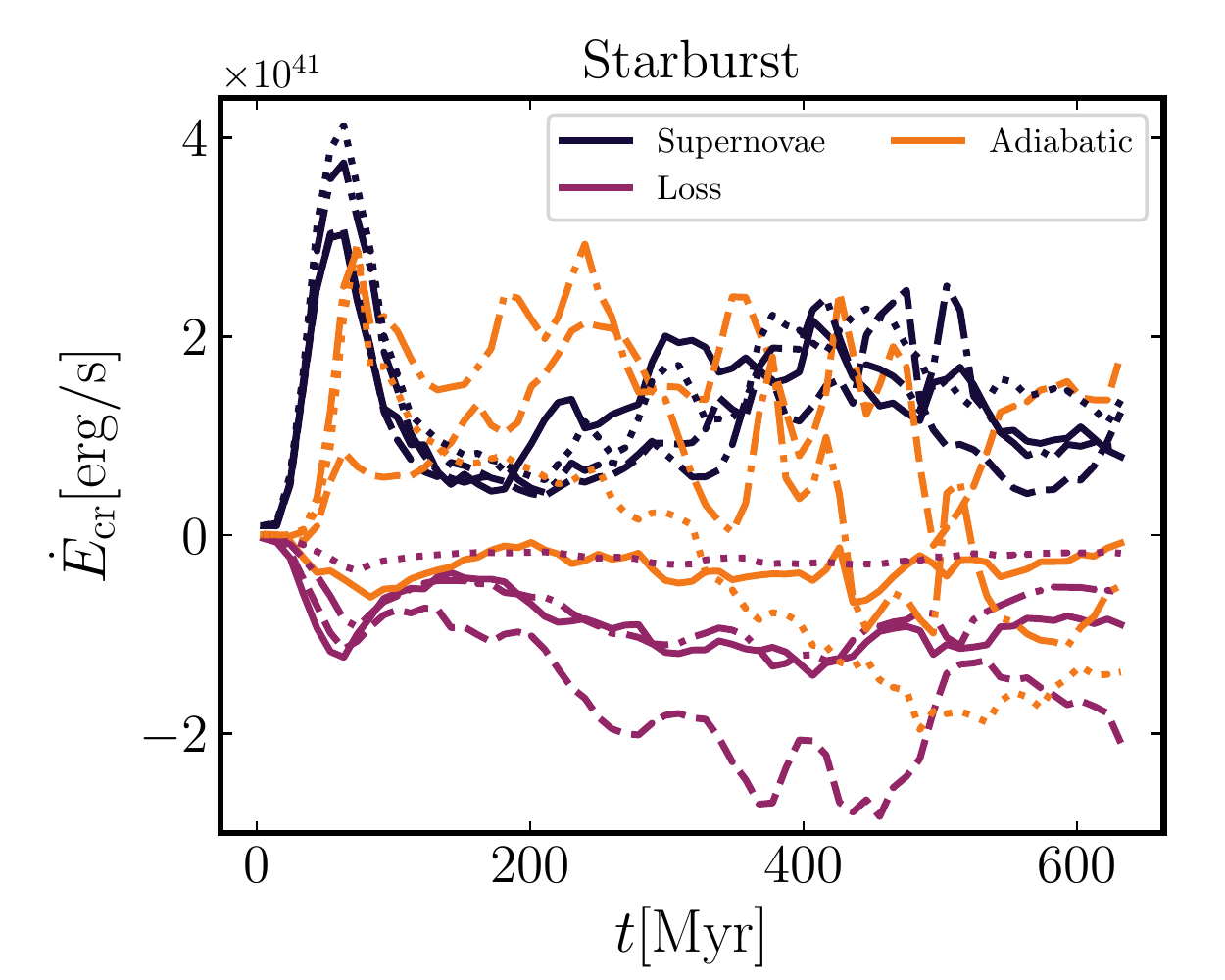}
    \includegraphics[width={0.95\columnwidth}]{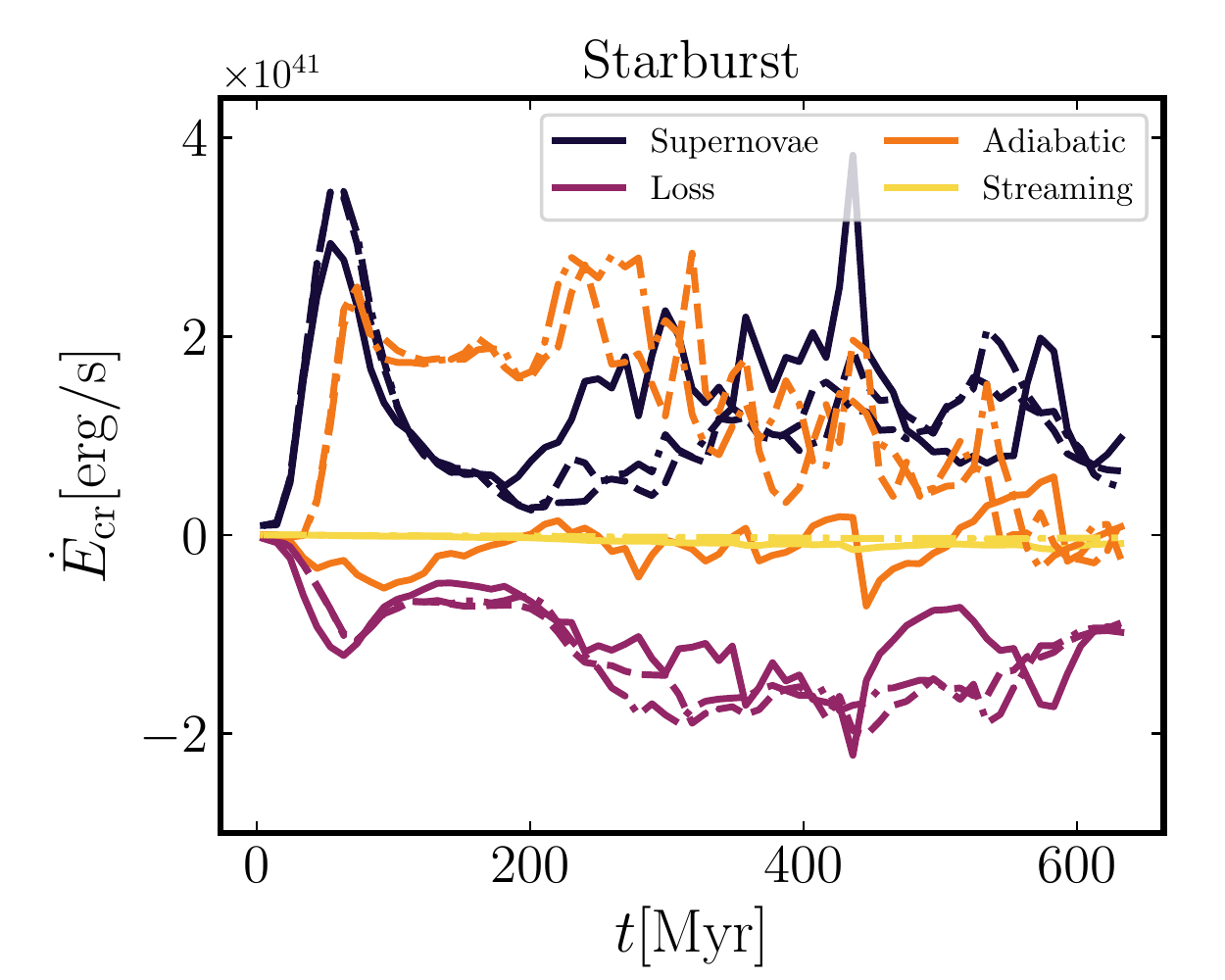}\\
\end{centering}
\vspace{-0.25cm}
\caption{Rate-of-change of total CR energy $\dot{E}_{\rm cr}$ (integrated over the box and averaged in $\sim 50\,$Myr intervals) in each simulation (labeled), owing to different gain (positive) or loss (negative) processes (see \S~\ref{method}). Left panels show runs without magnetic fields, whereas right panels show runs with magnetic fields. ``Supernovae'' (red) indicates injection from SNe. ``Loss'' includes the hadronic+Coulomb losses ($\Gamma_{\rm cr}$), ``Streaming'' the streaming loss term ($\Gamma_{\rm st}$). ``Adiabatic'' indicates the adiabatic (``PdV'') work term (includes work done by CRs on gas, and by gas on CRs; can be positive or negative). Faster transport (larger $\kappa$) means CRs spend less time in dense gas, reducing losses. While adiabatic terms are non-negligible, they rarely exceed SNe injection so do not boost $L_{\gamma}$ beyond a factor of $\sim 2$; they are also usually positive when $\kappa$ is low and CRs are trapped in dense gas (while they become negative at high $\kappa$).}
\vspace{-0.25cm}
\label{CRenergies}
\end{figure*}

\subsection{CR Energetics and the Importance of Different Gain/Loss Terms}
\label{sectionCRegy}

Fig.~\ref{CRenergies} shows the relative importances of various CR gain/loss terms in our simulations: SNe injection, collisional (hadronic+Coulomb) losses, ``streaming losses'' (energy loss to excitations of Alfven waves), and ``adiabatic'' terms (``PdV'' work lost pushing gas, or CR energy gain in compression). 

The initial injection from SNe is proportional to the SFR (with a few Myr delay), so it tracks the SFR and varies only by a relatively small amount in our different runs of a given galaxy model (even the highest/lowest SFR runs differ by at most a factor $\sim 2$). 

Collisional losses are important loss terms (within the galaxies) -- and we have already discussed these extensively as they are the origin of the $\gamma$-ray emission. Since they scale $\propto e_{\rm cr}\,n_{\rm n}$ they decrease with ``faster'' CR transport (higher $\kappa$) as CRs reach lower-density gas faster.\footnote{The (weak) exception to this rule is the  $\kappa=3\times10^{27}\,{\rm cm^{2}\,s^{-1}}$ run in our $L\star$ and {\bf Starburst} models, where collisional losses are slightly larger than in the corresponding ``{\bf Advection}'' runs. This is caused by the slightly stronger adiabatic compression term boosting $e_{\rm cr}$ in dense gas.} Streaming losses are comparatively small.

\begin{figure}
\begin{centering}
 \includegraphics[width={0.95\columnwidth}]{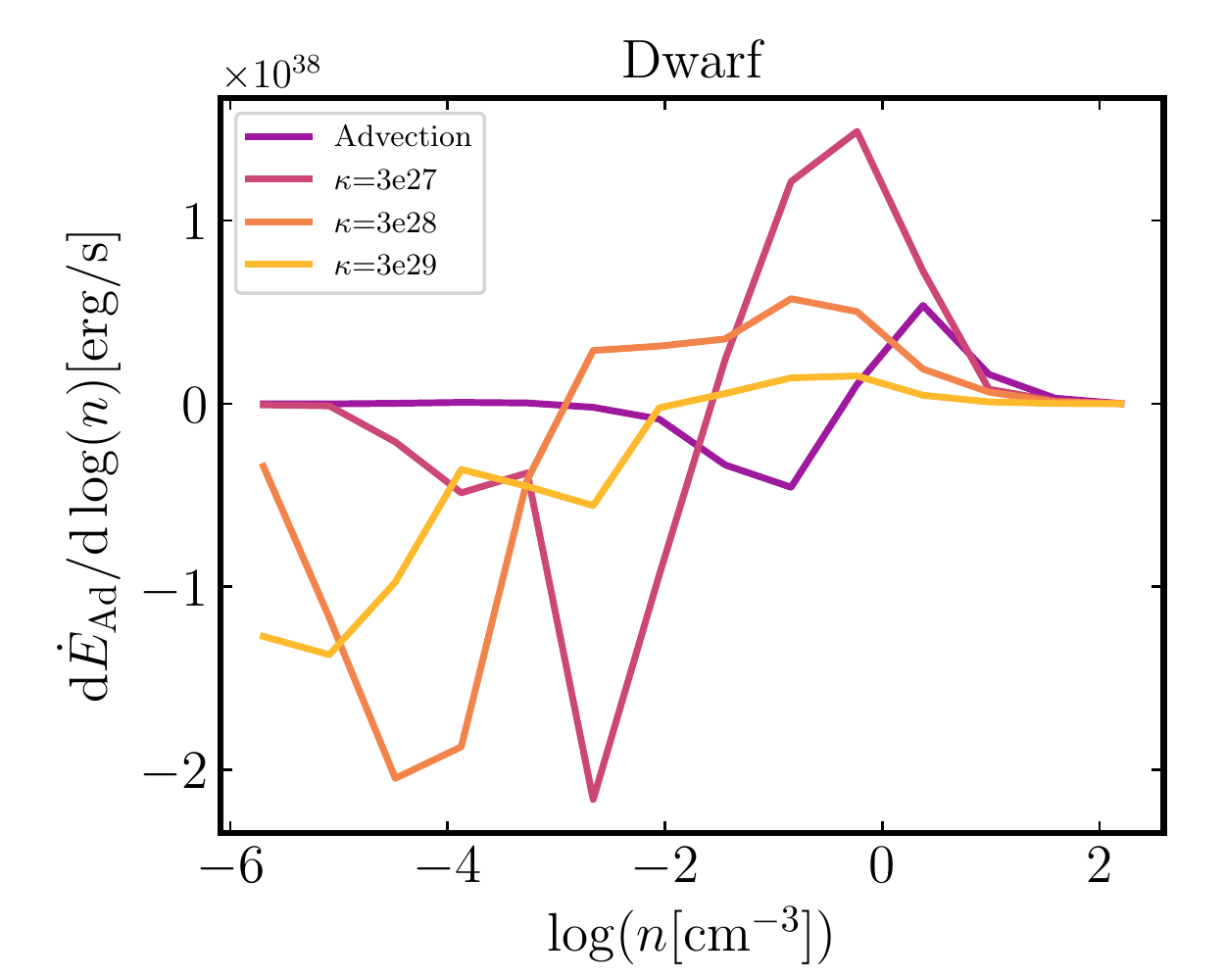}
  \includegraphics[width={0.95\columnwidth}]{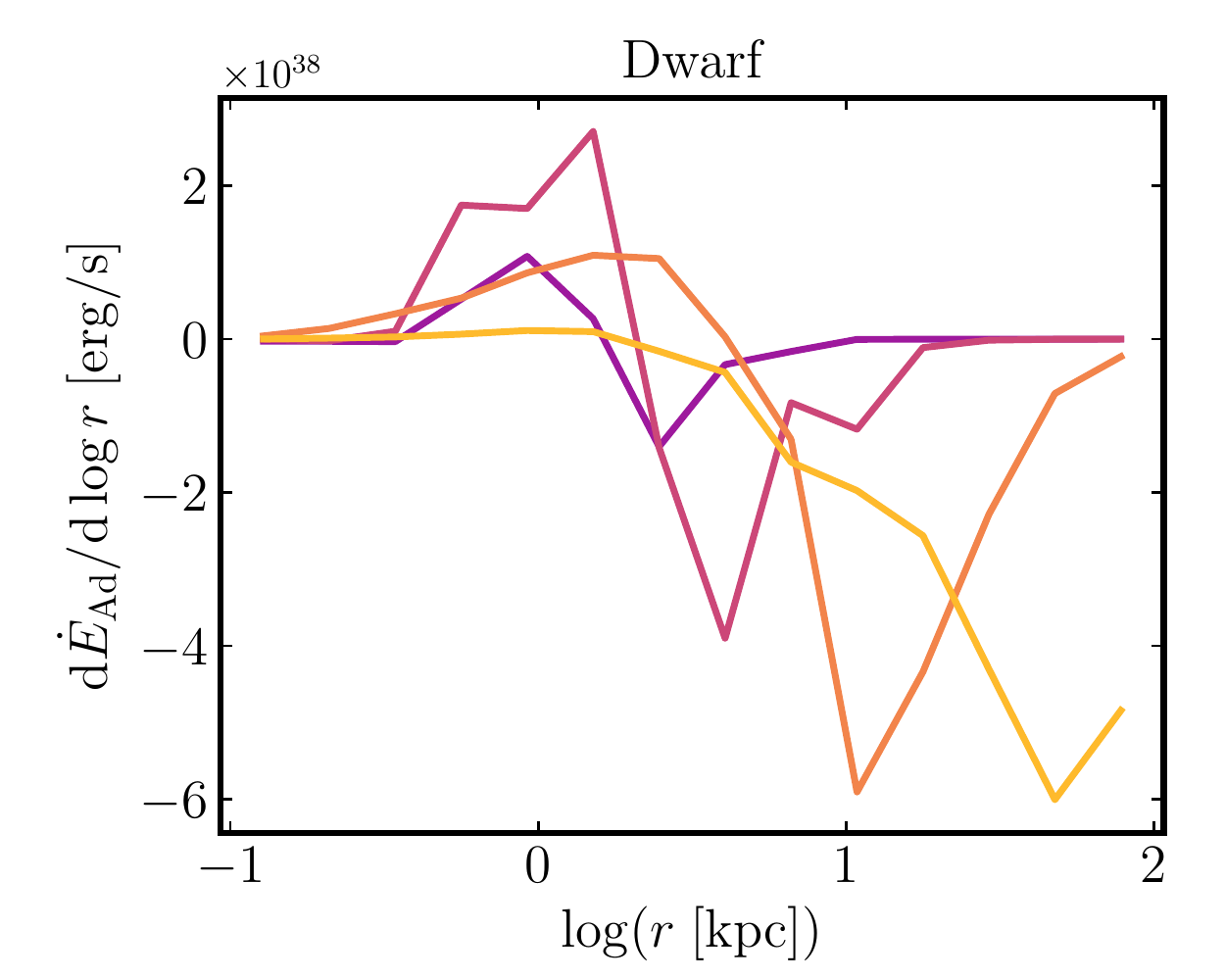}
\end{centering}
  \vspace{-0.25cm}
\caption{Contribution of CR energy at different local gas densities $n$ ({\em top}) or at different galacto-centric radii $r$ ({\em bottom}) to the total adiabatic work term in Fig.~\ref{CRenergies} (calculated for {\bf Dwarf} at $t=500\,$Myr). Gas at high-$n$ ($\gtrsim 0.01\,{\rm cm^{-3}}$) and low-$r$ ($\lesssim 5\,$kpc), i.e.\ within the disk, is primarily contracting, so the ``adiabatic term'' boosts CR energy (increasing $e_{\rm cr}$ and $L_{\gamma}$). Gas at low-$n_{\rm ISM}$ and high-$r$ is primarily expanding so the adiabatic work decreases $e_{\rm cr}$. In simulations with an explicitly-resolved multi-phase ISM like those here, CRs must {\em first} escape dense gas and the disk midplane, before adiabatic terms can significantly reduce $e_{\rm cr}$ or $L_{\gamma}$.
  \vspace{-0.25cm}}
\label{nismCRgl}
\end{figure}

\begin{figure}
\begin{centering}
 \includegraphics[width={0.95\columnwidth}]{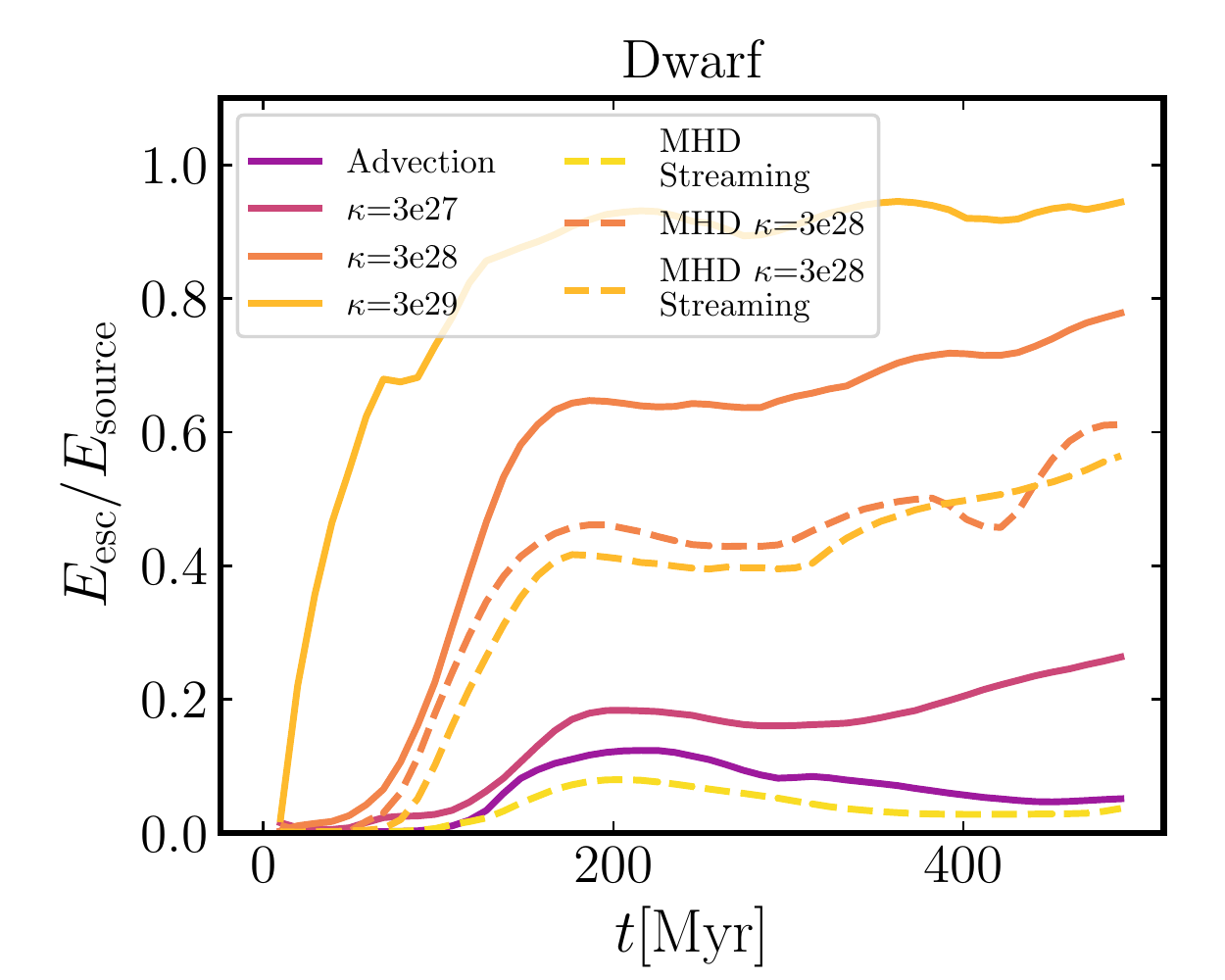}
  \includegraphics[width={0.95\columnwidth}]{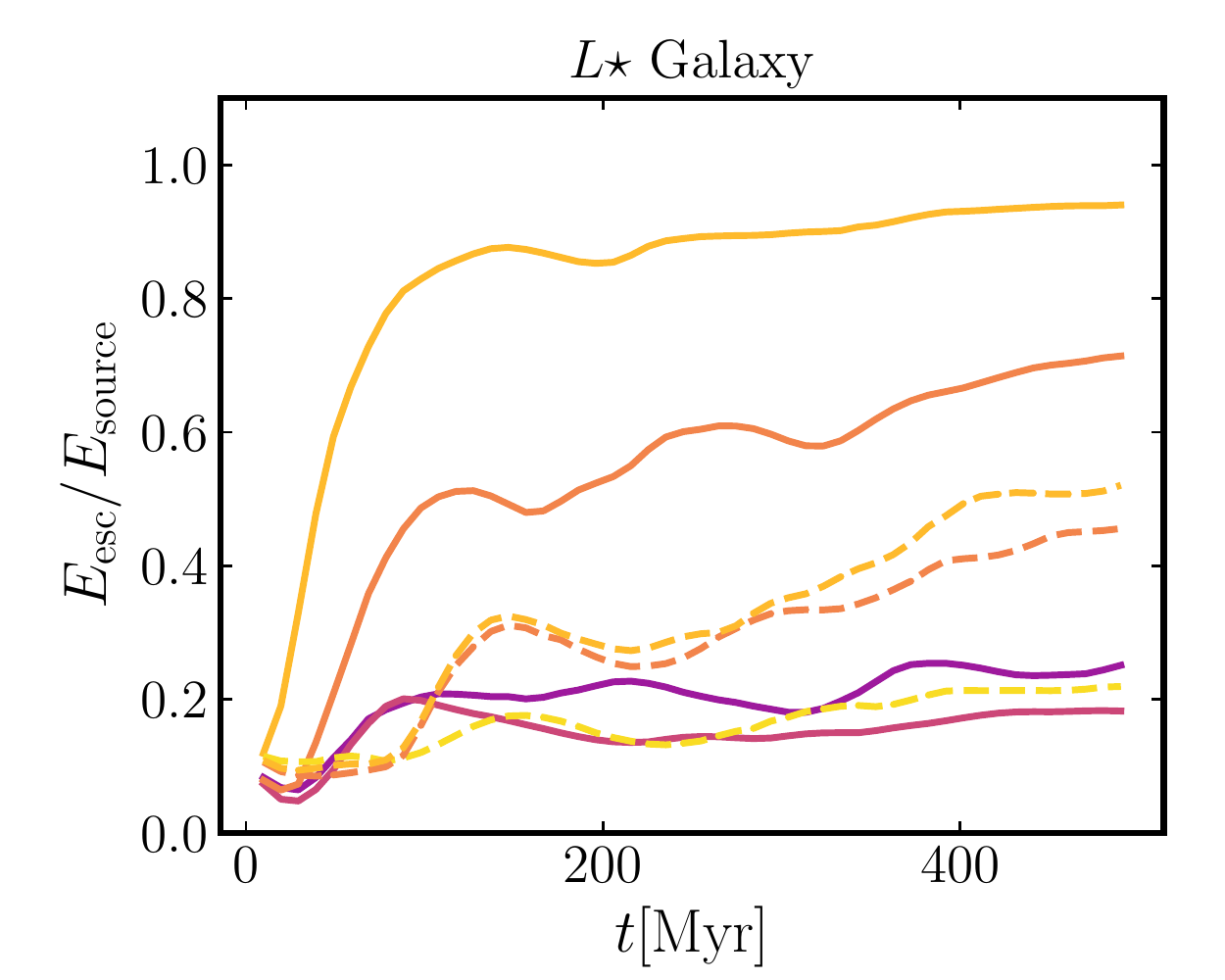}
\end{centering}
\vspace{-0.25cm}
\caption{Ratio between the cumulative CR energy escape ($E_{\rm esc}$) from the central region and the cumulative CR energy input in the central region ($E_{\rm source}$). The CR escape fraction, $E_{\rm esc}/E_{\rm source}$, increases with CR propagation speed: more than 90\% of CR energy leaves the central region for high $\kappa$, compared to only $\sim$10-20\% that leave for simulations with advection-only.}
\vspace{-0.25cm}
\label{EescEsou}
\end{figure}

The ``adiabatic'' term $\dot{E}_{\rm Ad}$ can be comparable to injection or collisional terms, but can be a gain or loss process. To better understand where the gains and losses occur, Fig.~\ref{nismCRgl} shows the contribution to the total $\dot{E}_{\rm Ad}$ from gas with different densities $n$ or at different galacto-centric radii $n$. For CRs at low ambient $n$ (or large $r$), $\dot{E}_{\rm Ad}$ tends to be a loss term (i.e.\ CRs are expanding or losing energy in rarefactions). For CRs in high ambient $n_{\rm ISM}$ and small $r$, it tends to be a gain (CRs are being compressed in converging flows). Recall, the ``adiabatic'' term is {\em defined} (Eq.~\ref{eqn:finite.volume}) by $-\int d^{3}{\bf x} P_{\rm cr}\,(\nabla\cdot {\bf v}) = -(\gamma_{\rm cr}-1)\,\int d E_{\rm cr}\,\nabla\cdot {\bf v}$ (where ${\bf v}$ is the gas velocity). So, combined with Fig.~\ref{nismCRgl}, this simply means that at high gas densities within galaxies, more of the ISM is collapsing/converging or being compressed (in e.g.\ shocks), while at low densities outside galaxies, more the gas is expanding in outflows. Whether one or the other term dominates depends on where most of the CR energy resides (shown in Fig.~\ref{nismcr}). 

So, unsurprisingly in Fig.~\ref{CRenergies}, our runs with the most efficient CR escape to large-$r$ and low-$n$ (all the highest $\kappa \sim 3\times10^{29}\,{\rm cm^{2}\,s^{-1}}$ runs, or most of the {\bf Dwarf} runs with even intermediate $\kappa$) show net $\dot{E}_{\rm Ad} < 0$, since CRs rapidly migrate to the expanding regions. In contrast, those with the least efficient escape (e.g.\ all the {\bf Starburst} runs and the $L\star$ runs with lower isotropically-averaged $\kappa$) show net $\dot{E}_{\rm Ad} > 0$.\footnote{Interestingly, if CRs do not preferentially stay in regions where the adiabatic term is mostly negative or positive, then the $\dot{E}_{\rm Ad}$ term will be relatively small, since adiabatic gains compensate adiabatic losses, as occurs in a couple of our ``{\bf Advection}'' runs.}

For either advection-only or low-$\kappa$ ($\sim 10^{28}\,{\rm cm^{2}\,s^{-1}}$), the qualitative behaviors of $\dot{E}_{\rm Ad}$ in both dwarf and MW-mass systems in \citet{Pfro17gamma} are similar to what we find here. 

However, for the reasons discussed in \S~\ref{comparisongamma}, our results do not support their conclusion that adiabatic losses are the dominant factor for the low $L_{\gamma}/L_{\rm SF}$ in dwarfs. At very low $\kappa$, $\dot{E}_{\rm Ad}$ is primarily a gain term. But even at higher $\kappa$ where $\dot{E}_{\rm Ad}<0$ is a loss term, it is insufficient (in itself) to explain the very low $L_{\gamma}/L_{\rm SF}$ observed in the SMC/LMC without significant CR leakage (the adiabatic+SNe terms are not enough to explain the loss terms in the top right panel of Fig.~\ref{CRenergies}). More importantly, the adiabatic losses arise only after the CRs have already escaped the dense gas, i.e. the regions which produce
most of $L_\gamma$.

To explicitly show that CR leakage is significant, in Fig. \ref{EescEsou} we plot the fraction of CR energy that escapes the central galactic region, $E_{\rm esc}/E_{\rm source}$. We define $E_{\rm source}$ as the total cumulative CR energy input within the central 6 kpc (10 kpc) and $E_{\rm esc}$ as the CR energy that leaves this central region in {\bf Dwarf} ($L\star$ {\bf galaxy})\footnote{In practice, we calculate $E_{\rm source}$ by summing up all positive CR energy gains within the central region, including SNe, adiabatic, and the small numerical error terms mentioned in footnote \ref{adaptiveerr}. To calculate $E_{\rm esc}$, we sum up all CR energies outside the central region and compensate for the collisional, streaming, and adiabatic losses.}. Outside of this central region the (hadronic) dissipation time is much longer than 50 Myr. At high CR propagation speeds (e.g. high $\kappa$), most of the CR energy indeed escapes the central region, where most $\gamma$ rays are produced. This shows that CR escape is the main reason for  reduced $\gamma$ ray emission in low-gas-density galaxies.

\section{Conclusions}
\label{conclusion}

We explore the effects of CRs on galaxies, in high-resolution, idealized (non-cosmological) (magneto-)hydrodynamic simulations of dwarf, $L\star$, and sub-$L\star$ starburst galaxies, using the FIRE-2 treatment of the multi-phase ISM, star formation, and stellar feedback, accounting for CR injection from SNe, collisional (hadronic+Coulomb) losses, and CR transport via diffusion and streaming. We focus on constraining CR propagation models (e.g.\ diffusion and streaming coefficients) using observations of GeV $\gamma$-rays from galaxies. Our main conclusions include:

\begin{enumerate}

\item We adopt a newly developed two moment method for CR transport, and show that it is computationally efficient and accurate, allowing us to simulate CR transport simultaneously including diffusion and streaming with diffusivities up to $\sim 3\times10^{29}\,{\rm cm^{2}\,s^{-1}}$ and $\sim$\,pc resolution.

\item The CR ``transport parameters'', in particular, the effective diffusivity $\kappa \equiv |{\bf F}_{\rm cr}|/|\nabla_{\|}e_{\rm cr}|$ (which can, microphysically, arise from a combination of streaming and diffusion), have a significant impact on galaxy properties and predicted $\gamma$-ray emission. With very slow propagation ($\kappa\lesssim 10^{28}{\rm cm^2/s}$), CRs are trapped in the disk and contribute to the mid-plane pressure gradients, so suppress SF (albeit only by modest factors $\sim 1.5-2$, {\em if} hadronic losses which limit the CR energy density are accounted for). However, these models are ruled out because they produce much larger $\gamma$-ray luminosities than the observed for dwarf or MW-like systems. At higher $\kappa \gtrsim 10^{28}\,{\rm cm^{2}\,s^{-1}}$, CRs form extended halos. This means they have weak effects on gas {\em within} the disk, but could help accelerate galactic winds or provide support via pressure gradients in the CGM.

\item The extent of the CR halo, and correspondingly the extent of the pionic $\gamma$-ray emission, increase with $\kappa$ as expected. For e.g.\ our $L\star$ galaxy, half the CR energy is located within 10\,kpc (30\,kpc) for $\kappa=3\times10^{28}\,{\rm cm^{2}\,s^{-1}}$ ($\kappa=3\times10^{29}\,{\rm cm^{2}\,s^{-1}}$). Correspondingly only about $\sim 50\%$ of the $\gamma$-rays are emitted from the central few kpc.

\item In our sub-$L\star$ starburst galaxies, the CR energy density reaches $\sim 10^{2-3}\,{\rm eV\,cm^{-3}}$ throughout the burst and is larger than thermal or magnetic pressure in the ISM (for any $\kappa$), but is still much smaller than the energy density in turbulent motions or that required to maintain hydrostatic equilibrium. This leads to weak CR effects at the central region of starburst, consistent with the results in \citetalias{Lack11}.

\item We constrain the average CR propagation speed/diffusivity with $\gamma$-ray ($>$\,GeV) emission from galaxies. The observed $L_{\gamma}-{\rm SFR}$ relation requires isotropically-averaged diffusivities $\kappa\sim 3\times 10^{29}{\rm cm^2/s}$ in dwarf and $L\star$ galaxies, and $\kappa \lesssim 3\times10^{29}{\rm cm^2/s}$ in sub-$L\star$ starburst galaxies.

 If CRs are transported only by gas advection, or streaming only at speeds which cannot exceed modestly super-Alfvenic values, or (equivalently) low isotropically-averaged effective diffusivities $\kappa < 3\times 10^{28}\,{\rm cm^{2}\,s^{-1}}$,then CRs escape galaxies too slowly and produce $\gamma$-ray luminosity close to the calorimetric limit. This over-predicts the observed $\gamma$-ray luminosities in dwarfs (e.g.\ the SMC, LMC, M33) and $L\star$ systems (M31, the MW) by an order of magnitude or more. However, for faster transport parameters (effective $\kappa \sim 3\times 10^{28-29}\,{\rm cm^{2}\,s^{-1}}$), CRs escape the dense regions rapidly and the $\gamma$-ray luminosity (which scales $\propto e_{\rm cr}\,n_{\rm gas}$) is reduced (especially in dwarf galaxies), predicting $\gamma$-ray luminosities in good agreement with those observed as a function of either gas surface density or SFR (see Figs. \ref{gassurFgFsf} and \ref{SFRLg}).

\item Given the transport parameters required to reproduce the observed $L_{\gamma}/L_{\rm SF}$, we find most CR protons escape from dwarf galaxies, i.e. low-gas-surface-density systems are not proton calorimeters, while our (sub-$L\star$) starburst models are (approximate) proton calorimeters.

\item  CR streaming at trans-Alfvenic speeds is relatively slow and cannot alone reduce $L_{\gamma}/L_{\rm SF}$ significantly below the calorimetric limit in our models (as required by observations), even if we allow modestly super-Alfvenic streaming (with $\sim 4\,v_{\rm A}$; see Appendix \ref{diffstr} ). For our favored effective $\kappa$, the equivalent streaming speed (using the fact that the CR flux is similar for a diffusivity $\kappa$ or streaming speed $v_{\rm st}  \sim  \kappa\,|\nabla P_{\rm cr}|/P_{\rm cr}$) is $\sim 10-100$ times the Alfven speed.

\item ``Adiabatic'' effects on CR energy densities (losses in expansion, or gains in compression) can be comparable to injection or collisional loss terms, but cannot alone reduce $L_{\gamma}/L_{\rm SF}$ close to the level required by observations of the MW/SMC/LMC/M33. In dense gas within the galaxies, the net effect of these terms is primarily to increase CR energy density (and $L_{\gamma}$), while in low-density gas outside galaxies, it is primarily to decrease the CR energy via expansion in outflows. This means that CR ``adiabatic losses'' are significant only {\em after} CRs already diffuse out of the dense ISM gas (where $\gamma$-rays are produced). 

\end{enumerate}

Our study only scratches the surface of the rich phenomena of CRs in galaxies and leaves out many important details. 
For example, it is clearly important to study the effects of CRs in cosmological galaxy simulations, which can treat CRs and magnetic field evolution consistently, explore the effects of CRs on magnetic field amplification, self-consistently generate starburst systems (in e.g.\ mergers), and (perhaps most importantly) explore the interaction of CRs with inflows and outflows in a ``live'' CGM/IGM environment. We will explore such cosmological runs in future work (Hopkins et al., in preparation).

Although we briefly mentioned the effects of CRs on galactic winds (which are ubiquitous in these simulations), we have not investigated them here. It has been proposed that CR-driven winds could have very different phase structure (compared to thermally-driven winds) and strongly modify the CGM properties \citep{Boot13,Sale16}. Although extensive literature on this topic exists, detailed study of CR winds in (cosmological) simulations that can already self-consistently drive galactic winds with stellar feedback \citep{Mura15,Angl17,Hafe18} is largely unexplored. Our simulations provide a unique combination of a high-resolution, multi-phase ISM, with explicit treatment of local star formation in self-gravitating substructures, individually time-resolved SNe and their thermal and momentum feedback combined with CR injection and transport. It will therefore be especially interesting to explore the effects of CRs on the development of galactic winds (Chan et al., in preparation).

We also do not study another important indirect CR constraint, the radio emission from synchrotron radiation, which has been observed in many galaxies \citep{Cond92}. These observations provide independent constraints on primary CRs, secondary CR electrons from CR protons, and magnetic fields. It is worth also exploring the observed FIR-radio correlation \citep{vand71,vand73} with galaxy simulations in a manner similar to our analysis of the connection between SFR and $\gamma$-ray emission. However, as mentioned in \cite{Lack10}, these correlation requires the consideration of secondary CRs, which we plan to incorporate in the future. 

Because this was an idealized parameter study, we have adapted a simple model with a constant isotropic/parallel diffusivity $\kappa$. But in essentially any physical model, this coefficient depends on local properties of the gas and CRs, in a manner which remains deeply uncertain both theoretically and observationally (see e.g.\ \citealt{Joki66,Enss03}). It would be interesting to investigate galaxy evolution and CR observables in studies where the CR transport coefficients vary dynamically and locally \citep[see e.g.][]{Farb18}, or with recently-developed models which attempt to actually predict the coefficients self-consistently \citep{Thom19}.

\section{ACKNOWLEDGEMENTS}
We thank the anonymous referee for the detailed comments that helped to improve this manuscript. We thank Patrick Diamond, Ellen Zweibel, Michael Norman, and Todd Thompson for insightful suggestions and advice, Eve Ostriker for pointing out a typo, and Bili Dong for his help with {\small YT}. We would like to thank the Simons Foundation and the participants of the {\it Galactic Superwinds} symposia for stimulating discussions. TKC and DK were supported by NSF grant AST-1715101 and the Cottrell Scholar Award from the Research Corporation for Science Advancement. Support for PFH was provided by an Alfred P. Sloan Research Fellowship, NSF Collaborative Research Grant \#1715847 and CAREER grant \#1455342, and NASA grants NNX15AT06G, JPL 1589742, 17-ATP17-0214. EQ was supported in part by a Simons Investigator Award from the Simons Foundation and by NSF grant AST-1715070. The Flatiron Institute
is supported by the Simons Foundation. CAFG was supported by NSF through grants AST-1517491, AST-1715216, and CAREER award AST-1652522, by NASA through grants NNX15AB22G and 17-ATP17-0067, and by a Cottrell Scholar Award from the Research Corporation for Science Advancement. 

The simulation presented here used computational resources granted by the Extreme Science and Engineering Discovery Environment (XSEDE), which is supported by National Science Foundation grant no. OCI-1053575, specifically allocation TG-AST120025. Numerical calculations were also run on the Caltech compute cluster ``Wheeler'', allocations from XSEDE TG-AST130039 and PRAC NSF.1713353 supported by the NSF, and NASA HEC SMD-16-7592. This work uses data hosted by the Flatiron Institute's FIRE data hub. This work also made use of {\small YT} \citep{YTproject}, matplotlib \citep{Hunt07}, numpy \citep{vand11}, scipy \citep{Jone01}, and NASA’s Astrophysics Data System.

\bibliographystyle{mn2e}
\bibliography{mn-jour,mybib}

\appendix

\section{A comparison of different streaming parameters}
\label{diffstr}
\begin{figure*}
\begin{centering}
\includegraphics[width={0.9\columnwidth}]{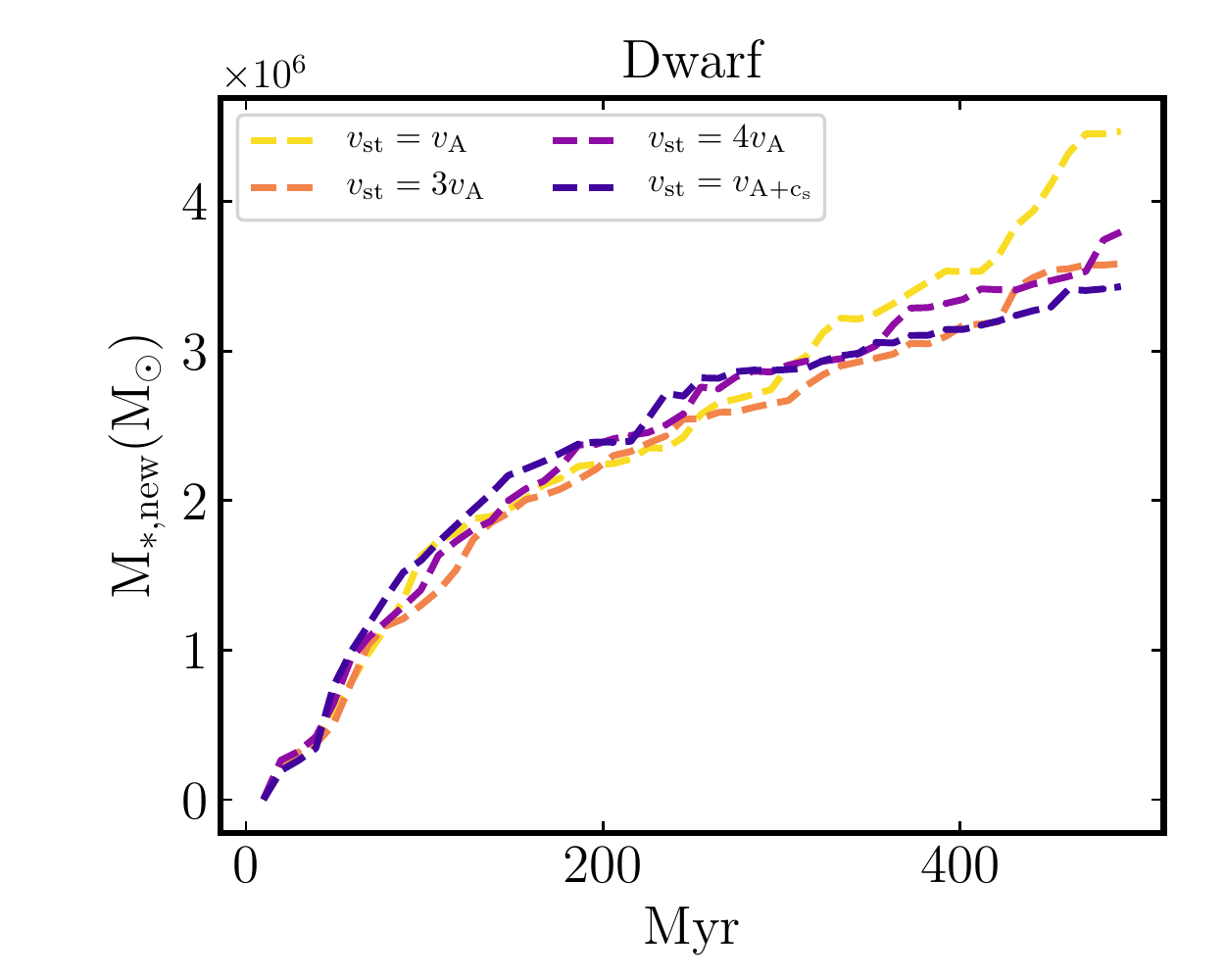}
\includegraphics[width={0.9\columnwidth}]{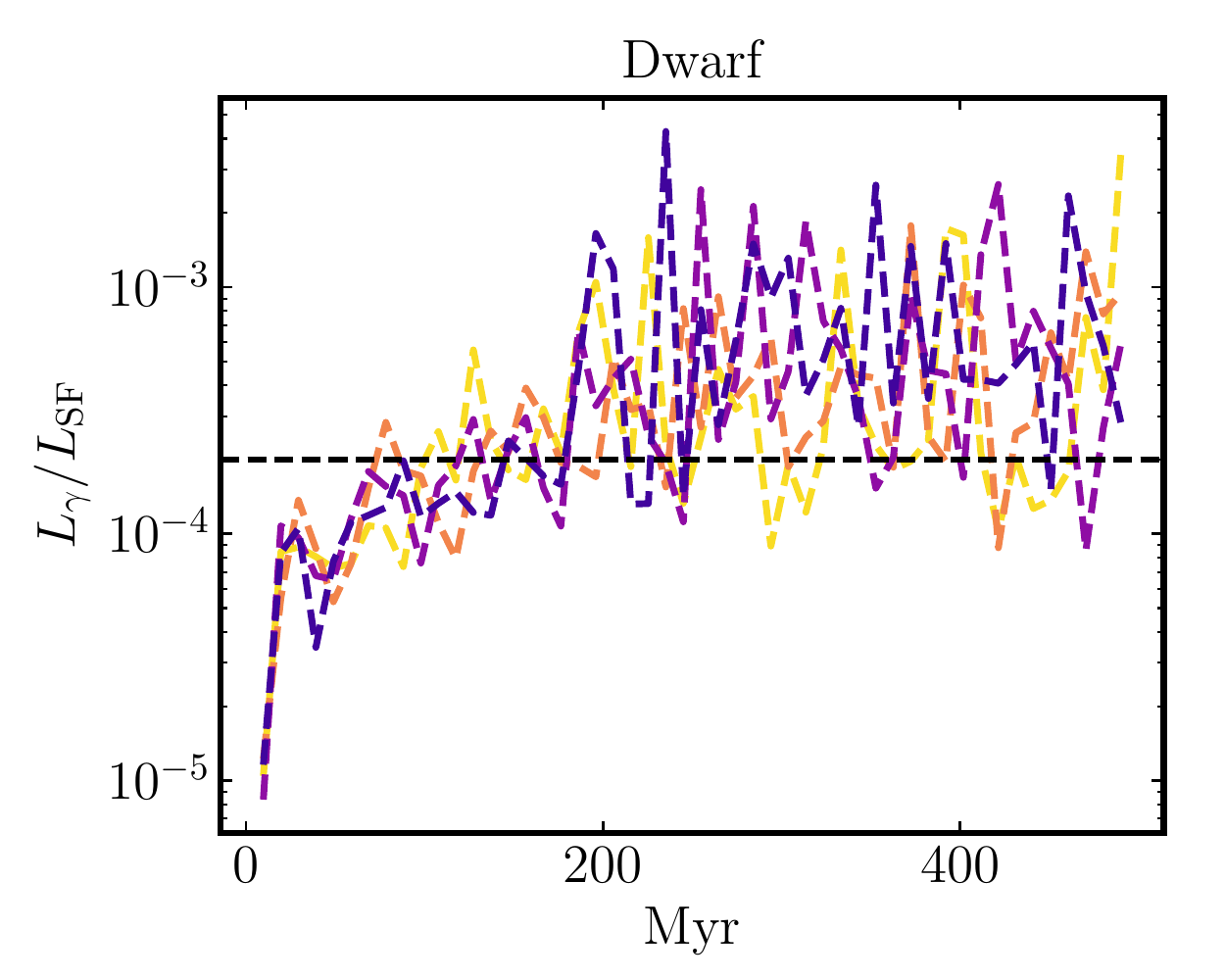}\\
  \includegraphics[width={0.9\columnwidth}]{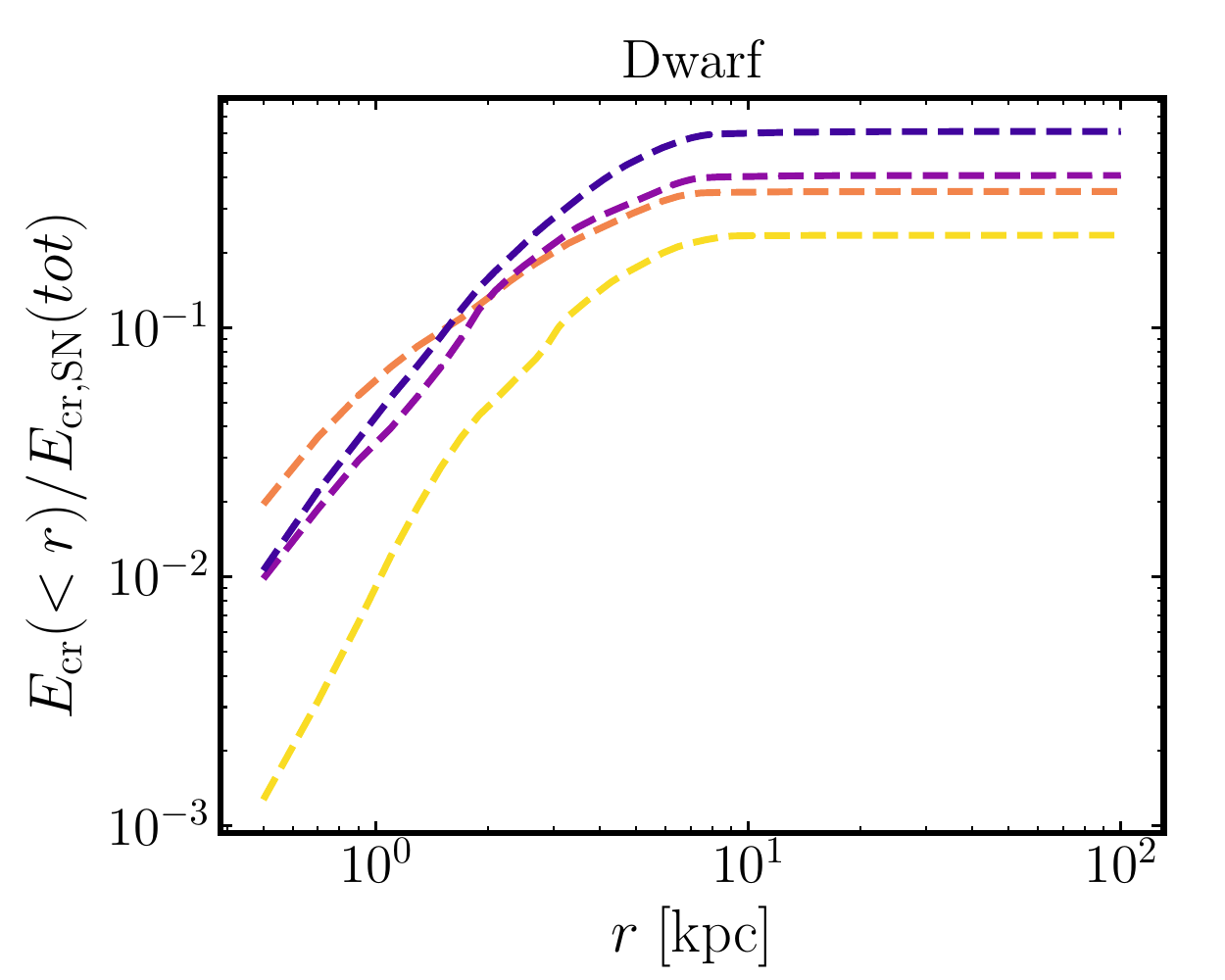}
  \includegraphics[width={0.9\columnwidth}]{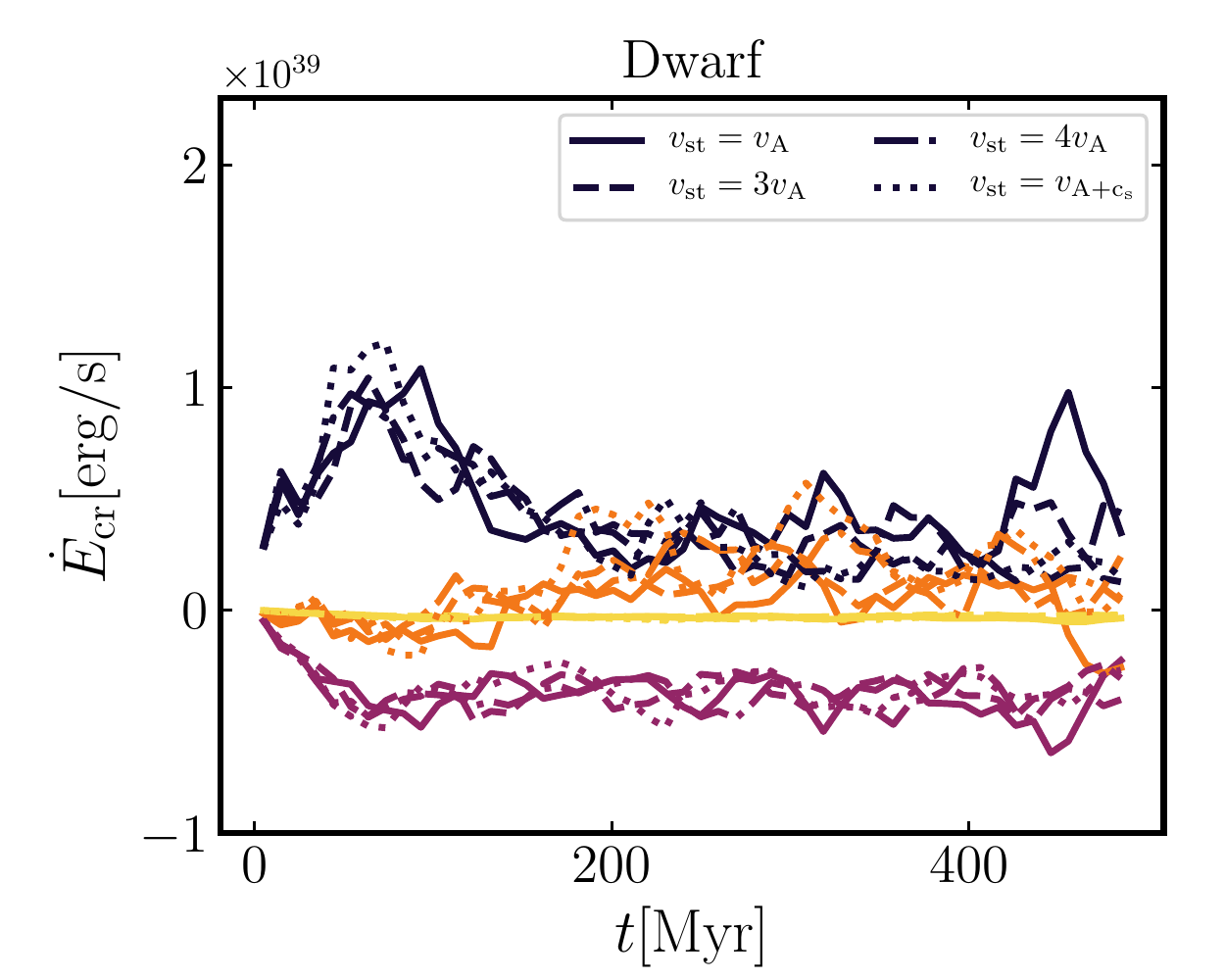}\\
  \end{centering}
\vspace{-0.2cm}
\caption{Comparison of effects of different CR streaming speeds and loss rates, described in Appendix~\ref{diffstr}. {\em Top Left:} Stellar mass vs.\ time (as Fig.~\ref{sfr}). {\em Top Right:} $\gamma$-ray luminosity relative to SF luminosity (as Fig.~\ref{Fgammasfr}). {\em Bottom Left:} Cumulative CR energy vs.\ radius (as Fig.~\ref{CRgasz}). {\em Bottom Right:} CR energy gain/loss rate (as Fig.~\ref{CRenergies}) via injection (black), adiabatic terms (orange), streaming losses (yellow), collisional losses (purple).
There are only small differences between streaming schemes, compared to e.g.\ the effects of changing the diffusion coefficient $\kappa$. Most importantly, the $L_{\gamma}/L_{\rm SF}$ ratio is around the calorimetric limit regardless of which streaming scheme we used (here there is no additional diffusion), which means streaming -- even when modestly super-Alfvenic -- is not effective in transporting CRs from the galaxies.}
\label{smcstreams}
\end{figure*}

In \S~\ref{strmethod}, we discussed uncertainties in the parameters describing CR streaming: both the streaming speed $v_{\rm st}$ and magnitude of the streaming loss term $\Gamma_{\rm st}=-{\bf v}_{\rm A}\cdot\nabla P_{\rm cr}$ owing to excited and thermalized high-frequency Alfven waves (independent of streaming speed; see \S\ref{strmethod}). Here we explore these more thoroughly. We consider four model variations:
\begin{enumerate}
\item{``$v_{\rm st}=v_{\rm A}$'': This is our default choice in the main text, with streaming speed equal to the Alfven speed.}

\item{``$v_{\rm st}=3v_{\rm A}$'': Here $v_{\rm st} = 3\,v_{\rm A}$, one of the super Alfvenic speeds considered in \cite{Rusz17}.}

\item{``$v_{\rm st}=4v_{\rm A}$'': Here $v_{\rm st} = 4 v_{\rm A}$, another super Alfvenic speed considered in \cite{Rusz17}}

\item{``$v_{\rm st}=v_{\rm A+c_{\rm s}}$'': Here $v_{\rm st} = \sqrt{v_{\rm A}^2+c_{\rm s}^2}$, the fastest MHD wavespeed (which has no particular physical motivation but resembles what might be inferred in hydrodynamic models or observations where a plasma $\beta\sim1$ is simply assumed).}
\end{enumerate}

Fig.~\ref{smcstreams} summarizes the results: the effects of this choice are much smaller than the variations of e.g.\ $\kappa$ discussed in the main text. Most importantly, we see {\em no} difference in averaged $L_{\gamma}/L_{\rm SF}$ (although there are significant fluctuations due to stochasticity), implying that streaming at these speeds -- even if modestly super-Alfvenic streaming is allowed -- is ineffective at transporting CRs from dense regions. Of course, if we limit the ``streaming losses'' to scale with ${\bf v}_{\rm A}\cdot  \nabla P_{\rm cr}$ and continue to increase $v_{\rm st}$ without limit, it  will eventually become ``fast enough.'' In fact, our Eq.~\ref{kappa.effective} shows that, given the manner in  which we approximate CR transport numerically, $\kappa$ and $v_{\rm st}$ are formally degenerate if we replace $\kappa$ with $v_{\rm st}\rightarrow \kappa\,|\hat{\bf B}\cdot \nabla e_{\rm cr}|/(e_{\rm cr}+P_{\rm cr})$. From this, using a typical CR gradient scale length $\sim 1-10\,$kpc in our galaxies and favored $\kappa\sim 3\times10^{29}\,{\rm cm^{-2}\,s^{-1}}$, we see that $v_{\rm st}  \sim 100-1000\,{\rm km\,s^{-1}} \gg v_{\rm A}$ (and $\gg c_{s}$ in most of the ISM) is required.

We have run the similar tests with our $L\star$ galaxy model (not shown) and find qualitatively identical results (with even smaller differences between streaming models).

\section{Numerical Tests}

\begin{figure*}
\includegraphics[width={0.32\textwidth}]{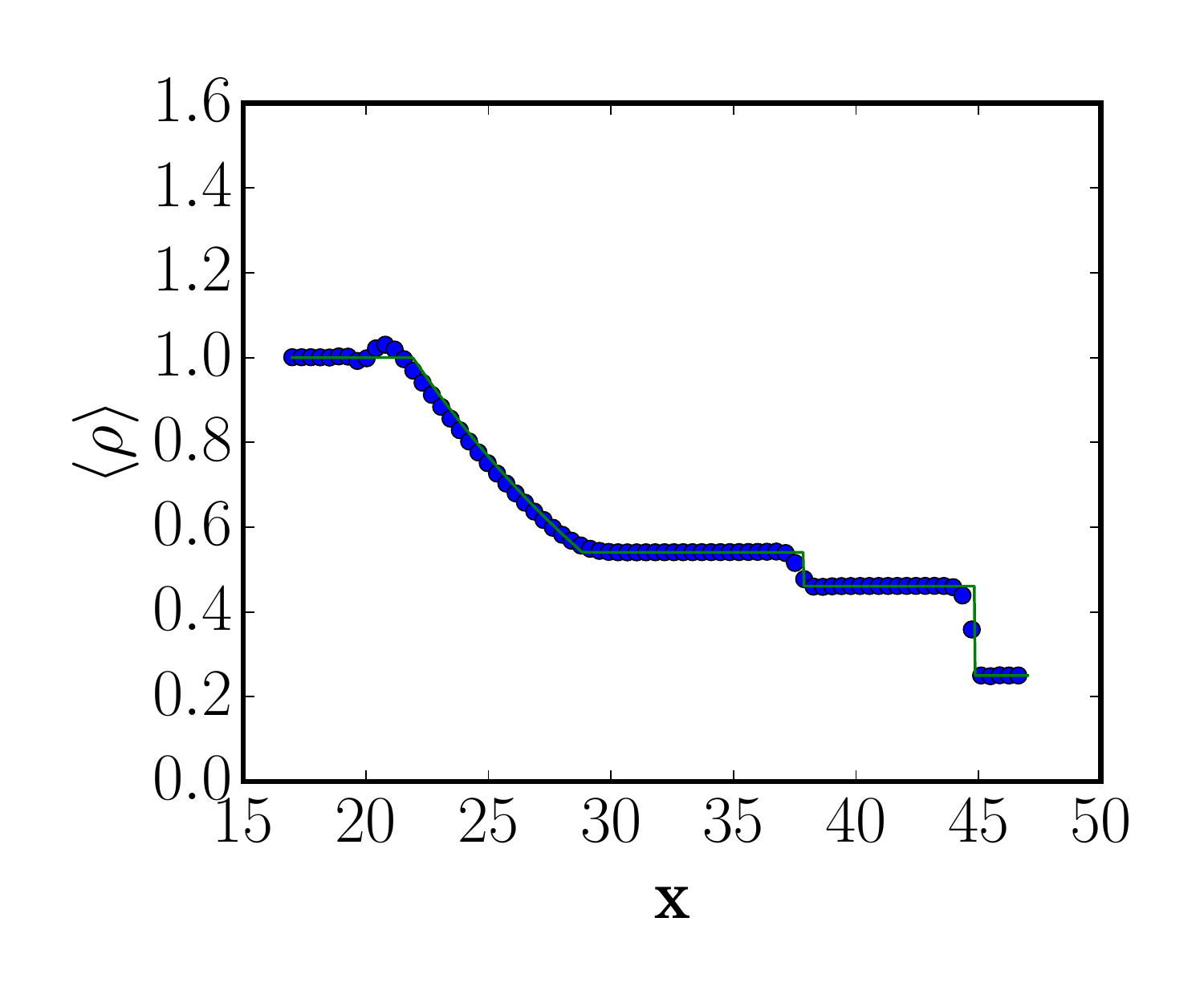}
\includegraphics[width={0.32\textwidth}]{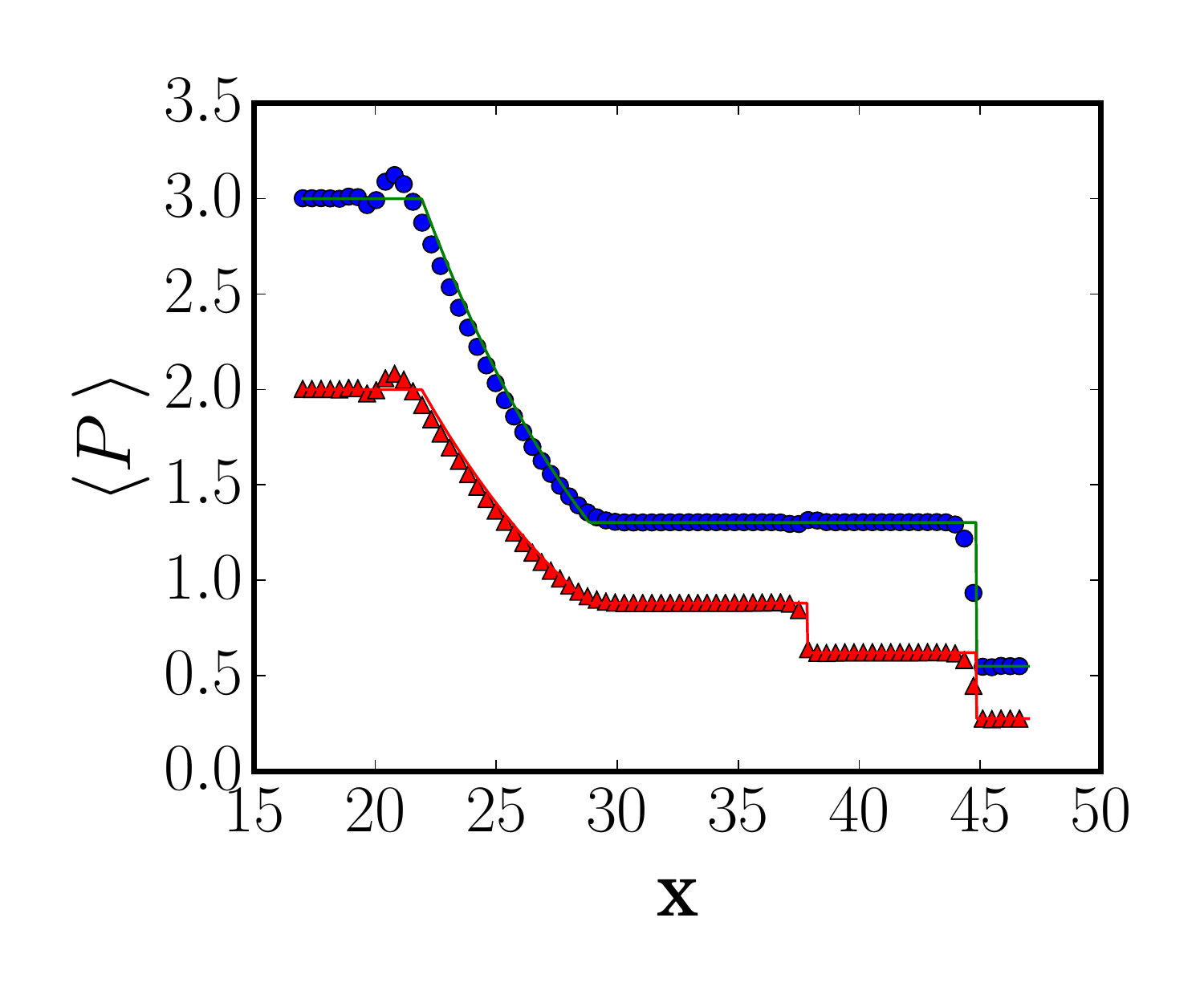}
\includegraphics[width={0.32\textwidth}]{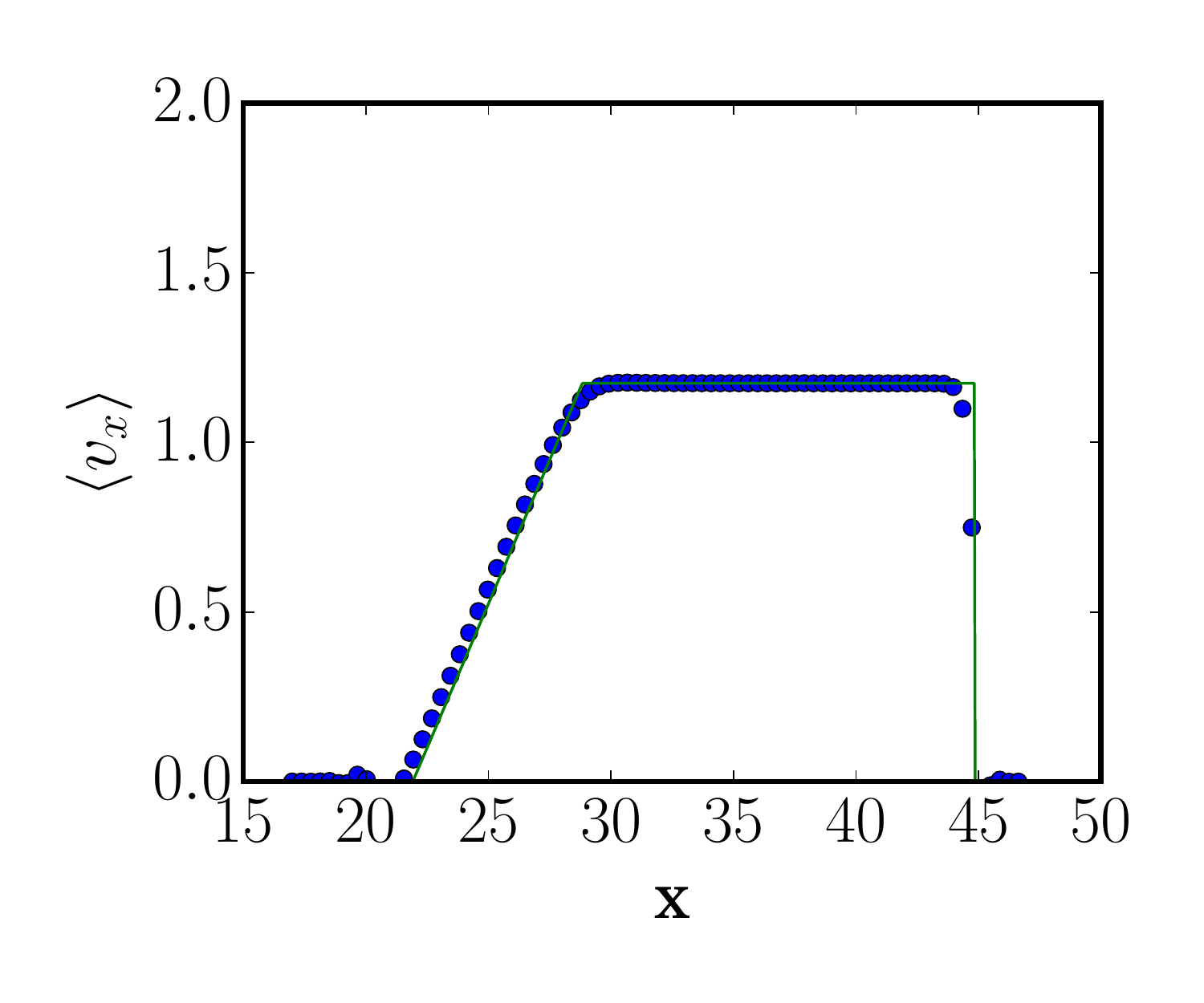}\\
\includegraphics[width={0.32\textwidth}]{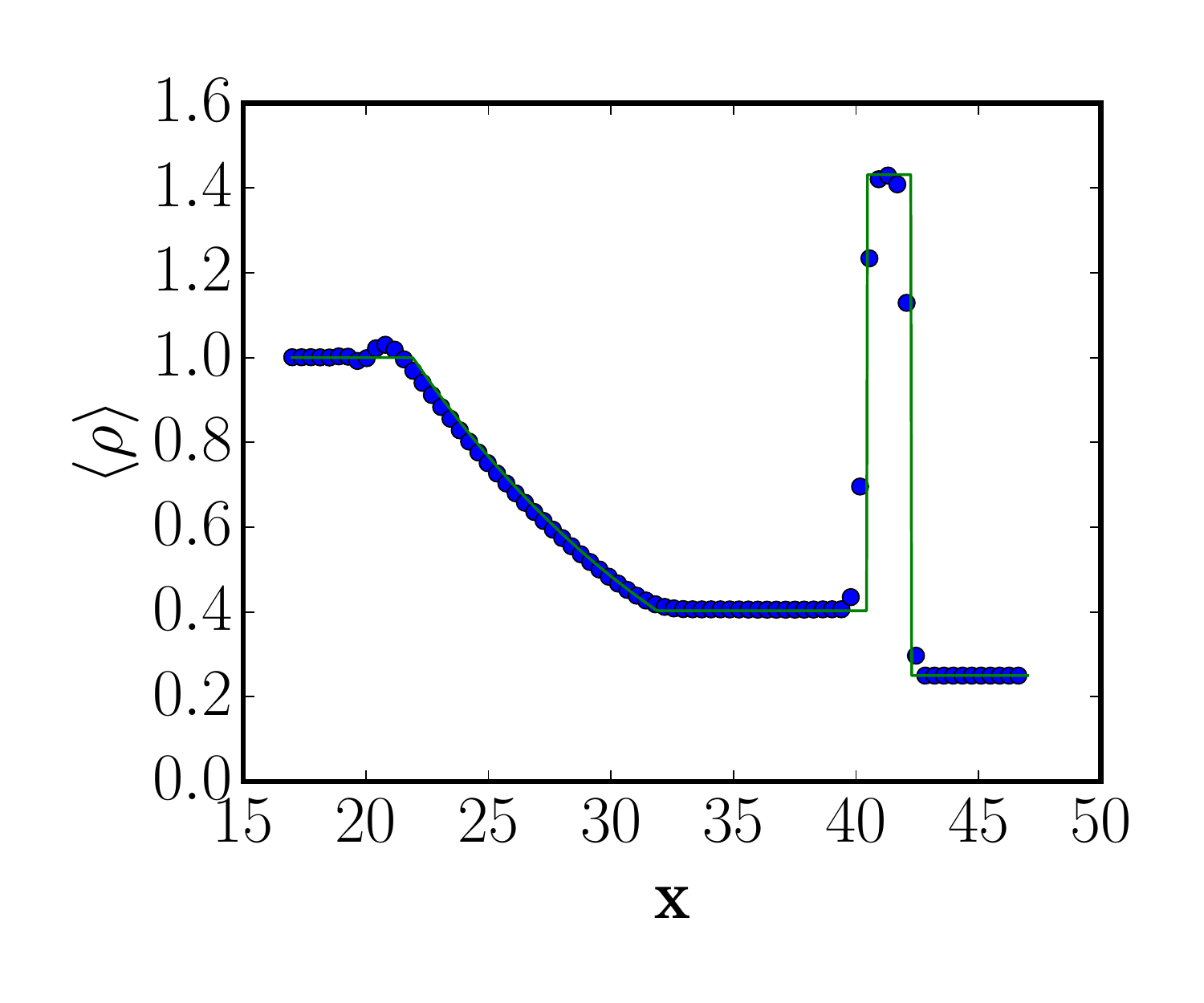}
\includegraphics[width={0.32\textwidth}]{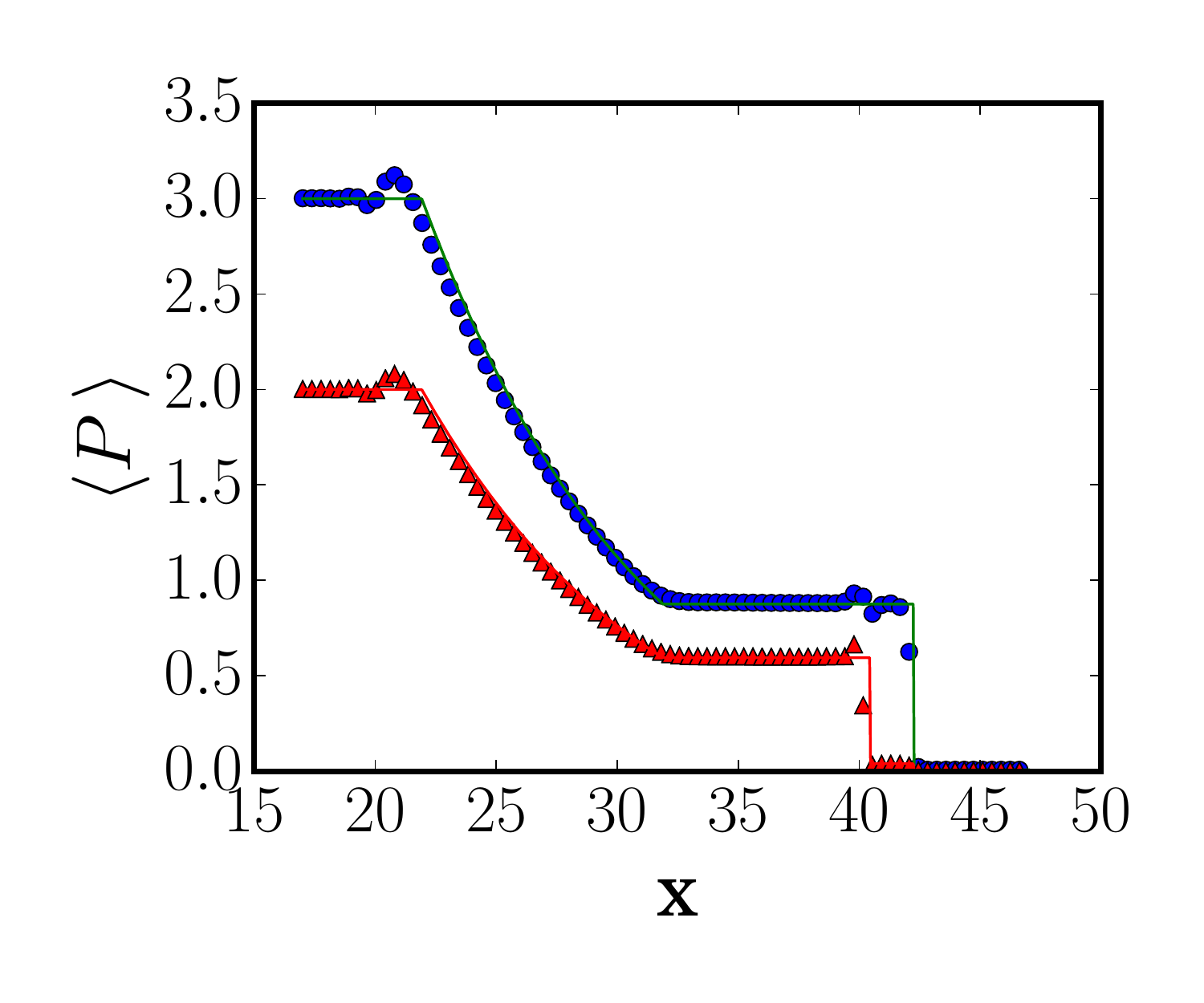}
\includegraphics[width={0.32\textwidth}]{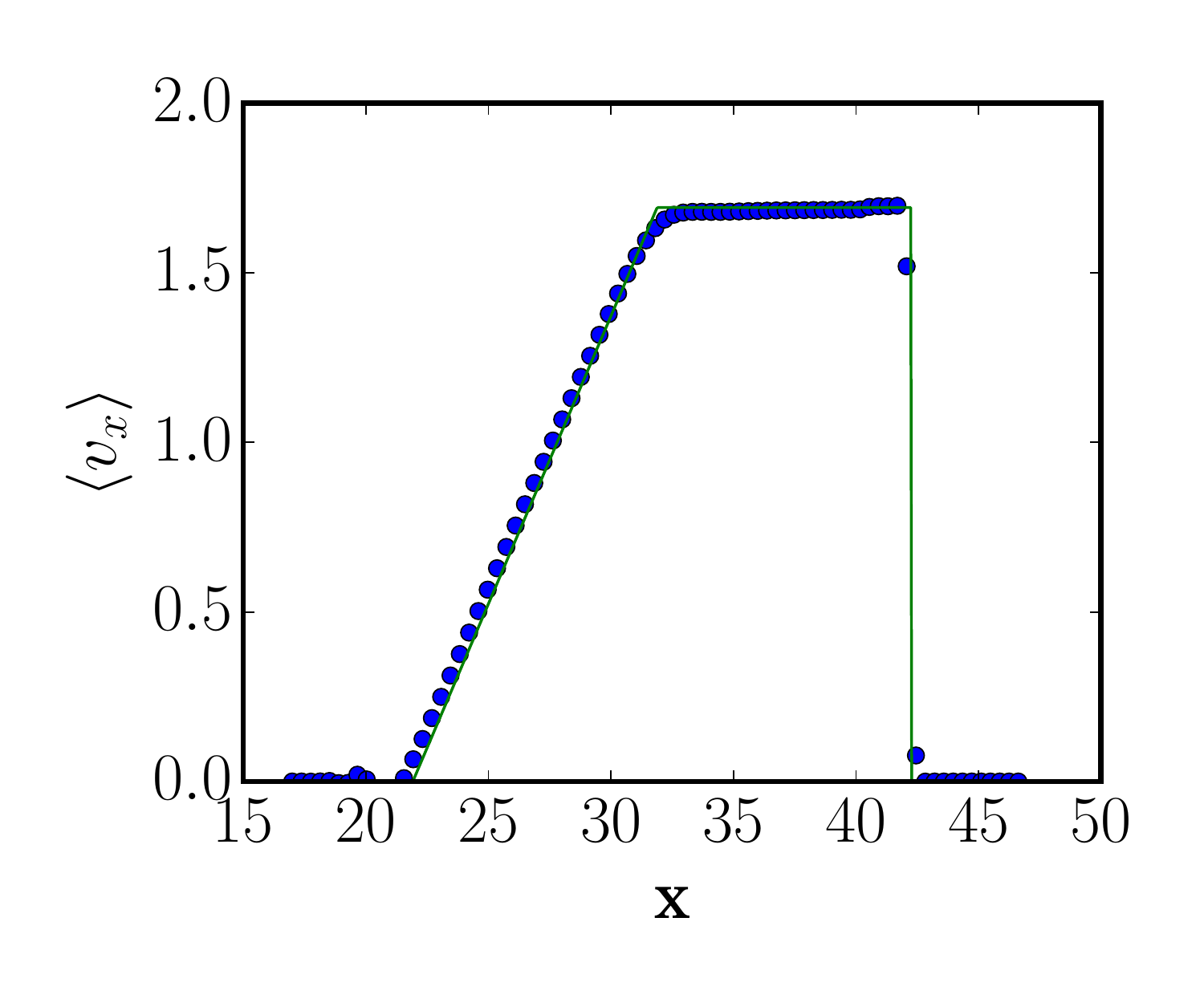}
\vspace{-0.35cm}
\caption{Density, pressure and gas velocity profiles of the Sod shocktubes with a composite of gas and CR (\S~\ref{shocktube}). We show a shocktube with Mach number $M=1.5$ in the top panels and $M=10$ in the bottom panels. In the left panels, we show the volume averaged gas densities from the analytic calculations ({\it green} lines; from Appendix B of \citealt{Pfro06}) and simulations ({\it blue} points). In the center panels, the green lines and blue points show the analytic and simulated volume-averaged total (gas + CR) pressure, respectively. Red lines and points show analytic and simulated volume-averaged CR pressure, respectively. In the right panels, we show the volume averaged gas velocities in the $x$ direction. The simulations are a good match to the analytic solutions.}
\label{SodCR}
\end{figure*}

Our CR transport implementation is described in \S~\ref{diffM1}. Here we present some numerical tests, including simple code validation problems, tests of the effect of the maximum CR free-streaming speed (or ``reduced speed of light'' $\tilde{c}$), comparison of our two-moment implementation to zeroth-moment ``pure diffusion'' solvers, and numerical resolution studies.

\subsection{CR Shocktube Test}
\label{shocktube}

We test our code implementation of CR coupling to adiabatic and advective terms and in the MHD Riemann problem using a variation of the \citet{Sod78} shocktube presented in \citet{Pfro06}. A 3D box of of dimensions $64\times16\times16$ is full of gas (adiabatic index $7/5$) and CRs. Half the box has initial $(\rho,\,v_{x}/c_{s},\,P_{\rm gas},\,P_{\rm cr})=(1,1,1,2)$ and the other half has $(0.25,1.5,0.275,0.275)$. We also consider another shocktube with $(1,1,1,2)$ and $(0.25,10,0.00384,0.00384$). In both our mass resolution is $0.004$. CRs have no diffusion or streaming (just advection). 

Fig.~\ref{SodCR} compares with analytic solutions from \citet{Pfro06} at $t=5$. The agreement is good (despite very small shock broadening and numerical oscillations near discontinuities). The small ``bumps'' on the left close to x=20 are due to the slope limiter and should converge away at a higher resolution in our MFM method (see the hydrodynamic Sod shocktube test in \citealt{Hopk15}).

\begin{figure}
\begin{centering}
\includegraphics[width={0.9\columnwidth}]{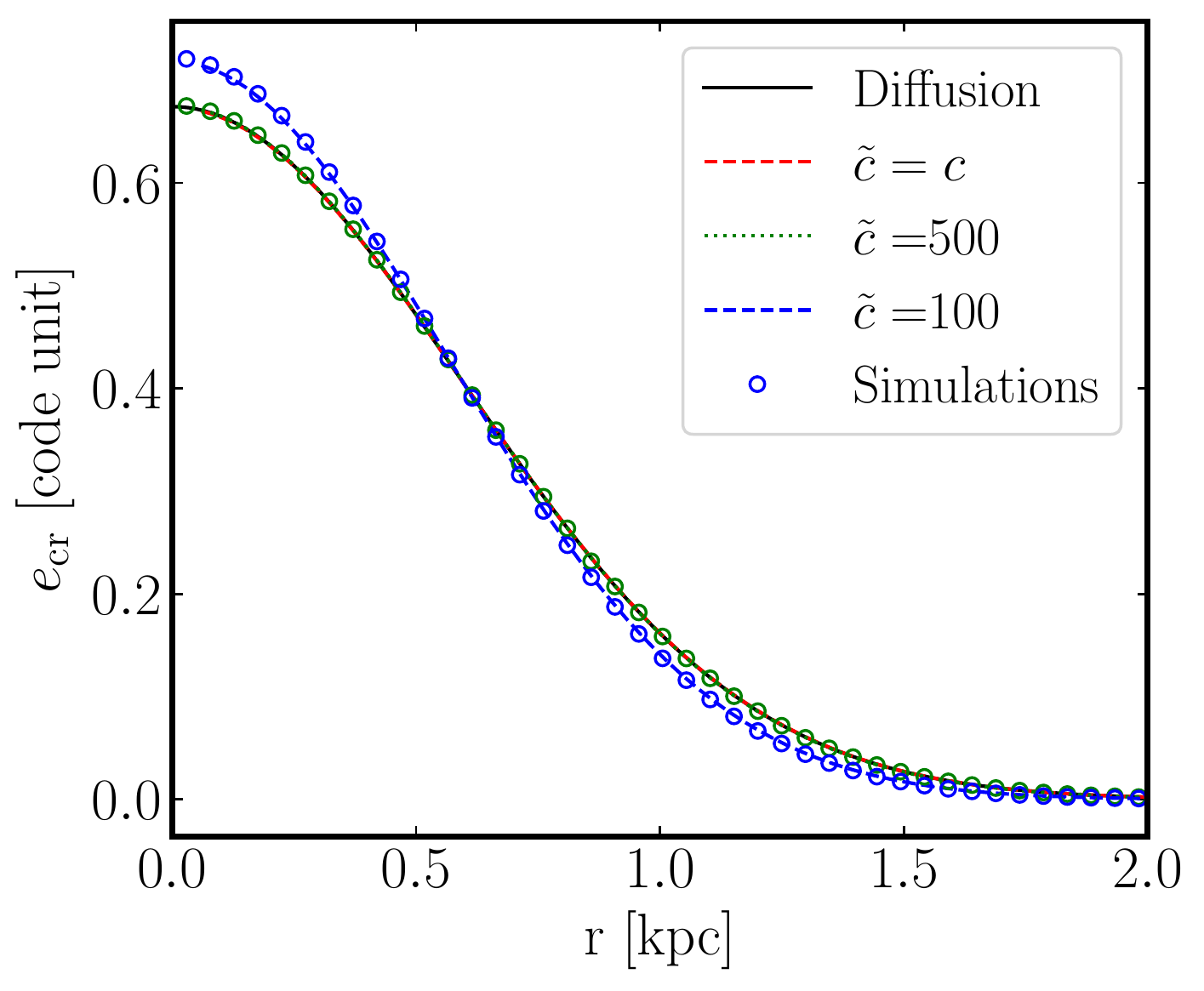}
\end{centering}
\vspace{-0.25cm}
\caption{Idealized 1D CR diffusion test (\S~\ref{advectdiffuse}), run for 5 Myr. We solve the two-moment CR transfer equation with $\tilde{c}$ given (in ${\rm km\,s^{-1}}$) and $\kappa=3\times10^{27}\,{\rm cm^{2}\,s^{-1}}$ given an initial Gaussian CR energy density with $\sigma=0.5\,$kpc. Lines show analytic solutions for both finite $\tilde{c}$ and the ``pure diffusion'' equation, whereas points show simulation results (colors represent $\tilde{c}$). At all $\tilde{c}$ the numerical solutions agree well with the analytic result. Also, given these spatial/timescales, the difference between solutions with finite $\tilde{c} \gtrsim 500\,{\rm km\,s^{-1}}$, $\tilde{c}=c$, and ``pure diffusion'' is extremely small, and even at $\tilde{c}= 100\,{\rm km\,s^{-1}}$ the solutions differ by less than $10\%$.}
\label{idealdiff}
\end{figure}

\subsection{Idealized Diffusion Test}
\label{advectdiffuse}

We now test a ``diffusion'' problem to validate the diffusive terms in our CR transport implementation, i.e.\ a problem where the gas does not move or respond to CRs (we disable the terms by which the CRs act on the gas), but the CRs are transported ($\tilde{\bf F}_{\rm cr}\ne 0$). The corresponding ``two moment'' equation is 
\begin{align}
\frac{\partial e_{\rm cr}}{\partial t} +\frac{\kappa}{(\gamma_{\rm cr}-1)\tilde{c}^2}\frac{\partial^2 e_{\rm cr}}{\partial t^2}  = \kappa \nabla^2{e}_{\rm cr},
\end{align}
which is a telegraph equation.

We initialize a 1D Gaussian distribution in $e_{\rm cr}^{0} \equiv e_{\rm cr}(r,\,t=0)$, centered at $r=0$, with total energy $E_{\rm cr}=1$, width $\sigma=0.5$\,kpc at the center of a $5\,$-kpc cube with $2048$ resolution elements, set constant $\kappa=3\times10^{27}\,{\rm cm^{2}\,s^{-1}}$, and set $\tilde{c}=100\,{\rm km\,s^{-1}}$ or $500\,{\rm km\,s^{-1}}$, and evolve the system for $5\,$Myr. We do not include magnetic fields so the diffusion is isotropic. Given the symmetry of the problem, this can be solved exactly via the usual separation of modes, giving solutions of the form:
\begin{align}
\label{twomomsol}
e_{\rm cr}({\bf r},\,t) =& \int d{\bf k}\, a^{\pm}_{\bf k}\,\exp{\{i\,[{\bf k}\cdot{\bf r} - \omega^{\pm}_{\bf k}\,t ] \}},
\end{align}
where $\kappa\omega_{\bf k}^{2}/(\gamma_{\rm cr}-1)/\tilde{c}^{2} + i\,\omega_{\bf k} = \kappa\,k^{2}$ and $\omega^\pm_{\bf k}$ are two roots of the previous equation,
\begin{align}
\label{omegacoeff}
\omega^\pm_{\bf k}=\frac{-i\pm\sqrt{4 \kappa^2k^2/(\gamma_{\rm cr}-1)/\tilde{c}^2  -1}}{2 \kappa/(\gamma_{\rm cr}-1)/\tilde{c}^2}.
\end{align}

In our ``diffusion'' test, the initial CR flux is set to zero. Together with the initial CR energy density, we can solve for
\begin{align}
\label{acoeff}
 a^{\pm}_{\bf k} = \frac{1}{(2\pi)^D}\frac{\exp(-\sigma^2k^2/2)}{1-\omega_{\bf k}^\pm/\omega_{\bf k}^\mp},
\end{align}
where D is the dimension of the Gaussian packet. With Eqs. \ref{omegacoeff} and \ref{acoeff}, the time evolution of CR energy density can then be calculated by integrating Eq. \ref{twomomsol} numerically.

This problem is entirely scale-free, and we can transform to solutions with any other value of $\kappa$ via suitable re-scaling. As $\tilde{c}^{2}\,t/\kappa \rightarrow \infty$, the solutions progressively approach the solution of the pure diffusion equation\footnote{Specifically, at this limit, the $a^{-}_{\bf k}$ term becomes exponentially small and the $a^{+}_{\bf k}$ term approaches the ``pure diffusion'' solution. Thus the solution to the two moment equation converges to the ``pure diffusion'' solution insensitive to the initial condition.}, which is:
\begin{align}
e_{\rm CR}({\bf r},\,t,\,\tilde{c}\rightarrow\infty) \rightarrow \frac{\exp\left[ -\frac{r^2}{2(2\kappa t +\sigma^2)} \right]}{[2\pi (2\kappa t+ \sigma^2)]^{3/2}}.
\end{align}

Fig.~\ref{idealdiff} shows the results of our simulation for varying $\tilde{c}$ at fixed $\kappa$ and $t$ (i.e.\ varying the dimensionless parameters $\tilde{c}^{2}\,t/\kappa$ and $r\,\tilde{c}/\kappa$ which determine the behavior of the problem)\footnote{We have turned off the HLL flux in the simulation to avoid small numerical diffusion.}. In all cases, the agreement with the exact solution is excellent, with numerical integration errors less than one percent. This validates that our two-moment implementation correctly solves the desired diffusion problem.

Moreover, Fig.~\ref{idealdiff} also gives us a practical estimate -- for typical units and spatial scales of our simulations -- of the rate at which solutions with lower $\tilde{c}$ converge to the solution with $\tilde{c} = c$ (the speed of light). In Fig.~\ref{idealdiff}, the test with $\tilde{c}=500\,{\rm km\,s^{-1}}$ is already effectively indistinguishable from the ``pure diffusion'' solution. Even at $\tilde{c}=100\,{\rm km\,s^{-1}}$, the solutions differ only by less than $10\%$. 

Of course, in real problems with bulk {\em gas} flows (e.g.\ galactic rotation), such a slow maximum CR transport speed would mean CRs would lag advection, in an unphysical manner, which motivates our additional tests below.

\begin{figure*}
\begin{centering}
\includegraphics[width={0.9\columnwidth}]{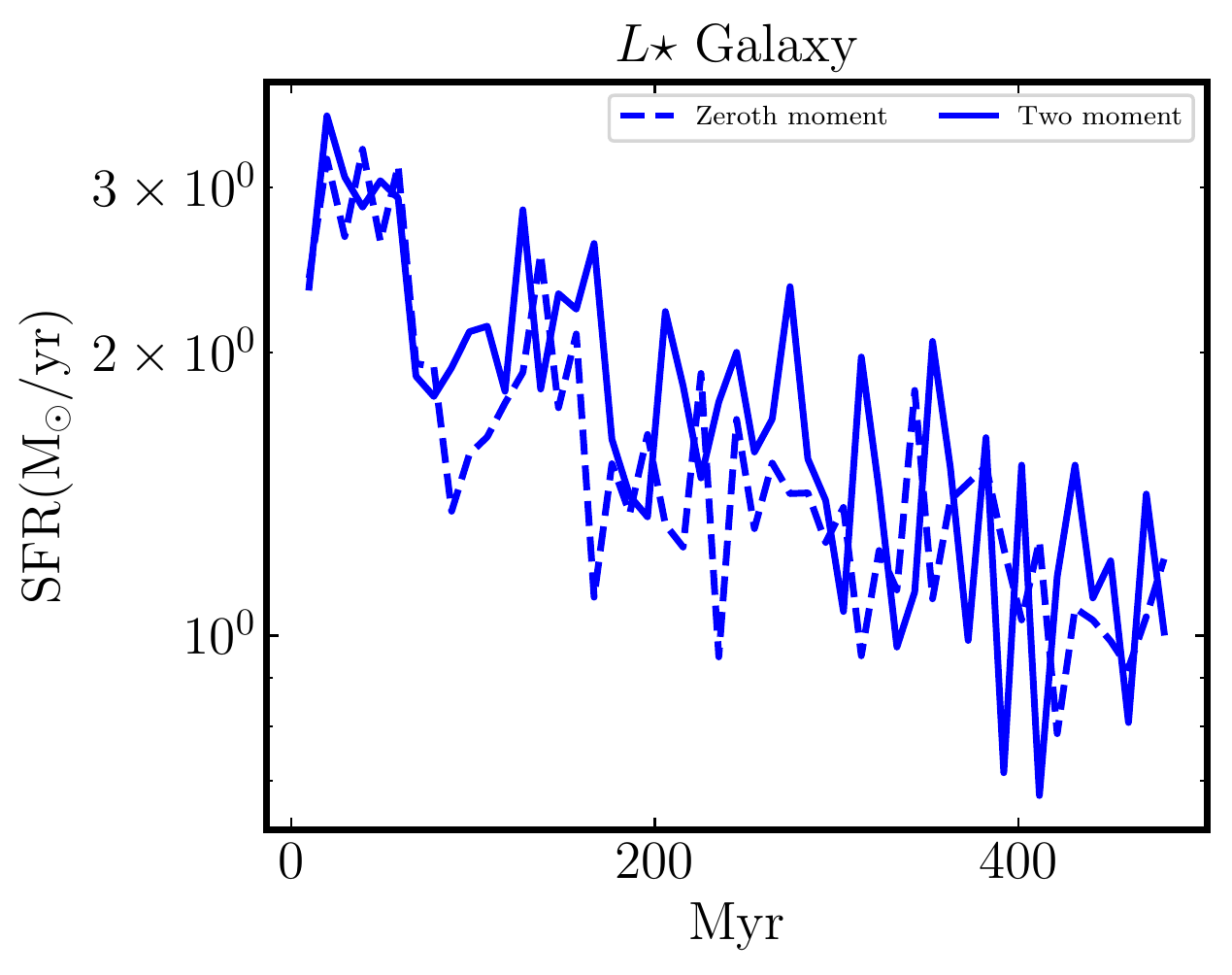}
\includegraphics[width={0.9\columnwidth}]{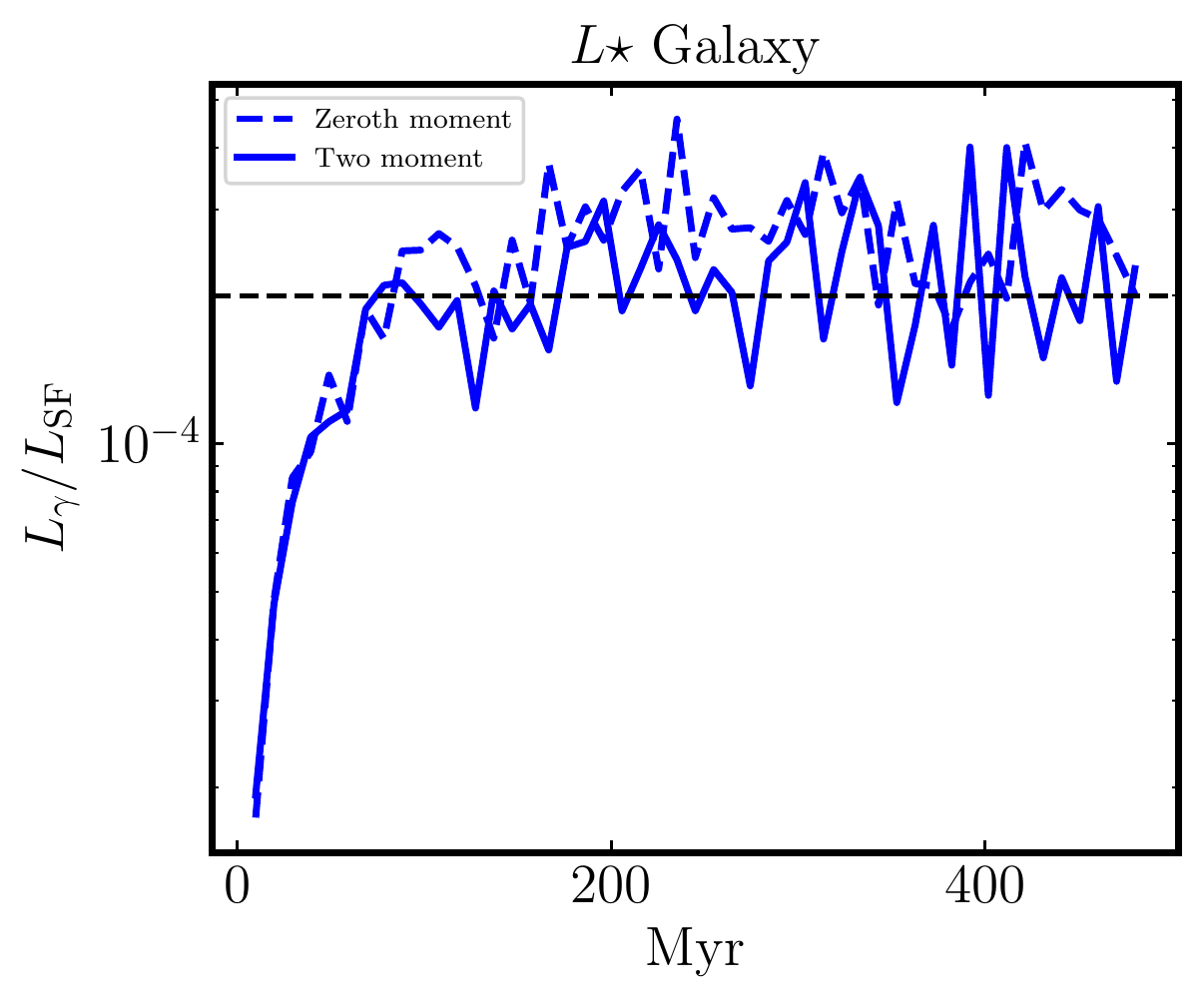}
\end{centering}
\vspace{-0.25cm}
\caption{SFR and $L_{\gamma}/L_{\rm SF}$ as Fig.~\ref{M1speed}, comparing runs with $\kappa=3\times10^{27}\,{\rm cm^{2}\,s^{-1}}$ using either (a) our default two-moment CR transport solver, with $\tilde{c}=400\,{\rm km\,s^{-1}}$, or (b) the ``zeroth-moment'' method (solve a pure single-diffusion equation, i.e.\ fixing ${\bf F}_{\rm cr} = -\kappa\,\nabla e_{\rm cr}$). The ``zeroth-moment'' solution is mathematically equivalent to $\tilde{c}\rightarrow \infty$. We see no meaningful systematic difference (if anything, $L_{\gamma}/L_{\rm SF}$ is slightly higher with $\tilde{c} \rightarrow \infty$, but this owes mostly to stochastic run-to-run variations here).}
\label{diffsol}
\end{figure*}

\begin{figure*}
\centering
\includegraphics[width={1.7\columnwidth}]{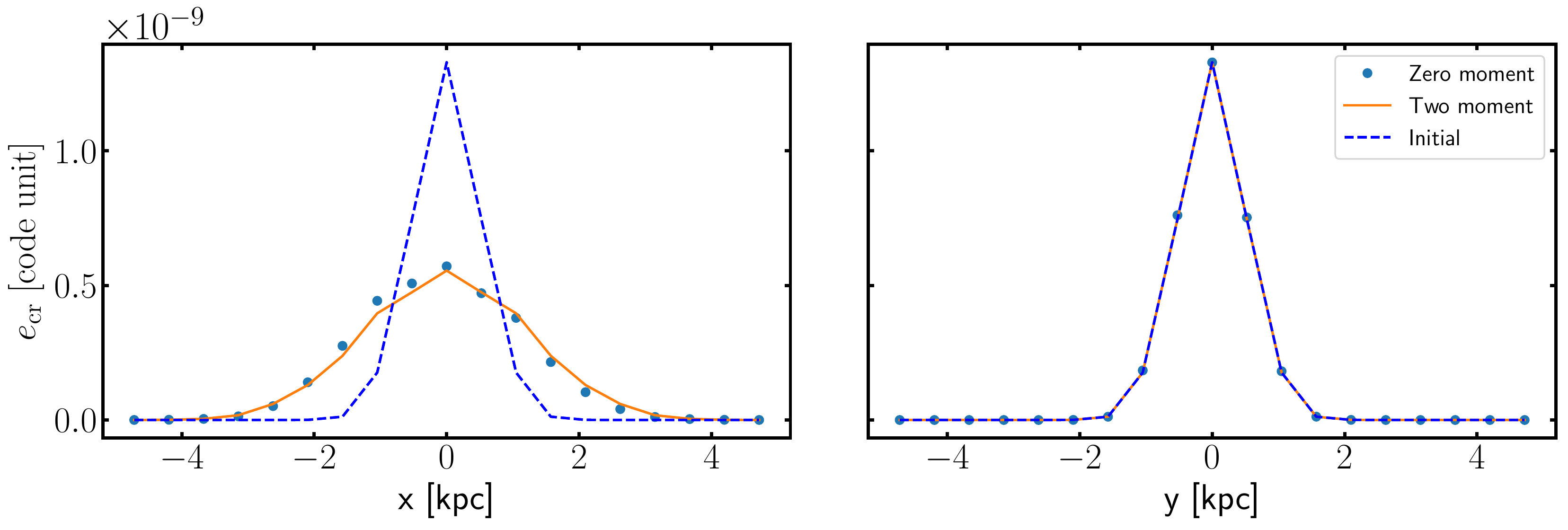}
\vspace{-0.25cm}
\caption{3D anisotropic diffusion tests as described in \S\ref{aniso}. We set a uniform magnetic field along the x direction. We consider CR energy density within thin slices along the z=0 (left) and x=0 (right) planes. CRs can only diffuse along the B field direction but not perpendicular to the B field in both schemes (but see caveats in \S\ref{sec:testanis}). The resultant CR energy densities in the two schemes are in very good agreement.}
\label{anisodiff}
\end{figure*}

\begin{figure*}
\includegraphics[width={1.7\columnwidth}]{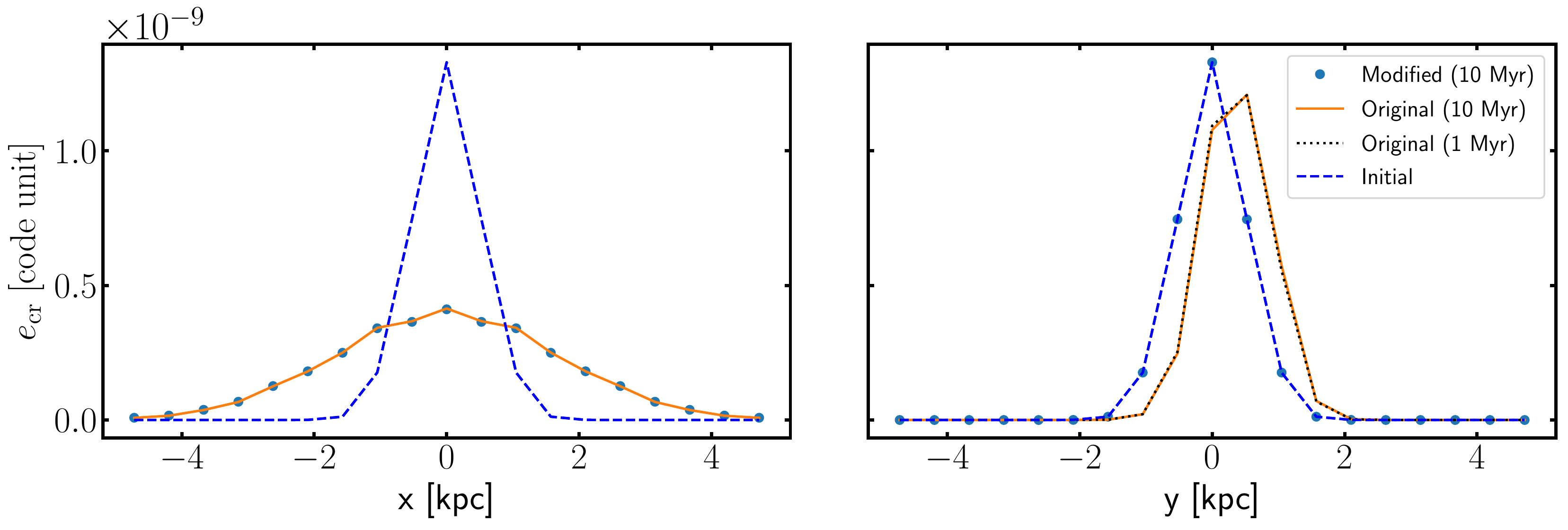}
\vspace{-0.25cm}
\caption{3D anisotropic diffusion tests as described in \S\ref{aniso} and Fig. \ref{anisodiff}, but with an initial CR flux pointing to the positive y direction (perpendicular to the B field, which points along the x direction; see \S \ref{sec:testanis}) and evolved over 10 Myr, except ``Original (1 Myr)'', which evolved over 1 Myr only. Here we consider two implementations of anisotropic diffusion, ``Original'' (Eq. \ref{Fcr}, default case in the main text) and ``Modified'' (Eq. \ref{Fcrll}; see the description in \S \ref{sec:testanis}). We also show the CR energy density (black dotted) at t = 1 Myr with ``Original'' on the right panel. We find that while the CR energy densities in both formulations agree very well along the B field (left panel), CRs can propagate across the B field in ``Original'', but not in ``Modified'' (right panel). However, the perpendicular CR flux is small and vanishes after 1 Myr, as shown by the overlap of the ``Original (10 Myr)'' and ``Original (1 Myr)'' lines. }
\label{anisodiffmod}
\end{figure*}

\begin{figure*}
\includegraphics[width={0.9\columnwidth}]{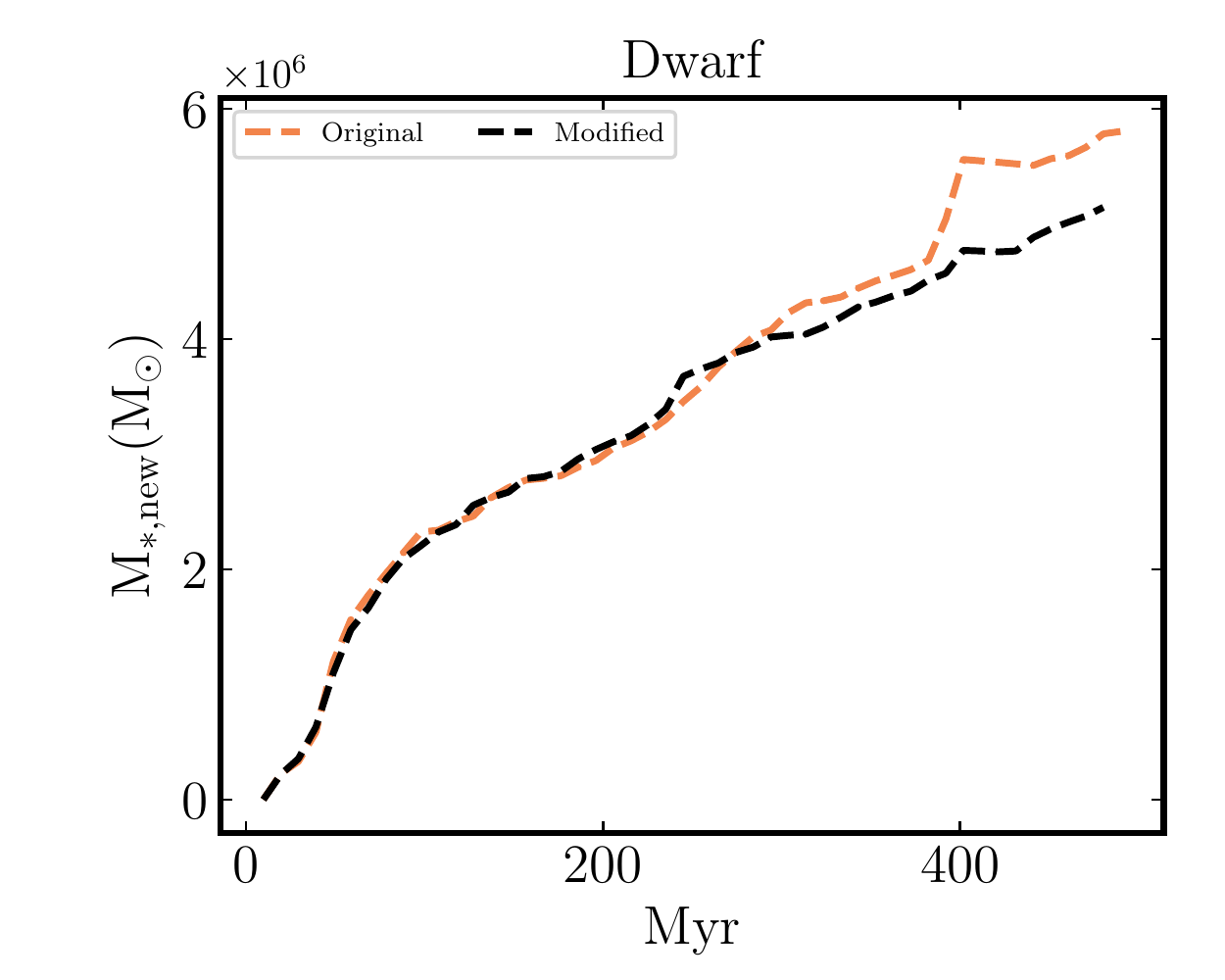}
\includegraphics[width={0.9\columnwidth}]{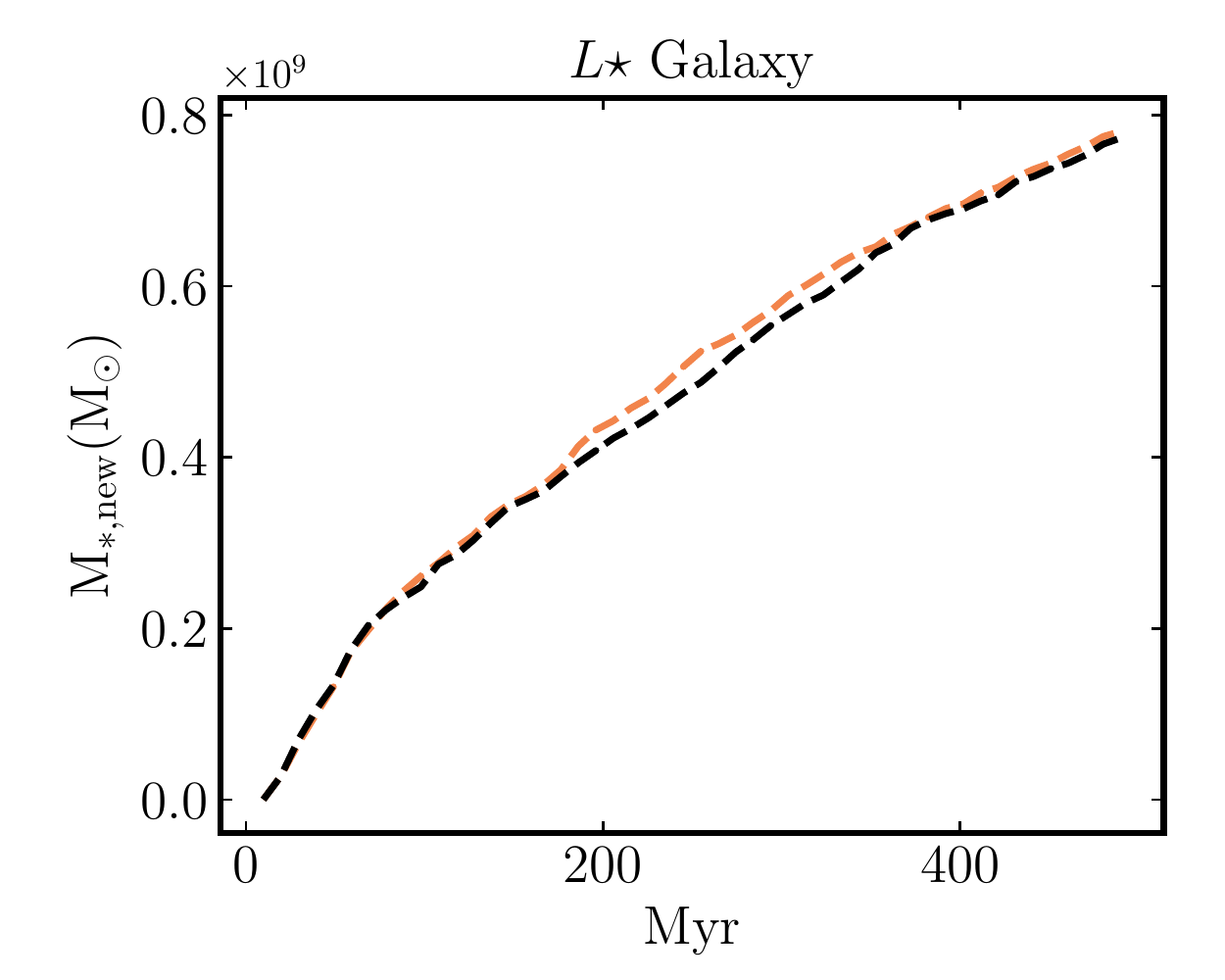}
\includegraphics[width={0.9\columnwidth}]{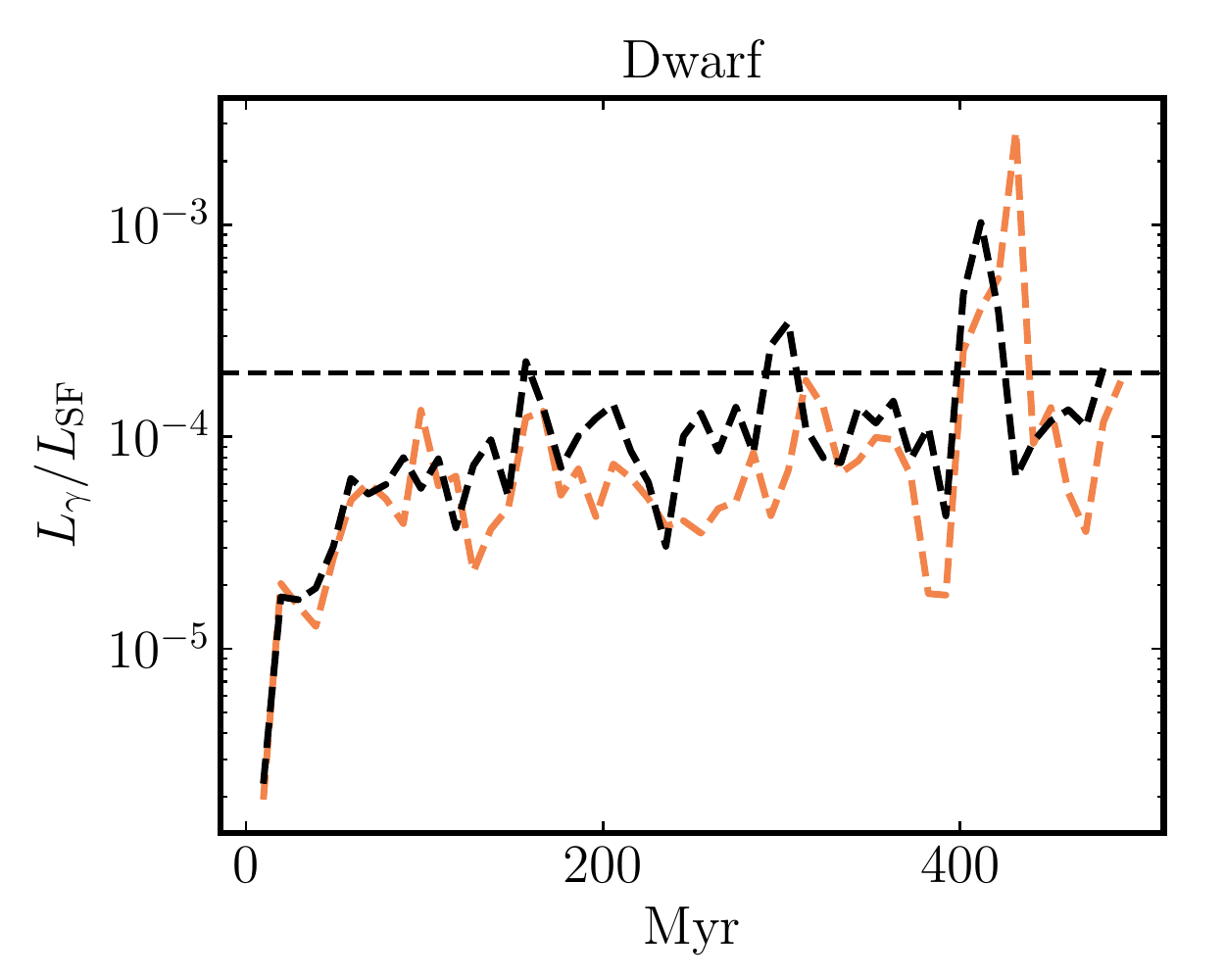}
\includegraphics[width={0.9\columnwidth}]{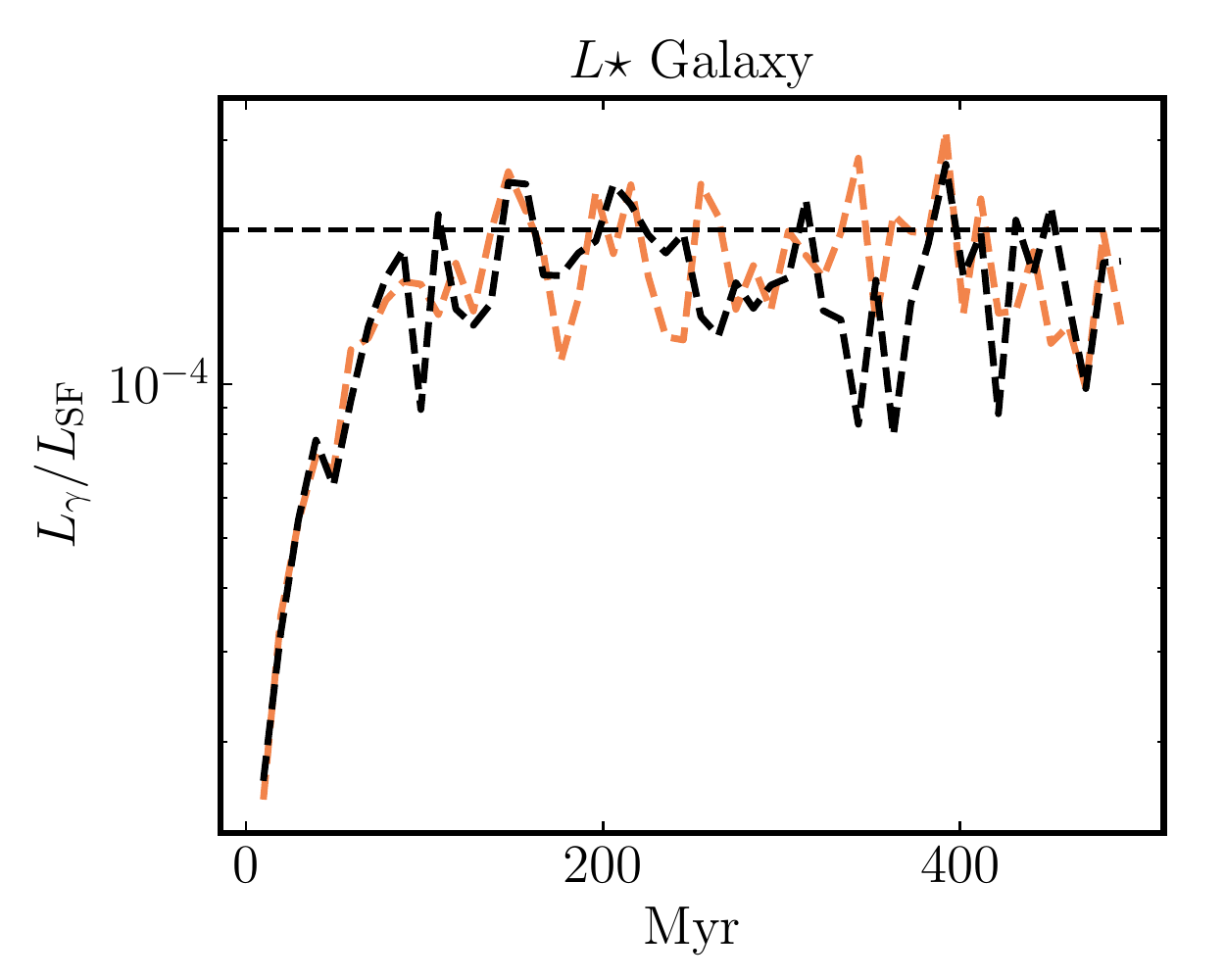}
\vspace{-0.25cm}
\caption{Time evolution of stellar mass accumulation ({\it top}) and ratio between pionic $\gamma$-ray luminosity and the total SF luminosity ({\em bottom}) with the ``Original'' (Eq. \ref{Fcr}) and ``Modified'' (Eq. \ref{Fcrll}) formulations (see \S\ref{sec:testanis}). Two formulations give similar star formation and $L_{\gamma}/L_{\rm SF}$, so the choice between two formulations should have negligible effects on our conclusions.}
\label{refereeplot}
\end{figure*}

\subsection{Comparison of Zeroth and Two Moment Approximations in Galaxy Simulations}
\label{purediff}

In Figure \ref{diffsol}, we compare results of a full galaxy simulation using, instead of our default two-moment expansion (where we explicitly evolve the CR flux $\tilde{\bf F}_{\rm cr}$ with a finite maximum free-streaming speed $\tilde{c}$, as discussed in \S~\ref{method}), the ``zeroth moment'' or ``pure diffusion'' method  (for detailed tests of our pure isotropic/anisotropic diffusion solver, we refer to \citealt{Hopk17diff}). In the equations of motion, the latter simply replaces the explicitly-evolved flux vector with the value $\tilde{\bf F}_{\rm cr} = -\kappa\,\nabla e_{\rm cr}$. This is mathematically equivalent to taking $\tilde{c} \rightarrow \infty$, and imposes a number of numerical difficulties discussed in \S~\ref{method} (not least of which is a much smaller timestep limit). However, Fig.~\ref{diffsol} shows there are only small differences in SFR and $L_{\gamma}/L_{\rm SF}$ between these two approaches. We find the same in all other galaxy properties studied here.

\subsection{Anisotropic Diffusion Test}
\label{aniso}
 We next test our scheme with an intrinsically multi-dimensional anisotropic diffusion test. We consider an initial 3D (256 particles on a side) spherically symmetric Gaussian profile with an anisotropic diffusion coefficient $\kappa=3\times 10^{28}{\rm cm^2/s}$ and a fixed B field pointing in the x direction. We set $\tilde{c}=1000\,{\rm km/s}$, initial ${\bf F}_{\rm cr}=0$, and evolve over 5 Myr with the zeroth moment (``pure diffusion'') and two moment schemes. Fig. \ref{anisodiff} show the resultant CR energy densities through the x or y planes. We found a good agreement between two schemes and that CRs only diffuse along the B field direction (if there is no initial ${\bf F}_{\rm cr}$ perpendicular to the B field; see \S \ref{sec:testanis}). We have also checked that the total CR energy is conserved (with deviations smaller than one in a thousand) in both schemes.

\subsection{Test of a variation in the CR flux equation (Eq. \ref{Fcr})}
\label{sec:testanis}

Eq.~\ref{Fcr} governs the CR flux evolution, and obviously approaches the desired anisotropic diffusion+streaming equation (with $\tilde{\bf F}_{\rm  cr}\rightarrow -\kappa_{\ast}\,\nabla_{\|}e_{\rm cr}$) when $\tilde{c}\rightarrow \infty$ is large, or in local quasi-steady-state ($\partial \tilde{\bf  F}_{\rm cr}/\partial t\rightarrow 0$), or on spatial scales $> \kappa/\tilde{c}$ (timescales $>\kappa/\tilde{c}^{2}$). However, on small spatial/time scales when out-of-equilibrium, small fluxes perpendicular to the magnetic fields can arise from either integration error or rapid small-scale fluctuations in the magnetic field direction\footnote{We thank the anonymous referee pointing out this potential issue.}. If we divide the flux into parallel and perpendicular components $\tilde{\bf F}_{\rm cr} = \tilde{\bf F}_{\rm cr,\|} + \tilde{\bf F}_{\rm cr,\perp}$, then we see the perpendicular component obeys $\tilde{c}^{-2}\,\partial \tilde{\bf F}_{\rm cr,\perp}/\partial t = -[(\gamma_{\rm cr}-1)/\kappa_{\ast}]\,\tilde{\bf F}_{\rm cr,\perp}$ (taking ${\bf v}=0$, for simplicity, though this  does not change our conclusions below). So an initial $\tilde{\bf F}_{\rm cr,\perp}$ can formally propagate but will be exponentially damped on a timescale $\sim \kappa_{\ast}/\tilde{c}^{2}$. For $\kappa^*$ needed to match the $\gamma$ ray luminosity of nearby galaxies ($\sim 10^{29}$) and $\tilde{c}\sim 1000\,{\rm km/s}$, this timescale is $\sim 0.2$\,Myr, much shorter than the hadronic interaction time in the MW's ISM ($n_{\rm ISM}\sim 1\,{\rm cm^{-3}}$), so our $\gamma$ ray constraint will not be affected with or without $\tilde{\bf F}_{\rm cr,\perp }$ (well within the uncertainties). Likewise since this is much shorter than relevant dynamical times, essentially no effects of CR pressure and other dynamics should be altered.

If desired, we can trivially set $\tilde{\bf F}_{\rm cr,\perp}=0$ every timestep, and evolve only the parallel component of Eq.~\ref{Fcr}: 
\begin{align}
\frac{1}{\tilde{c}^{2}}\left[ \frac{\partial\tilde{\bf F}_{\rm cr}}{\partial t} + \nabla\cdot({\bf v}\otimes\tilde{\bf F}_{\rm cr}) \right]_{\|} + \nabla_{\|} P_{\rm cr}  = -\frac{(\gamma_{\rm cr}-1)}{\kappa^{\ast}}\,\tilde{\bf F}_{\rm cr,\|}
\label{Fcrll}
\end{align}
where ${\bf X}_{\|}\equiv (\hat{\bf B}\otimes\hat{\bf B})\cdot{\bf X}$. We have implemented Eq.~\ref{Fcrll} and compare it (in Fig.~\ref{anisodiffmod}) to our default Eq.~\ref{Fcr} in an anisotropic  diffusion test (as \S~\ref{aniso}) with an initial (intentionally erroneous) pure-perpendicular flux $\tilde{\bf F}_{\rm cr} = \tilde{\bf  F}_{\rm  cr,\perp} = (2000\,{\rm km/s})e_{\rm cr} \hat{y}$. This shows that the ``original'' Eq.~\ref{Fcr} does formally allow CRs to propagate perpendicular to the B field, although the effect is rapidly damped, while the ``modified'' Eq.~\ref{Fcrll} does not.
 
Fig.~\ref{refereeplot} compares two galaxy simulations ({\bf Dwarf} and {\bf $L\star$ galaxy}, with the latter at one level lower resolution than the fiducial  main-text case) run with MHD (anisotropic diffusion), $\kappa_\|$=3e28, and either Eq.~\ref{Fcr} or Eq.~\ref{Fcrll}. We find essentially no difference in their star  formation rates or gamma-ray luminosities (apart from a small stochastic deviation associated with one slightly-stronger burst in the dwarf run).  
This is expected given the arguments above. Moreover, note that all the differences between Eq.~\ref{Fcr} and Eq.~\ref{Fcrll} appear  on  scales $\ll \kappa/\tilde{c}$; but \S~\ref{purediff} showed that taking $\tilde{c}\rightarrow \infty$ does not change our conclusions.

Of course, CRs {\it can} physically propagate across mean magnetic fields due to unresolved (micro-scale) magnetic turbulence \citep{Zwei17,Farb18}, so a small $\tilde{\bf F}_{\rm cr,\perp }$ might not be unreasonable. But in the main text, we show that even much larger perpendicular diffusivities given  by assuming {\em isotropic} diffusion (much  larger than the fluxes that arise from  the numerical or physical effects described above) do not strongly alter our conclusions regarding either the effects of CRs or the observationally-favored effective $\kappa$.

It is worth noting that several  recent studies including this one and \citet{Zwei17}, \citet{Jian18}, and \citet{Thom19} have adopted slightly-different forms of the CR flux equation (although the CR energy equation is consistent in all cases). If we impose some desired scattering ($\kappa_{\ast}$), all of these can be written (after some algebra) in the form:
\begin{align}
    \frac{1}{\tilde{c}^{2}}\,\mathbb{D}_{t}\tilde{\bf F}_{\rm cr} + \nabla_{\|}P_{\rm cr} &=  -\frac{(\gamma_{\rm cr}-1)}{\kappa_{\ast}}\,\tilde{\bf F}_{\rm cr},
\end{align}
where the differences are contained in the operator $\mathbb{D}_{t}$.  For our default Eq.~\ref{Fcr}, $\mathbb{D}_{t}{\tilde{\mathbf{F}}_{\rm cr} } = \partial  {\tilde{\mathbf{F}}_{\rm cr} }/\partial t + \nabla\cdot ({\bf v}\otimes{ \tilde{\mathbf{F}}_{\rm cr}})$.  For the zeroth-moment or ``pure diffusion+streaming'' equation in \S~\ref{purediff}, $\mathbb{D}_{t}{\tilde{\mathbf{F}}_{\rm cr} }=\mathbf{0}$. For \citet{Jian18}, $\mathbb{D}_{t}{\tilde{\mathbf{F}}_{\rm cr} } = [\partial\{ {\tilde{\mathbf{F}}_{\rm cr} }+{\bf v}\,(e_{\rm cr}+P_{\rm cr}) \}/\partial t]_{\|}$, and for \citet{Thom19}, $\mathbb{D}_{t}\tilde{\mathbf{F}}_{\rm cr}  = \hat{\tilde{\mathbf{F}}}_{\rm cr} \,[\partial|{\tilde{\mathbf{F}}_{\rm cr} }|/\partial t + \nabla\cdot({\bf v}\,|{\tilde{\mathbf{F}}_{\rm cr} }|) + {\tilde{\mathbf{F}}_{\rm cr} }\cdot\{(\hat{\tilde{\mathbf{F}}}_{\rm cr} \cdot\nabla){\bf v} \}]   =  \partial  {\tilde{\mathbf{F}}_{\rm cr} }/\partial t + \nabla\cdot ({\bf v}\otimes\tilde{\mathbf{F}}_{\rm cr} ) + ({\tilde{\mathbf{F}}_{\rm cr} }\cdot\nabla)\,({\bf v}_{\|}-{\bf v}_{\perp})$, with $\tilde{\bf F}_{\rm cr,\perp}=\mathbf{0}.$\footnote{For \citet{Jian18}, we have formally taken the limit of the diffusion/interaction tensor $\mathbb{K} \equiv \kappa_{\ast}\,(\hat{\bf B}\otimes\hat{\bf B}) + \kappa_{\perp}\,(\mathbb{I} - \hat{\bf B}\otimes\hat{\bf B})$ ($\mathbb{I}$ is the identity tensor) as $\kappa_{\perp} \rightarrow 0$. For \citet{Thom19}, we use $\tilde{\mathbf{F}}_{\rm cr} =|\tilde{\mathbf{F}}_{\rm cr} |\,\hat{\tilde{\mathbf{F}}}_{\rm cr} =f_{\rm cr}\,\hat{\bf B}$ in their notation, and replace their scattering term $\partial f/\partial t|_{\rm scatt}$ (which explicitly attempts to account for dynamically-evolving, forward-and-backward propagating gyro-scale Alfven waves) with $-[(\gamma_{\rm cr}-1)/\kappa_{\ast}] \tilde{\bf F}_{\rm cr}$ for a locally-stationary $\kappa_{\ast}$.}

What is important for our purposes here is to note that all of these expressions differ only inside the term suppressed by $\tilde{c}^{-2}$. This means they all relax to the same diffusion+streaming equation in quasi-steady state and/or as $\tilde{c}\rightarrow\infty$, and differ only on scales below the CR mean  free path  $\sim \kappa/\tilde{c}$. To the extent that our galaxy-scale results are converged with  respect  to the value of $\tilde{c}$, and are not changed if we take $\tilde{c}\rightarrow\infty$ (\S~\ref{purediff}), these differences in $\mathbb{D}_{t}\tilde{\bf F}_{\rm cr}$ cannot alter  our conclusions. Moreover, if  we adopt a ``reduced speed of light'' $\tilde{c}\ll  c$, then by definition the flux equation does not exactly represent reality on scales $\ll \kappa/\tilde{c}$, {\em regardless}  of the form of $\mathbb{D}_{t}$ (i.e.\ the regime where these expressions differ is exactly where they all  become inexact when $\tilde{c} \ll c$ is adopted). We have explicitly verified that our conclusions are robust to these choices of flux equation, running limited galaxy-scale simulations with the forms of  $\mathbb{D}_{t}\tilde{\bf F}_{\rm cr}$ from \citet{Jian18} or \citet{Thom19} as defined above: the results are similar to those  in Fig.~\ref{refereeplot}.

\begin{figure}
\begin{centering}
\includegraphics[width={0.9\columnwidth}]{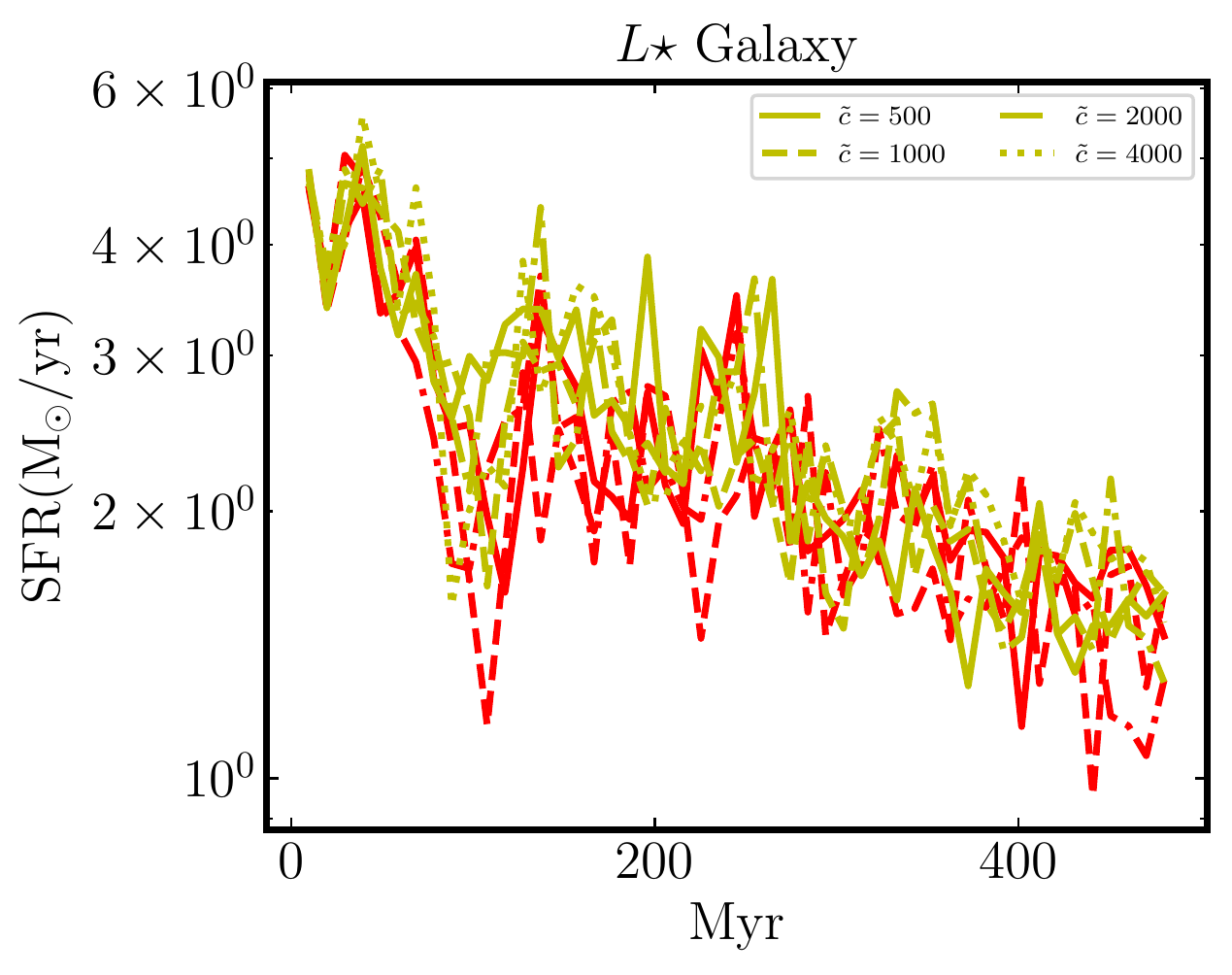}
\includegraphics[width={0.9\columnwidth}]{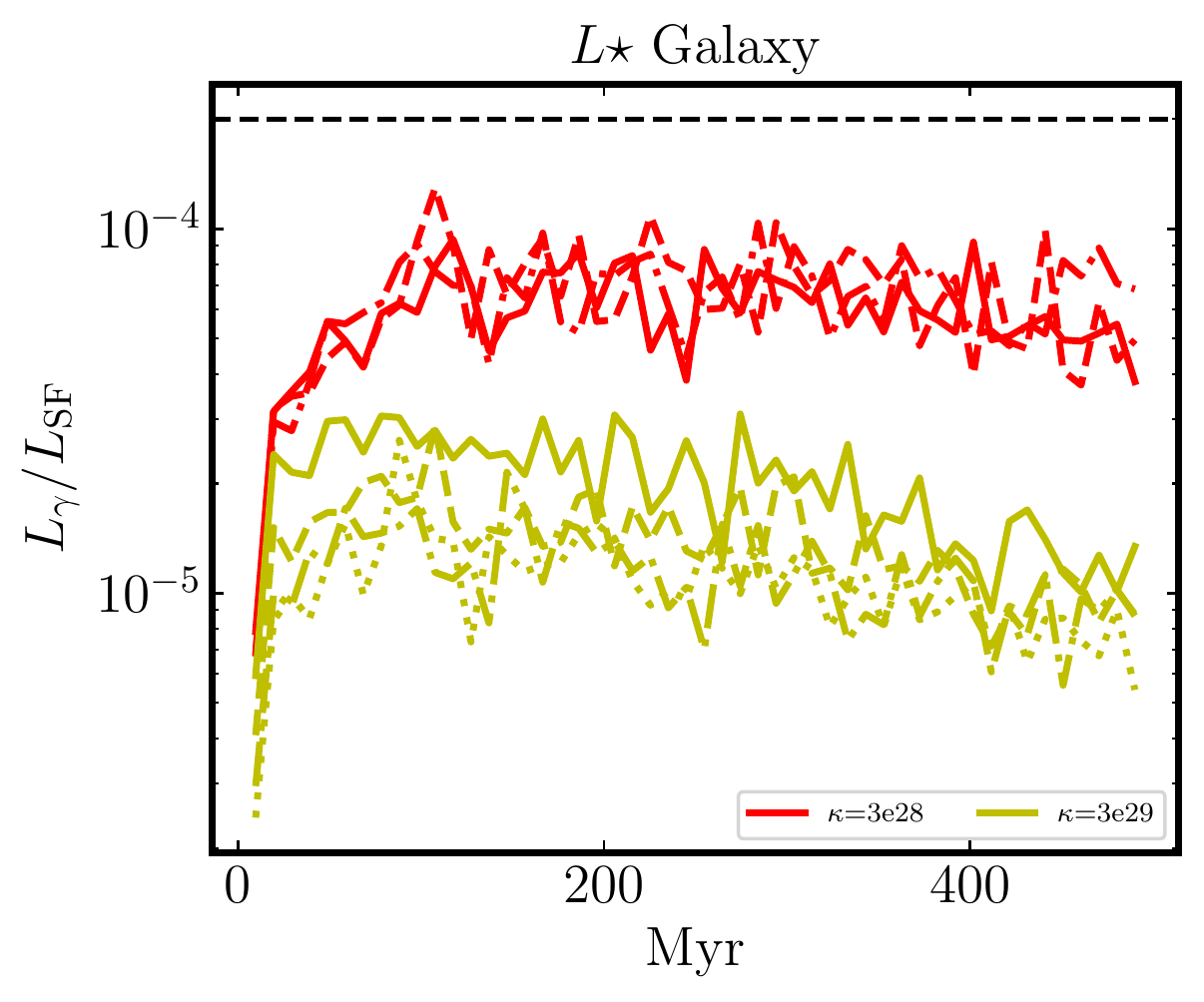}
\end{centering}
\vspace{-0.25cm}
\caption{Time evolution of SFR ({\em top}, as Fig.~\ref{sfr}) and ratio between pionic $\gamma$-ray and total SF luminosity ({\em bottom}, as Fig.~\ref{Fgammasfr}, with different ``maximum CR free-streaming speeds'' $\tilde{c}$ (equivalently, ``reduced speed of light'' for CRs) in ${\rm km\,s^{-1}}$, and different $\kappa$ (see \S~\ref{reducedSOL}). We show the $L\star$ galaxy without magnetic fields or streaming (dependence on $\tilde{c}$ is smaller with these added). One $\tilde{c}$ is faster than bulk transport and diffusive CR escape velocities from the disk, the results should be independent of it, and we confirm this. For $\kappa<10^{29}\,{\rm cm^{2}\,s^{-1}}$, we see no systematic differences for any $\tilde{c} \gtrsim 500\,{\rm km\,s^{-1}}$. For larger $\kappa$ these same values produce no detectable difference in galaxy properties, but the ``slowest'' ($\tilde{c}\sim 500\,{\rm km\,s^{-1}}$) produces a slightly larger (factor $\sim 1.4$) $L_{\gamma}/L_{\rm SF}$ owing to slightly slower CR escape.
\label{M1speed}}
\end{figure}

\subsection{Comparison of Different (Finite) Maximum CR Propagation Speeds in Galaxy Simulations}
\label{reducedSOL}

We now examine the effect of varying $\tilde{c}$ on a full simulation (our $L\star$ model). For $\tilde{c} \gtrsim 500\,{\rm km\,s^{-1}}$, which is generally faster than the bulk rotation and outflow speeds (at least those containing most of the gas) in the galaxies, Fig.~\ref{M1speed} shows there is a small impact of $\tilde{c}$ on the SFR (we find the same for all other galaxy properties, not shown here). For $L_{\gamma}/L_{\rm SF}$, we find almost no impact of $\tilde{c}$ in simulations with $\kappa< 10^{29} {\rm cm^2/s}$ for any values $\tilde{c} \gtrsim 500\,{\rm km\,s^{-1}}$.

Because the effective ``advective velocity'' of CRs under pure diffusion is $\sim \kappa / \ell $ (where $\ell $ is some gradient scale-length), at much larger diffusion coefficients, e.g.\ our $\kappa=3\times10^{29}\,{\rm cm^{2}\,s^{-1}}$, where {\em most} of the CRs escape diffusively, the value of $L_{\gamma}$ is slightly larger for $\tilde{c}=500\,{\rm km\,s^{-1}}$ compared to much-higher $\tilde{c}$ (because escape is slightly slower). But once $\tilde{c} \ge 1000\,{\rm km\,s^{-1}}$, we see no detectable difference. Moreover by the latter half of the time we run for, the differences even for $\tilde{c}= 500\,{\rm km\,s^{-1}}$ (compared to $\tilde{c}\sim 4000\,{\rm km\,s^{-1}}$) at $\kappa=3\times10^{29}\,{\rm cm^{2}\,s^{-1}}$ are factors of $\sim 1.4$, not large enough to change any of our conclusions.

In tests run for shorter duration and tests of our {\bf Dwarf} galaxy (not shown), we have also verified similar conclusions, and found that runs with magnetic fields (since these slow down the transport) and finite streaming velocities (since these dominate the transport over diffusion in some regimes) exhibit even weaker dependence on $\tilde{c}$ within the range $\tilde{c} \sim 500-4000\,{\rm km\,s^{-1}}$, even at $\kappa=3\times10^{29}\,{\rm cm^{2}\,s^{-1}}$.

\subsection{Idealized Streaming Test}
\label{idealstreamtest}
Here we test our streaming implementation with an initial 1D symmetric triangular CR profile with $e_{\rm cr} = 2-|x|$ (and ${\bf F}_{\rm cr}={\bf v}_{\rm str}e_{\rm cr}$), where we disable the gas motion and CR streaming loss ($v_{\rm st}\cdot \nabla P_{\rm cr}$), but CRs can move across gas particles with  a constant $v_{\rm st}=1\; {\rm km/s}$. In the numerical test, we consider evenly spaced 2048 particles over 10 kpc, the reduced speed of light  $\tilde{c}=1000 \,\rm{km/s}$, and a small diffusion coefficient $3\times 10^{22}\;{\rm cm^2/s}$.\footnote{We include a very small diffusion coefficient to avoid potential (numerical) overflows in Eq. \ref{Fcr}, i.e. when the CR pressure profile is flat or, equivalently, $\nabla P_{\rm cr}$ is huge.} Here, we also consider a variant test  in which we relax the limiter on CR pressure gradient\footnote{The limiter is necessary to avoid divergence and other numerical issues in realistic galaxy simulations.} in order to correctly capture the plateau region and also turn off the HLL flux to avoid numerical diffusion smoothing the discontinuity (so  we can  cleanly separate the effects of numerical dissipation, which are resolution-dependent, and the effects of the actual form of the flux equation and its implementation of streaming).\footnote{Because the exact solution to this test contains dis-continuous first derivatives, any numerically stable method will introduce some dissipation at the cusps, even at infinite resolution.} We also show the ``original'' test run with the original limiter and the HLL flux included, to illustrate the version that we used in the main text. 

\begin{figure}
\begin{centering}
\includegraphics[width={0.9\columnwidth}]{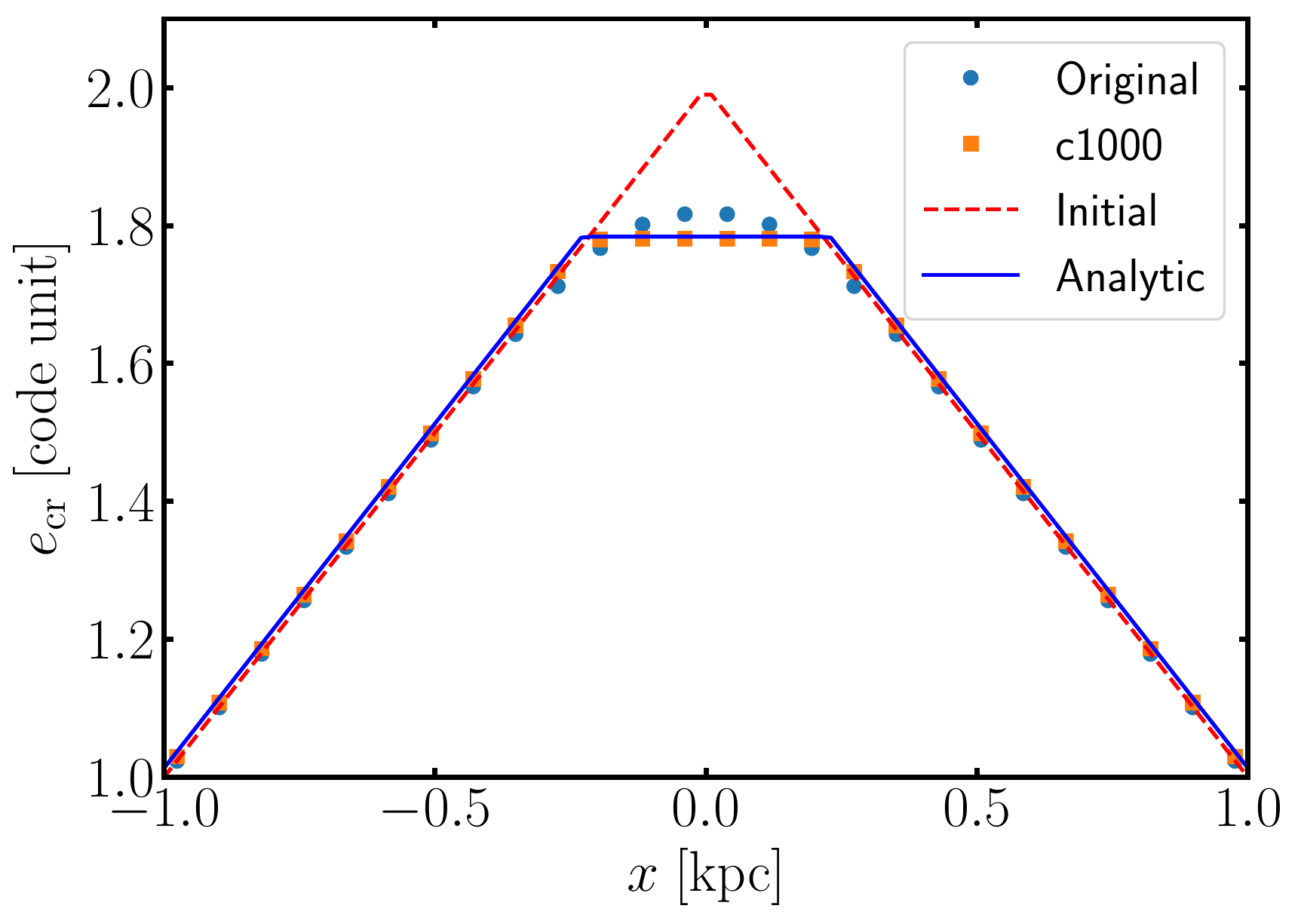}
\end{centering}
\vspace{-0.25cm}
\caption{Idealized 1D CR streaming test (\S~\ref{idealstreamtest}) with an initial triangular CR distribution evolved over 10 Myr. Red and blue lines show the analytic solutions at t = 0 Myr and t = 10 Myr respectively. Orange squares (blue circles) show the numerical results with (without) CR pressure gradient limiters and HLL flux. The analytic and numerical results agree well in overall shapes.}
\label{idealstream}
\end{figure}

The analytic solution, calculated in \citet{Jian18}, has a flat CR distribution between $x = \pm x_{\rm m}$ and two inclined distributions for $|x| > x_{\rm m}\; \& \; |x| < 1$  with
\begin{align}
e_{\rm cr}(x, t) = 2 + 4v_{\rm st}t-|x|,
\end{align}
and $x_{\rm m}$ is determined from the energy conservation:
\begin{align}
x_{\rm m} = \sqrt{\left ( 1+\frac{4}{3}v_{\rm st}t \right )^2+\frac{8v_{\rm st}t}{3}-1}.
\end{align}

Fig. \ref{idealstream} shows the analytic solution agrees well with our numerical solution at $t = 10\;{\rm Myr}$, although the ``original'' run has a round top rather than a plateau mainly because of the limiter on the CR pressure gradient.

\subsection{Resolution study}
\label{resostudy}
\begin{figure}
\begin{centering}
\includegraphics[width={0.9\columnwidth}]{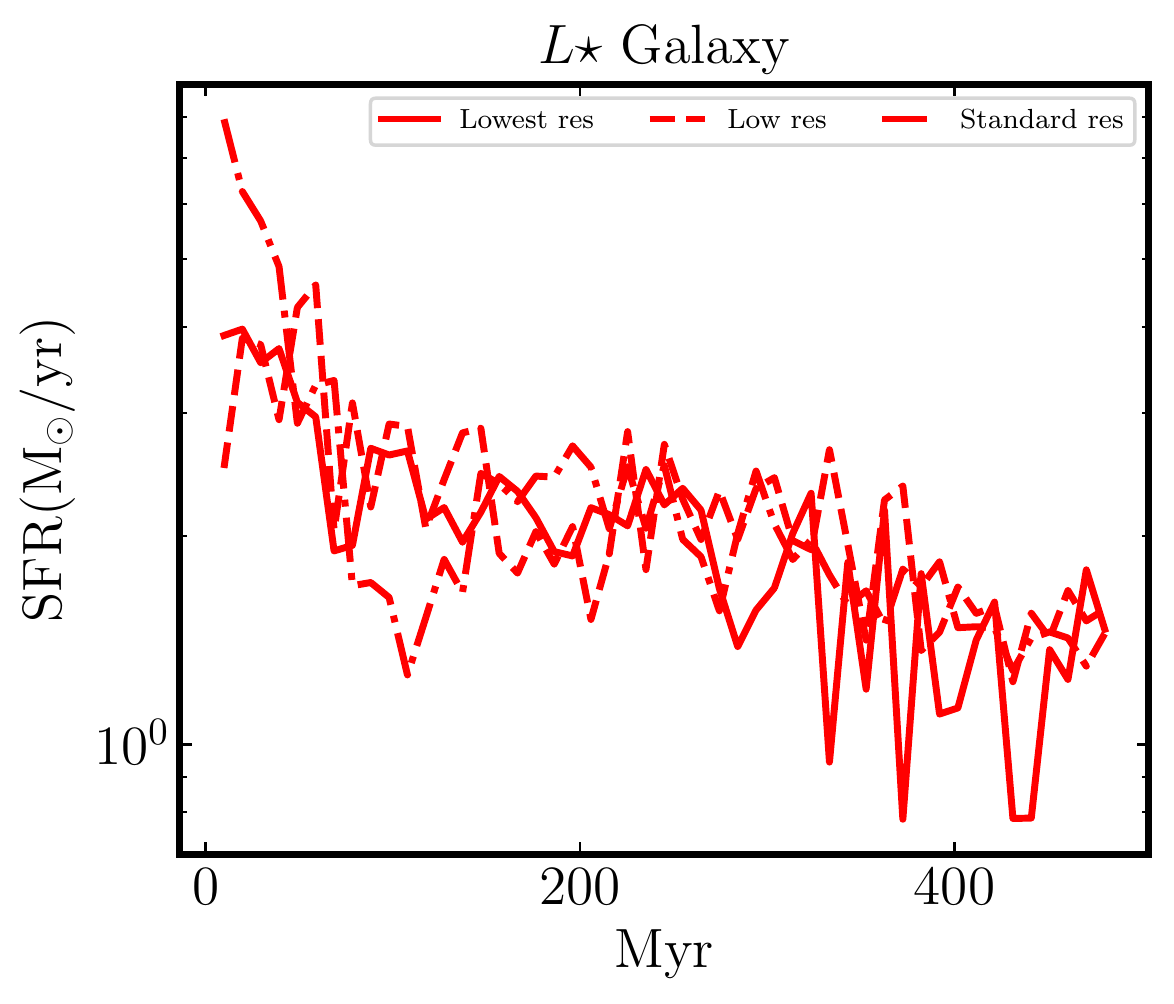}
\includegraphics[width={0.9\columnwidth}]{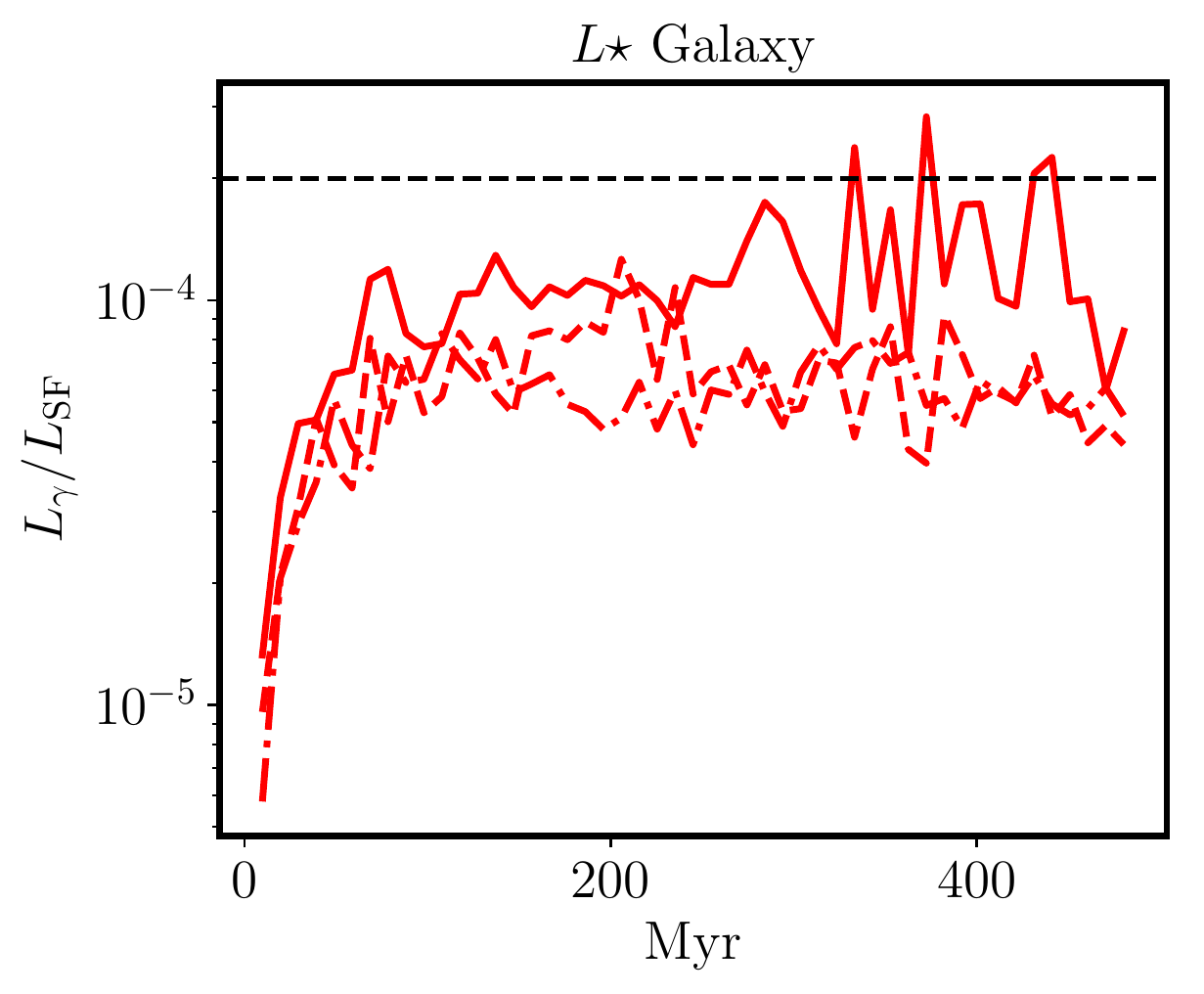}
\end{centering}
\vspace{-0.25cm}
\caption{SFR and $L_{\gamma}/L_{\rm SF}$ as Fig.~\ref{M1speed}, in a resolution survey. We consider runs without magnetic fields, with $\kappa=3\times 10^{28}{\rm cm^2\,s^{-1}}$. ``Standard'' (dash-dotted) is the resolution used in the main text, ``Low'' (dashed) is 10x poorer mass resolution, and ``Lowest'' (solid) is 100x poorer. Between ``low'' and ``standard'' resolution we see no difference in any property studied. Even at ``lowest'' resolution our qualitative conclusions are similar, although the artificially poorly-resolved ISM leads to noticeable biases in e.g.\ $L_{\gamma}/L_{\rm SF}$.}
\label{reso}
\end{figure}

In Fig.~\ref{reso}, we show the properties of {\bf $L\star$ Galaxy} runs, at three different resolution levels. Our baseline is the fiducial resolution listed in Table \ref{SIC}, but we compare runs with 10x and 100x poorer mass resolution. Even at 10x poorer resolution, we find very similar SFRs, $L_{\gamma}/L_{\rm SF}$, and all other galaxy properties studied here; the same is true in the {\bf Dwarf} runs (not shown). However systematic offsets do begin to appear at 100x poorer resolution. Other more detailed properties (e.g.\ phase structure of galactic winds) may require much higher resolution - this will be explored in future work.

\section{Comparison with $L_{\rm 0.1-100GeV}$ observations}

\label{Fermicom}
For completeness, we also compare our simulations with the broader, 0.1-100 GeV, observed energy range of $\gamma$ rays from \citet{Acke12} and \citet{Roja16}.

In the main text, we estimate the $L_{\rm > 1 GeV}$ following the L11 approach with the hadronic loss rate taken from \citet{Guo08} (see Eq. \ref{eqlackigamma} and the related text), with an implicit assumption that all CR energy is at > 1GeV to simplify the calculation. However, to properly account for the lower energy $\gamma$ rays and enable direct comparison to both the observations and simulations from \cite{Pfro17gamma}, we follow \cite{Pfro17gamma} to calculate  $L_{\rm 0.1-100 GeV}$.

\begin{figure}
\begin{centering}
\includegraphics[width={0.90\columnwidth}]{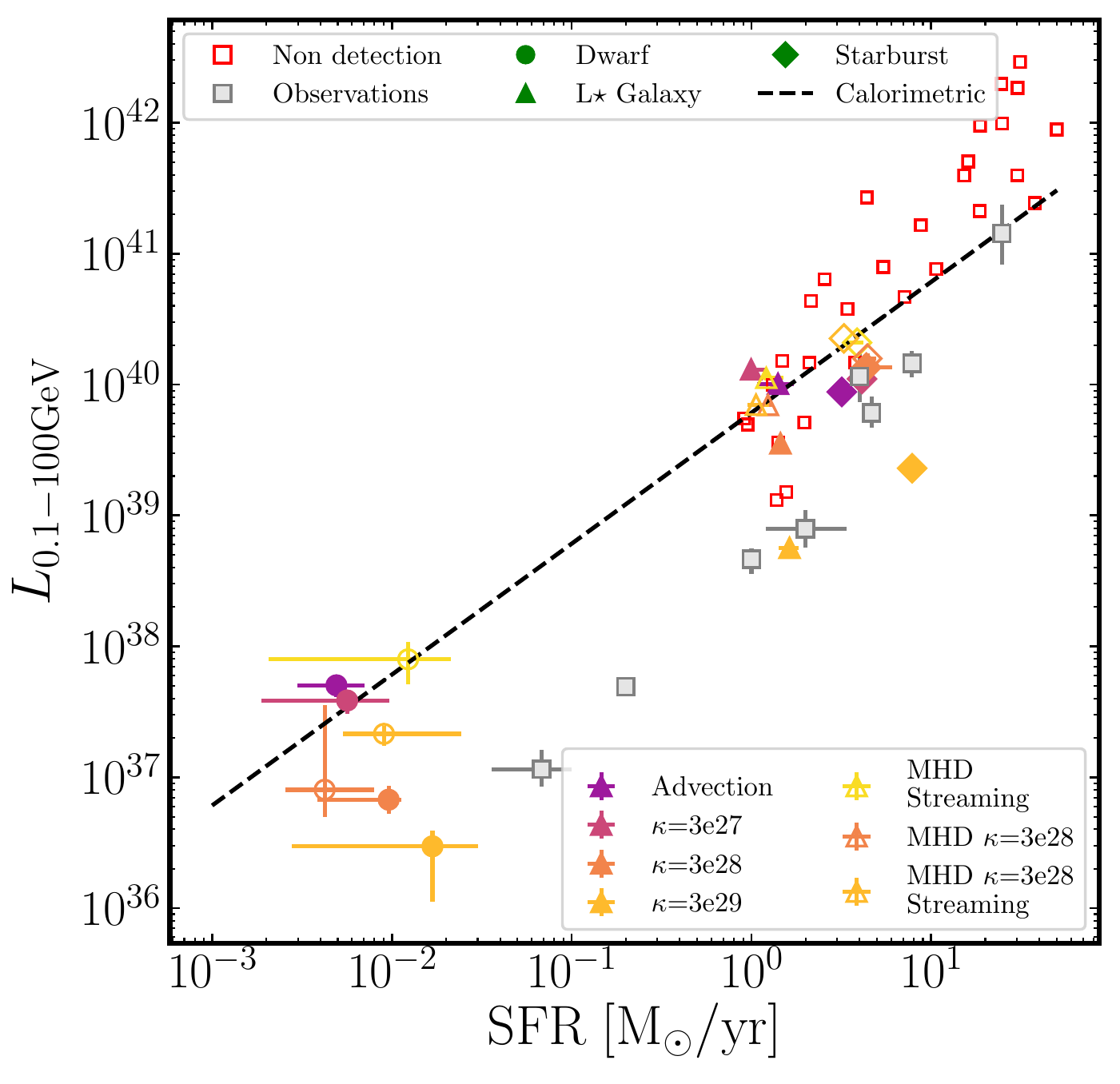}
\end{centering}
\caption{$\gamma$-ray luminosity $L_{\rm 0.1-100 GeV}$ ($0.1 {\rm GeV}<E_{\gamma}< 100 {\rm GeV}$) vs SFR (averaged over $\sim 10\,$Myr) from our simulations, compared to observations. Grey squares show the observed $L_{\rm 0.1-100 GeV}$ from \citet{Acke12} but we use the SFRs described in \S\ref{comobs}. Red empty squares show the upper limits of non detection of the $\gamma$ ray fluxes by Fermi LAT (galaxies without AGN), calculated in \citet{Roja16}. Their SFRs were estimated with the Chabrier IMF, so we convert them assuming the Kroupa IMF to be consistent with our simulations. For the {\bf Starburst} models we restrict to times ``during starburst'' (SFR $>3\,\msun\,{\rm yr^{-1}}$) and take 5-Myr averaged SFR.}
\label{SFRLgfermi}
\end{figure}

We assume that CRs follow a power-law spectrum with a constant spectral index 2.2 (as used in the main text) and the low momentum cutoff $q = P_p/(m_pc)=1$, where $m_p$ is the proton mass. The normalization is determined by integrating the spectrum over energy and comparing to the local CR energy density. Then we calculate the energy-integrated $\gamma$-ray emissivity from pion decay for the energy band 0.1-100 GeV with Eq. 6 in \cite{Pfro17gamma}, and then integrate to get $L_{\rm 0.1-100 GeV}$. While the secondary IC emission from high energy CR electrons also contributes to $\gamma$-ray luminosity in this range, we find it is relatively unimportant compared to emission from pion decay, since (1) in high B field regions, CR electrons cool preferentially through synchrotron radiation; (2) in low B field regions, IC luminosity is only a small fraction (< 20\%) of the $\gamma$-ray luminosity. Thus,
we simply neglect its contribution in Fig.\ref{SFRLgfermi}. 

Fig. \ref{SFRLgfermi} shows that $\kappa\sim 1{\rm e}29-3{\rm e}29\;{\rm cm^2/s}$ provides the best match to the observations for {\bf Dwarf} and {\bf L$\star$ Galaxy} runs,
consistent with our conclusions in the main text.

\bsp
\label{lastpage}

\end{document}